\documentclass[final,12pt]{elsarticle}
\usepackage[margin=1in]{geometry}

\usepackage{hyperref}
\usepackage{mathtools}
\usepackage{amsmath}
\usepackage{amssymb}
\usepackage{amsbsy}
\usepackage{bm}
\usepackage{textcomp}
\usepackage{xcolor}
\usepackage{fancyhdr}
\usepackage{float}
\usepackage{dsfont}
\usepackage[normalem]{ulem}

\newcommand{\sixj}[6]{\left\{\begin{array}{ccc} #1 & #2 & #3 \\ #4 & #5 & #6 \\ \end{array}\right\}}

\newcommand{\half}{\frac{1}{2}}

\newcommand{\SU}[1]{\ensuremath{\mathrm{SU}( #1 )}}

\newcommand{\SpR}[1]{\ensuremath{\mathrm{Sp}( #1,\mathbb{R} )}}

\newcommand{\adagg}{\ensuremath{ {a^\dagger} }}

\newcommand{\RedCG}[3]{\ensuremath{\langle#1;#2\|#3\rangle}}
\newcommand{\Wigsixj}[6]{\ensuremath{\left\{\begin{matrix}#1 & #2 & #3\cr
#4 & #5 & #6 \end{matrix}\right\}}}
\newcommand{\Wigninej}[9]{\ensuremath{\left\{\begin{matrix}#1 & #2 & #3\cr
#4 & #5 & #6 \cr #7 & #8 & #9 \end{matrix}\right\}}}

\newcommand{\braketop}[3]{\ensuremath{\left\langle #1 \right| #2 \left| #3 \right\rangle}}

\newcommand{\RedME}[3]{\ensuremath{\langle #1 \| #2 \| #3 \rangle}}

\newcommand{\bra}[1]{\ensuremath{\left\langle #1 \right|}}
\newcommand{\ket}[1]{\ensuremath{\left| #1 \right\rangle}}
\newcommand{\braket}[2]{\ensuremath{\langle #1 | #2 \rangle}}

\newcommand{\hw}{\ensuremath{\hbar\Omega}}

\newcommand{\iu}{\mathrm{i}\mkern1mu}

\newcommand {\red} [1]{\textcolor{red}{#1}}

\newcommand{\bea}{\begin{eqnarray}}
\newcommand{\eea}{\end{eqnarray}}
\newcommand{\bitz}{\begin{itemize}}
\newcommand{\eitz}{\end{itemize}}

\newcommand{\eg}{{e.g.\ }}

\newcommand{\etcx}{{etc.}}

\begin{document}

\title{
\textit{Ab initio} symmetry-adapted approaches to nuclear reactions
}
\author[LSU]{Kristina D Launey}
\author[FRIB]{Grigor H. Sargsyan}
\author[LSU]{Alexis Mercenne}
\author[LLNL]{Jutta E. Escher}
\author[LSU]{Darin C. Mumma}
\address[LSU]{Department of Physics and Astronomy, Louisiana State University, Baton Rouge, LA 70803, USA}
\address[FRIB]{Facility for Rare Isotope Beams, Michigan State University, East Lansing, Michigan 48824, USA}
\address[LLNL]{Nuclear and Chemical Sciences Division, Lawrence Livermore National Laboratory, Livermore, CA 94550, USA}

\begin{abstract}
In this review, we discuss recent applications of the {\it ab initio}  symmetry-adapted no-core shell-model (SA-NCSM) theory for study and prediction of structure and reactions of stable and unstable nuclei from light to medium mass range. We explore structure properties of neutron-rich He, Mg, and Li isotopes, with a focus on nuclear collectivity, clustering, and spectroscopic factors, as well as  multi-particle excitations of utmost significance in the proximity of the drip lines. In addition, we present extensions of the SA-NCSM with continuum for determining the microscopic structure of reaction fragments, which enables calculations of reaction cross sections for targets from the lightest $^{4,6}$He to $^{40}$Ca, rooted in first principles. We illustrate this for neutron and proton elastic scattering, deuteron and alpha capture reactions, and alpha knock-out reactions. Furthermore, we discuss 
microscopic optical potentials with uncertainty quantification, a critical ingredient in many reaction models, and reaction observables with uncertainties that stem from the underlying chiral potential.  We also discuss the impact of alpha clustering on reactions of significance to nuclear astrophysics, as well as on beta decays and beyond-the-standard-model physics.  

\end{abstract}

\maketitle
\tableofcontents

\section{Introduction }

Modeling atomic nuclei and their reactions, starting from the nuclear constituents, protons and neutrons, represents one of the most challenging complexity in physics today. 
Historically, multiple theoretical approaches have been used to capture the rich variety of experimentally observed nuclear phenomena and properties. Complementary methods have been dedicated to describe specific properties of the empirical data, covering the range
from light to heavy isotopes, from structure to reaction observables,  from spherical to deformed nuclei, and from short-range physics to long-range collective correlations and clustering. 

\begin{figure}[t]
    \centering
\includegraphics[width=0.75\linewidth]{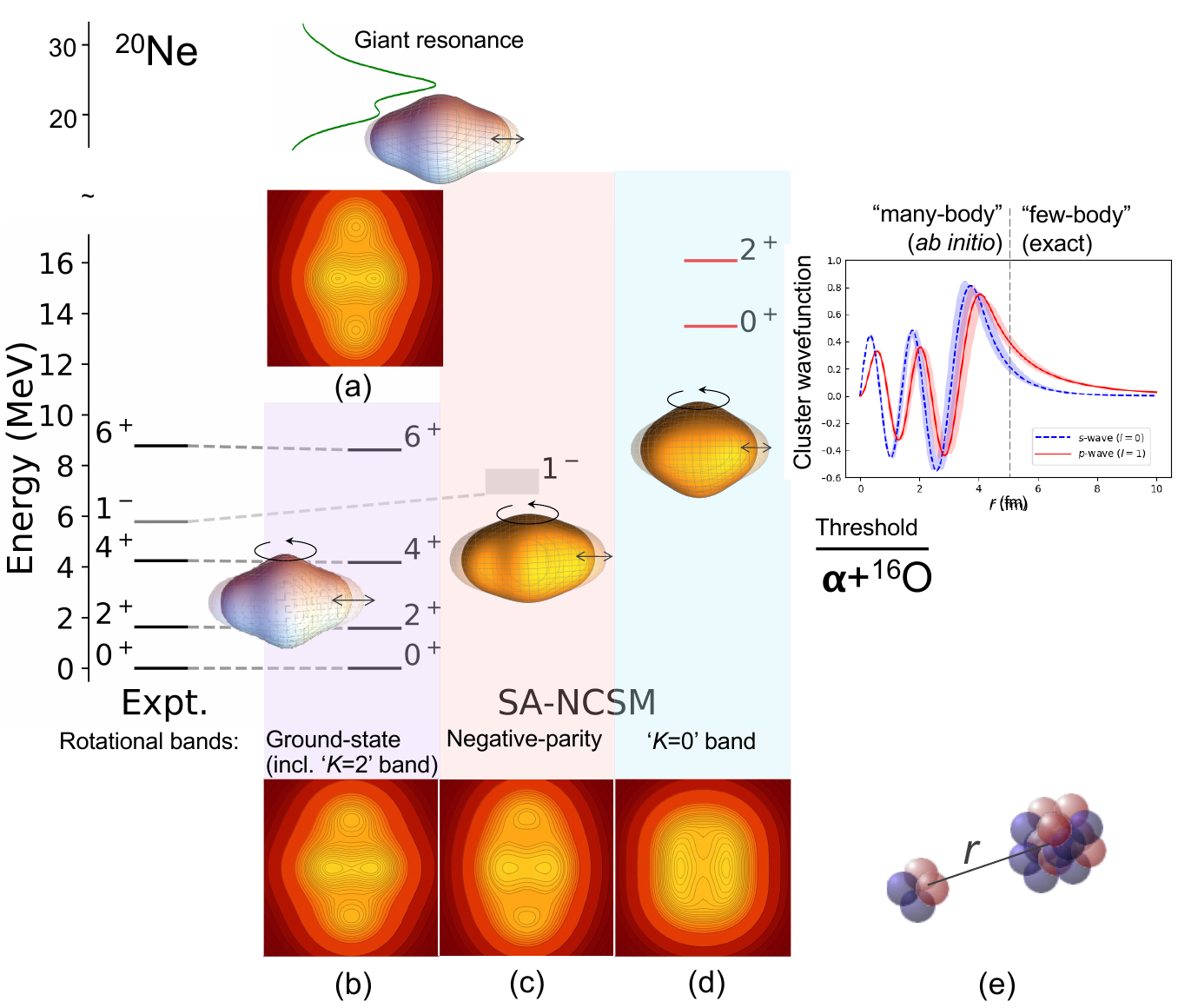} 
    \caption{
Microscopic description of deformation, rotations, vibrations, and clustering in the \textit{ab initio} SA-NCSM framework with chiral EFT interactions, illustrated for $^{20}$Ne: (a) giant-resonance mode; rotational bands above (b) the ground state (g.s.) (which may also include a band above another $2^+$ state often referred to as ``$K=2$"),
(c) a negative-parity state (only the $1^-$ member of the band is shown), and an illustrative excited $0^+$ state of different deformation (sometimes referred to as a ``$K=0$" band), together with (e) the $^{16}$O$+\alpha$ wavefunction with exact asymptotics at large distances. Excitation energies, along with the one-body density profiles for the most dominant shape in the \emph{body-fixed frame}, are shown for a model space of $145$ million basis states (13 harmonic oscillator shells),
    which includes hundreds of nuclear shapes (the giant resonance shown by the monopole response function from Ref.~\cite{Burrows_2025};
    the g.s. rotational band described in Ref.~\cite{DytrychLDRWRBB20,LauneyMD_ARNPS21}; the first excited $1^-$ state from Ref.~\cite{DreyfussLESBDD20} with extrapolated energy;  inset adapted from Ref.~\cite{LauneyMD_ARNPS21}).
    }
    \label{fig:SAfeatures}
\end{figure}

Fortunately, we have now reached state-of-the-art theoretical developments, which coupled with frontier computational capabilities, allow for a fully microscopic (time-independent) approach that unifies all of the above within a single framework. One such example is the \textit{ab initio} symmetry-adapted no-core shell-model (SA-NCSM) theory \cite{LauneyDD16,DytrychLDRWRBB20,LauneyMD_ARNPS21}, the focus of this review. As conceptually illustrated in Fig.~\ref{fig:SAfeatures} for the intermediate-mass $^{20}$Ne nucleus, the approach can capture deformation, rotational bands, multi-particle excitations, single-particle effects and unnatural parity, clustering substructures and giant monopole resonances (i.e., broad energy peaks observed above 20 MeV), along with bound and continuum physics; in addition, it is applicable to spherical and deformed nuclei, from light isotopes all the way to the medium-mass region.

The SA-NCSM theory, detailed in Sec.~\ref{sec:SANCSM}, enumerates all possible many-body configurations of the protons and neutrons inside a nucleus (or within all reaction fragments involved in a nuclear reaction). The nuclear force among the nucleons is modeled with the chiral effective field theory (EFT) (e.g., \cite{BedaqueVKolck02,EntemM03,Epelbaum:2014sza,RevModPhys.92.025004}), which starts from nucleon and pion degrees of freedom (dof), the relevant dof for the low-energy regime of nuclei, while
accounting for the symmetry and symmetry-breaking
patterns of the underlying theory of quantum chromodynamics (QCD). The SA-NCSM theory builds upon the successful concepts of the symplectic shell model \cite{RosensteelR77,DreyfussLTDB13,DytrychDSBV09} and the no-core shell model \cite{NavratilVB00,BarrettNV13}. In addition, through 
a physically relevant basis that respects the inherent symmetries of nuclei, hence called a symmetry-adapted (SA) basis,
it expands the reach of practically exact models to the medium-mass region, namely, around the Calcium isotopes and mass $A\approx 50$ \cite{LauneyMD_ARNPS21,Burrows_2025,Baker24_PhysRevC.110.034605} (e.g., the calculations in Fig.~\ref{fig:SAfeatures} require $3.64 \times 10^{12}$ basis states, which is beyond the current reach, but become feasible in the SA-NCSM). 

Indeed, the medium-mass region 
is of utmost importance, as it is a major milepost between the light and heavy nuclei. In this region,
nuclear systems exhibit properties very similar to those characteristic of heavy nuclei. This represents a unique opportunity to utilize \textit{ab initio} nuclear simulations to inform the modeling of heavier species. While the heavy nuclei have been reached through a subset of the SA-NCSM theory, namely, by utilizing  microscopic interactions (e.g., see $^{166}$Er in Ref. \cite{Rowe_book16}), 
it has come as a surprise that the same physics of one or two deformed shapes governs even lower-mass nuclei, such as $^6$Li, $^8$He, $^8$Be, $^{12}$C, $^{20}$Ne and $^{48}$Ti, as unveiled by large scale \textit{ab initio} calculations \cite{DytrychDSBV09,DytrychLDRWRBB20,sargsyanlbgs2022,LauneyMD_ARNPS21}.

In addition, the symmetry-adapted concept allows not only to reach heavy nuclei, but also larger spaces in which nucleons reside (referred to as ``model spaces"). This makes it ideal for achieving accurate descriptions of nuclear wavefunctions within the nuclear interaction range (the so-called interior), including coupling to the continuum and accounting for spatially expanded deformed modes and clustering. This is critical to reproducing electric quadrupole ($E2$) transitions and moments without effective charges (that is, we use the bare proton charge and zero neutron charge, eliminating the need for charge parameters that are typically fitted to data)~\cite{Henderson:2017dqc,Ruotsalainen19,LauneyMD_ARNPS21,sargsyanlbgs2022}. This is also critical to providing descriptions of giant resonances (Fig.~\ref{fig:SAfeatures}a) \cite{DytrychLDRWRBB20,BurrowsLMBSDL24}, as well as of clustering substructures (Fig.~\ref{fig:SAfeatures}b \& c) that play an important role not only in determining the nuclear structure and related observables (see. e.g., \cite{RevModPhys.87.1067,FreerHKLM18,LynnTGL19,Elhatisari:2022zrb, Elhatisari:2017eno, Shen:2022bak}), but also in addressing physics beyond the standard model \cite{sargsyanlbgs2022,BurkeySGSC2022, GallantSSC2023,LongfellowGSB2024}. Furthermore, in the framework of the SA-NCSM with continuum, the wavefunction asymptotics can be straightforwardly recovered by using the exact Coulomb wavefunctions at large distances (Fig.~\ref{fig:SAfeatures}e).
In this way, the SA-NCSM framework can be utilized to provide structure observables, and in addition, reaction observables, such as partial widths 
(determining the probability for decay) and amplitudes of the wavefunction tail at large distances, the so-called asymptotic normalization coefficients (ANC) 
\cite{DreyfussLESBDD20,sargsyanlbgs2022,sargsyan:23}, response functions and sum rules \cite{BakerLBND20,Burrows_2025}, as well as alpha capture and knock-out reaction cross sections \cite{DreyfussLESBDD20,SargsyanYOLELT25}, detailed in Sec.~\ref{sec:alpha}. Furthermore, given accurate wavefunctions in the interior, the framework can subsequently treat localized clusters, and -- through the Green's function (GF) technique or the resonating group method (RGM) -- can deduce effective potentials~\cite{MercenneLDEQSD21,BurrowsLMBSDL24,SargsyanPKE2024}, which are needed as input to few-body approaches to reactions (Sec.~\ref{sec:scatt}). We note that reaction fragments (or clusters) have been widely used and are still used as degrees of freedom in few-body or effective theories, often without taking into account the cluster intrinsic structure, while championing modeling of multi-cluster reactions and descriptions currently inaccessible to many-body approaches due to the numerical complexity.

Furthermore, different models have historically been used for spherical and deformed nuclei. Specifically, nuclei adjacent to closed shells, which are not too ``rigid", have been regarded as spherical and effectively described by low-energy quadrupole vibrations. Whereas deformed nuclei have been modeled by low-energy, normal-mode, vibrations of the quadrupole-deformed shape, the so-called $\beta$-vibrations and $\gamma$-vibrations collective modes~\cite{Wood16}. However, 
these vibrational modes have been recently confronted in light of
the observed large deformation across the nuclear chart, even in nuclei with practically spherical ground states where shapes of different deformation co-exist in the low-lying energy spectrum~\cite{HeydeW11}. This prevalence of deformation is now evident from the vast body of experimental data, especially with the increasing amount of measurements of electric monopole ($E0$) and quadrupole ($E2$) transitions (see the review \cite{garrett2022103931} and references therein). In fact, multi-nucleon transfer reaction spectroscopic data have exposed multi-particle excitations, or equally enhanced deformation,
within low-lying $0^+$ states in singly closed-shell nuclei, such as the Ca isotopes and $N=20$ isotones (see, e.g., \cite{Wood16,physics4030048}). 

To this end, the SA-NCSM framework provides a unified description of vibrations, rotations, and multi-particle  deformed configurations. Most importantly, it utilizes a physically relevant basis that can shed new light onto the physics that governs the nuclei. While it is well understood that wavefunctions are not observables and that every complete basis yields the same observables, 
nuclear wavefunctions in the low-energy spectrum ($\lesssim 30$ MeV) have been shown from full large-scale calculations \cite{DytrychLDRWRBB20} to be described by typically one or two SA-NCSM basis configurations, each of which corresponds to a microscopic nuclear shape.
This points to the physical relevance of this basis. A very important consequence of this is that we now understand that low-lying nuclear states take on one or two predominant shapes, which determine the nuclear deformation and rotations (Fig.~\ref{fig:SAfeatures}b \& d), and exhibit high-energy monopole and quadrupole vibrations of the giant-resonance type. 
These modes differ from the above-mentioned low-energy ($\beta$ and $\gamma$) vibrations:
in low-lying states, they are manifested as ``surface" vibrations, as illustrated in Fig.~\ref{fig:SAfeatures}(b)--(d), while states in which they dominate appear around 25--30 MeV as giant resonances (Fig.~\ref{fig:SAfeatures}a). 
Most importantly, they are critical to fully developing a shape and decreasing its energy.  That is, without these vibrations, deformed shapes appear high in the energy spectrum. Furthermore, through the same mechanism, highly deformed shapes become energetically favored and appear as low-lying excited $0^+$ states, with the corresponding rotational band built upon them (Fig.~\ref{fig:SAfeatures}). This implies that vibrations are indeed critical to the description of low-lying excited $0^+$ state, however, these vibrations are energetic and -- in many low-lying $0^+$ states similar to $0^+_2$ in $^{12}$C and $^{16}$O -- are built upon very deformed shapes of multi-particle nature \cite{RoweTW06,DreyfussLTDB13}. These features are further discussed in Sec.~\ref{sec:SANCSM}, with a focus on neutron-rich light and medium-mass nuclei. 

To summarize, as illustrated in Fig.~\ref{fig:SAfeatures}, low-lying states can be organized in rotational bands of different shapes, with each band having states with practically the same deformation and vibrations (including spherical shapes, which have zero deformation, implying no rotational band and only energetic vibrations).  We emphasize that 
the SA-NCSM basis is a complete basis and calculations use as many basis states as dictated by the physics at hand. 
Given the current code implementation of the SA-NCSM theory, calculations have provided accurate observables up through the medium-mass nuclei \cite{LauneyMD_ARNPS21,Henderson:2017dqc, Ruotsalainen19,PhysRevC.100.014322,PhysRevLett.125.112503,DreyfussLESBDD20,Burrows_2025}, from binding energies and beta decays to the long-range electric quadrupole moments and $E2$ transitions. The richness and quality of the SA-NCSM model thus allows one to determine various observables, driven by short-range to long-rage physics, including the impact of short-range contact interaction on quadrupole moments \cite{BeckerLEDLSD25}, as well as, for example, beta decays \cite{SargsyanThesis21}, electron-scattering form factors \cite{DytrychHLDMVLO14}, and octupole deformation.

The paper is organized as follows. An overview of reaction approaches, along with their important role for understanding the physics of nuclei, is presented in Sec.~\ref{sec:reactions}. The \textit{ab initio} symmetry-adapted no-core shell model with continuum is discussed in Sec.~\ref{sec:SANCSM} with a focus on structure calculations and the role of multi-particle configurations toward the drip line. Section~\ref{sec:SFovlps} discusses single-nucleon spectroscopic factors that can provide information about the target structure in reactions. The next section presents modeling of neutron and proton elastic scattering, with a focus on the SA-NCSM Green's function approach, SANCSM/GF (Sec.~\ref{sec:GF}), the SA resonating group method, SA-RGM (Sec.~\ref{sec:RGM}), and a Feshbach projection approach to optical potentials (Sec.~\ref{sec:selfie}). Proton elastic scattering at intermediate energies for $^{20}$Ne and $^{40}$Ca targets is discussed in the framework of the multiple scattering theory in Sec.~\ref{sec:mst}. The role of alpha clustering in reactions and decays is explored, with a focus on alpha capture reactions (Sec.~\ref{sec:alphacapture}), 
alpha knock-out reactions (Sec.~\ref{sec:alphaknockout}), and beta-decay strengths, and for probing beyond-the-standard-model physics (Sec.~\ref{sec:A8beta}).
Uncertainty quantification for reaction observables from the underlying potential is discussed in the last section (Sec.~\ref{sec:UQ}).

\section{Overview of reaction approaches}
\label{sec:reactions}

Achieving reliable predictions of reaction properties, such as phase shifts, cross sections, and decay widths, with quantified uncertainties, is one of the key goals of nuclear physics. A robust understanding of reaction mechanisms is important for planning and interpreting experiments that probe nuclear structure aspects including the emergence and evolution of shell structure, collectivity, and clustering properties.
In turn, successful predictions of reaction data provide stringent tests of the theories that have been developed to describe static and dynamic properties of nuclei.
Finally, reaction cross sections are important ingredients for multi-physics simulations used in the areas of astrophysics, national security, medicine, and nuclear energy~\cite{Arcones:17, Arnould:20, Hayes:17}.

Low-energy nuclear reaction physics is a field with a rich variety of reaction types and phenomena.
Representative examples include: 
\bitz
\item Phase shifts and scattering cross sections for light nuclei~\cite{Quaglioni:20,BurrowsLMBSDL24} and for intermediate-mass nuclei~\cite{MercenneLDEQSD21};
\item Direct capture of nucleons on light nuclei, such as the $^7$Be(p,$\gamma$) reaction, which plays an important role in understanding the generation of energy and neutrinos in the sun~\cite{Adelberger:98, Adelberger:11, Acharya:24}, as well as alpha capture reactions for oxygen targets \cite{DreyfussLESBDD20} of significance to x-ray burst (XRB) simulations \cite{Cyburt10};
\item Single and double
charge-exchange reactions, which are
important probes of electron capture, beta and double-beta decays that can inform nuclear structure, nuclear astrophysics and beyond-the-standard-model physics \cite{LENSKE2019103716,CAPPUZZELLO2023103999,LangankeMZ2021, GiraudZZBA2023};
\item Compound nuclear reactions proceeding through isolated resonances, such $^{22}$Ne($\alpha$,n)$^{25}$Mg, which produces neutrons in asymptotic giant branch stars and contributes to heavy-element nucleosynthesis via the astrophysical s-process~\cite{Longland:12,Massimi:17,Adsley:21};
\item Compound neutron captures on medium-mass and heavy nuclei which are responsible for the production of the heavy elements in the s-, i-, and r-processes~\cite{Capote:09, Rochman:17, Arnould:20, Cowan:21rmp, Koning:23};
\item Elastic and inelastic scattering for nuclear data evaluations~\cite{Baba:90, Brown:18endf8, Dupuis:19, Kerveno:21, Thapa:24, Escher:25CNRb}, and for experimental studies of pygmy resonances, giant resonances, and collectivity~\cite{Savran:13, Savran:18, Bracco:19a, Lanza:23, Tsoneva:24, Holl:21, Chen:22};
\item Transfer and knock-out reactions for probing shell evolution and clustering properties of nuclei~\cite{Hansen03, Otsuka:20rmp, Aumann:21, Chen:24,SargsyanYOLELT25}.
\eitz

Ideally, \emph{a many-body theory} can be used to describe and predict these phenomena within one unified framework, starting from a set of basic ingredients, such as nucleons and their interaction with each other.  In practice, however, computational expenses and conceptual questions make this a challenging endeavor, and one often relies on various 
specific \emph{few-body theory} frameworks
to describe the few fragments in the reactions of interest, including the examples given.  Moreover, the underlying approximations typically make it necessary to adjust (scale) the calculated results to reproduce measured data.
To advance the predictive power of reaction theory, the goal is to
revisit the approximations made in treating the reaction mechanisms considered and in calculating the reaction ingredients.  This is a significant challenge for nuclear theory.

\vspace{8pt}
\noindent
\textit{For light and medium-mass nuclei} ($A \lesssim 50$), recent years have seen remarkable progress in the development of many-body approaches 
to scattering and nuclear reactions from first principles, that is, with inter-nucleon interactions typically derived in the chiral effective field theory, without the need to fit interaction parameters in the nuclear medium (see Refs. \cite{1402-4896-91-5-053002,0954-3899-41-12-123002,FRIBTAwhite2018,LauneyMD_ARNPS21, navratil:22handbook} for reviews).
Efforts have focused on treating structure and reactions on the same footing in \textit{ab initio} approaches, which have primarily been applied to light and intermediate-mass nuclei,
including, e.g.,  studies of elastic scattering \cite{NollettPWCH07,HagenDHP07,quaglioni08,ElhatisariLRE15,Lazauskas2018,MercenneLDEQSD21,BurrowsLMBSDL24}, photoabsorption \cite{PhysRevC.90.064619}, transfer \cite{navratil:12}, capture reactions \cite{PhysRevLett.105.232502,DreyfussLESBDD20}, as well as thermonuclear fusion \cite{HupinQN19}.
Next, we describe key
reaction approaches in this mass region, which have been, in addition, benefited from using symmetry-adapted bases.

\vspace{4pt}
\noindent
{\bf Green's function (GF) formalism.}
Recent applications  have built upon an earlier theoretical framework introduced by Feshbach \cite{Feshbach:58}, leading to the Green's function formulation \cite{capuzzi2000223} and to the successful dispersive optical model \cite{Mahaux:1986zz,Mahaux1991,DickhoffBook}. These applications have provided \textit{ab initio} nucleon-nucleus potentials and associated reaction observables for elastic scattering for closed-shell nuclei at low projectile energies ($\lesssim20$ MeV per nucleon, relevant to the astrophysical regime) based on the Green's function technique with the coupled-cluster method \cite{RotureauDHNP17,PhysRevC.98.044625}, the self-consistent Green's function method \cite{idini19}, and the SA-NCSM~\cite{BurrowsLMBSDL24} (the SA-NCSM/GF is reviewed in Sec.~\ref{sec:GF}). In addition, the Feshbach projection has been utilized for neutron scattering cross sections for a $^{24}$Mg target with calculations from the valence-shell model \cite{SargsyanPKE2024}, and now expanded through SA-NCSM calculations (Sec.~\ref{sec:selfie}).  An important outcome of these approaches is a nonlocal, energy dependent, and dispersive optical potential (see also Ref.~\cite{FRIBTAwhite2018}), which, in turn, can be utilized by  few-body methods. These potentials -- rooted in first principles -- provide cross sections for elastic proton or neutron scattering, and in addition can be used as input to modeling (d,p) and (d,n) reactions \cite{Rotureau_2020}.

\vspace{4pt}
\noindent
{\bf The resonating group method (RGM).}
Another way to address the long-distance behavior of nuclear wavefunction and to introduce coupling to the continuum is by using a basis that explicitly considers cluster degrees of freedom. This can be done through a combination of the NCSM or SA-NCSM with the resonating group method \cite{quaglioni08,quaglionin2009,navratil:12,MercenneLEDP19}.
In the RGM \cite{WildermuthT77}, nucleons are organized within different groups, or clusters, ``resonating'' through the inter-cluster exchange of nucleons. 
This ensures the Pauli exclusion principle and along with the consideration of the cluster internal structure are two of the most important features of the approach.
 A hybrid basis approach, the no-core shell model with continuum (NCSMC) \cite{BaroniNQ13,Baroni_PRC_2013}, uses mixed shell-model and RGM basis to achieve a faster convergence \cite{PhysRevLett.114.212502}. 
Another way to include the continuum in the  NCSM framework  is to start with a continuum single-particle basis, such as the Berggren basis \cite{Papadimitriou2013,Li2019,Michel2023,Hu2020}.

A major advantage of modern implementations of the RGM is that one can treat structure and reactions on the same footing, by employing state-of-the-art nuclear interactions including nucleon-nucleon (NN) and 3-nucleon (3N) forces and utilizing high-performance computational capabilities.
A significant challenge, however, is the exponential growth of the size of the NCSM or NCSMC bases with increasing number of nucleons and the space in which nucleons reside, limiting 
both the structure and reaction applications to lighter systems.  
This has motivated the exploration of 
various truncation schemes (e.g., see Ref.~\cite{Kravvaris:24b}).

A recent major improvement uses the symmetry-adapted basis in the RGM approach (SA-RGM) \cite{MercenneLDEQSD21}, through the integration of the SA-NCSM with the RGM approach. It  enables the reach of intermediate-mass targets (illustrated for $^{20}$Ne in~\cite{MercenneLDEQSD21,LauneyMD_ARNPS21}) and medium-mass nuclei (illustrated here for $^{40}$Ca), as discussed in more detail in Sec.~\ref{sec:RGM}.

\vspace{4pt}
\noindent
{\bf Multiple scattering theory (MST).} Other studies that account for the structure of the target have built upon the framework pioneered by Watson~\cite{Watson1953a,KMT} for elastic scattering of a single-nucleon projectile, leading to the spectator expansion of the multiple scattering theory~\cite{Siciliano:1977zz}. Recent applications include \textit{ab initio} nucleon-nucleus potentials at leading order and associated reaction observables for elastic scattering in the intermediate-energy regime ($\gtrsim 65$ MeV per nucleon) using the MST with one-body densities from the \textit{ab initio} NCSM for light nuclei \cite{BurrowsBEWLMP20,VorabbiGFGNM2022}, and the \textit{ab initio} SA-NCSM for $^{20}$Ne and $^{40}$Ca targets~\cite{LauneyMD_ARNPS21,Baker24_PhysRevC.110.034605} (Sec.~\ref{sec:mst}).

\vspace{8pt}
\noindent
\textit{For intermediate-mass and heavy nuclei} ($A \gtrsim 20$),
most current descriptions of reactions still
make significant simplifications.  
Typically, we distinguish between direct and compound nuclear reactions and treat them in different formalisms, as discussed below.
The coupled reactions channels (CRC) formalism is widely used to treat direct inelastic and transfer reactions.  While the internal structure of the interacting nuclei is often approximated with simple models, the formalism allows for microscopic nuclear structure inputs.
Compound nuclear reactions, on the other hand, are treated either in an R-matrix approach (for reactions involving isolated resonances) or the statistical Hauser-Feshbach formalism.
Significant efforts have been devoted to integrating advanced structure descriptions in both direct and statistical reaction theories 
for medium-mass and heavy nuclei, which can be aided by exploiting symmetry-adapted bases.

\vspace{4pt}
\noindent
{\bf
Direct reactions: coupled reaction channels (CRC) formalism.}
\label{sec:reactions_crc}
The coupled reactions channels model starts with an ansatz~\cite{Thompson:88,Thompson:09book,Satchler-83}: 
\bea
\Phi = \displaystyle\sum_c \Psi_{{\rm p}, c} \Psi_{{\rm t}, c} \psi(\vec{r}_c)
\label{eq:crc}
\eea
where $\Psi_{{\rm p}, c}$
and $\Psi_{{\rm t}, c}$
are projectile and target wavefunctions, respectively, and $\psi(\vec{r}_c)$ describes their relative motion, $c=1,2,\ldots$ enumerates the channels taken into account, and $\vec{r}_c$ is the relative coordinate between the center-of-mass of the projectile and target nuclei.
For transfer reactions, one includes combinations 
$\Psi_{{\rm p}, c} \Psi_{{\rm t}, c}$
in different mass partitions, and for elastic/inelastic scattering one considers excited states of the projectile and/or target. Full CRC calculations allow for both.

Using the Schr\"odinger equation, one can derive the CRC equations, which -- in their most general form -- contain local coupling, nonlocal couplings, and nonorthogonality terms.

For the example of inelastic scattering with local potentials, they can be written as a set of coupled radial equations:
\bea
\left\{ \frac{d^2}{dr^2}\!-\!\frac{l_c(l_c+1)}{r^2}\!-\!\frac{2\mu_c}{\hbar^2} V^J_{cc} (r)\!+\!k_c^2\right\} r u_c(r)= \displaystyle\sum_{c' \neq c} \frac{2\mu_{c'}}{\hbar^2} V^J_{cc'}(r) r u_{c'}(r) \:\: ,
\eea
where $u_c(r)$ is the relative-motion wavefunction in channel $c$, $r$ the distance between projectile and target, $\mu_c$ the reduced mass, $l_c$ denotes the relative angular momentum, and $k_c$ the relative momentum; $V^J_{cc'}$ is the potential that couples the ground states of target and projectile to the relevant excited states, and the diagonal term $V^J_{cc}$ is the optical potential for channel $c$. 

The coupling potentials $V^J_{cc'}$ can be expressed in a manner where they can make use of microscopic nuclear structure information. For the example of nucleon-nucleus scattering, they are written as a sum over angular-momentum recoupling coefficients and folding integrals of the form
\bea
&& 4\pi\sqrt{2I_c+1} \int_0^{\infty} dr_t r^2_t v_{L_0}^{S_0T_0}(r,r_t) \rho_{L_0S_0J_0}^{T_0 T_{0z},cc'}(r_t)  \:\: .
\eea  
Here $I_c$ denotes the target spin, $v_{L_0}^{S_0T_0}(r,r_t)$ is the effective interaction between the projectile nucleon at $r$ and a target nucleon at $r_t$ for a partial wave $L_0$ and total spin (isospin) $S_0$ ($T_0$), and $\rho_{L_0S_0J_0}^{T_0 T_{0z},cc'}(r_t)$ is the (radial) transition density that is obtained from the structure calculation for the target nucleus. 
The quantities $L_0, S_0, J_0, T_0$ 
(and $T_{0z}$) determine
the change, respectively, in orbital angular momentum, spin, total angular momentum, isospin of the target nucleus (and isospin projection, e.g., for charge-exchange reactions). 
For scattering by a composite projectile, this generalizes to a double-folding expression.

Recent applications of this approach have used structure information from the quasiparticle random-phase approximation (QRPA) and employed an effective nucleon-nucleon interaction of the Jeukenne-Lejeune-Mahaux (JLM) type~\cite{Bauge:00a,Bauge:00b,Bauge:01,Bauge:98} to describe inelastic nucleon-nucleus scattering~\cite{Dupuis:19, Chimanski:24wip, Thapa:24}.  Simpler effective interactions, folded with (Q)RPA structure transition densities, have been used in CRC calculations that include specific two-step transfer reactions to calculate reaction cross sections~\cite{Nobre:10,Nobre:11} and the population of doorway states~\cite{Escher:22tr}.  

The CRC formalism is able to accommodate breakup effects by introducing discrete $\Psi_{{\rm p}, c}$ and $ \Psi_{{\rm t}, c}$ that are representations of continuum levels for projectile and target, respectively.  This leads to the continuum-discretized coupled channels (CDCC) implementation~\cite{Thompson:88, Thompson:09book, Hagino:22, Timofeyuk:20, Descouvemont:18, Watanabe:21}.

Calculations of transfer reactions require overlaps between projectile and ejectile, as well as target and remnant nuclei, providing another opportunity for input from microscopic structure theories. 
One-body overlap functions from the Green's function Monte Carlo (GFMC) and variational Monte Carlo (VMC) methods, for instance, have been used in descriptions of transfer (and knock-out) reactions for light nuclei~\cite{Brida:11, Grinyer:12}. Another example includes alpha overlap functions from the SA-NCSM for alpha knock-out reactions with intermediate-mass targets~\cite{SargsyanYOLELT25}.
Indeed, \textit{ab initio} approaches applicable in this mass region can provide microscopic inputs for CRC calculations.  For example, transition densities from the SA-NCSM can be used for inelastic scattering calculations.  Similarly, overlap functions can be used for transfer calculations (for details, see Ref.~\cite{Timofeyuk:20}).  
Integrating such structure information into the CRC formalism can potentially establish a bridge between 
light nuclei and heavier species.

\vspace{4pt}
\noindent
{\bf Compound nuclear reactions.}
\label{sec:reactions_statistical}
Compound nuclear reactions proceed via the formation of an intermediate, equilibrated ``compound'' nucleus (CN).  At very low energies they produce narrow, isolated resonances and their cross sections can be formulated in the framework of the R-matrix formalism~\cite{Lane:58, Azuma:10, Descouvemont:10, Thompson:09book}. 
With increasing projectile energy, the resonances begin to broaden and overlap, forming the unresolved-resonance region and, at even higher energies, a region in which the statistical Hauser-Feshbach formalism is applicable~\cite{HauserFeshbach:52, Feshbach:80, Feshbach:92, Capote:09, Carlson:14, Koning:23}. The transition between the various regimes depends on the projectile type and on the level density of the CN formed. 

While R-matrix theory, often referred to as ``calculable R-matrix'',  provides a simple and elegant way to solve coupled-channels problems~\cite{Descouvemont:10}, for medium-mass and heavy nuclei there is typically not sufficient information to set up an appropriate set of equations for a given resonance reaction.  An example is the $^{22}$Ne($\alpha$,n)$^{25}$Mg reaction.  To obtain an evaluated cross section in such cases, one has to rely on the ``phenomenological R-matrix" approach~\cite{Descouvemont:10}, which provides an efficient way for parameterizing measured reaction data that exhibits resonance structure.  Phenomenological R-matrix codes have become a standard tool for nuclear data evaluations~\cite{Azuma:10, Thompson:2019aa}.

In situations where the resonances strongly overlap, the cross section for a reaction $a+A \rightarrow$ $B^*$ $\rightarrow c+C$ proceeding through the compound nucleus $B^*$ can be described in Hauser-Feshbach statistical reaction theory~\cite{HauserFeshbach:52, Feshbach:80, Feshbach:92, Capote:09, Carlson:14, Koning:23}:
\begin{equation}
\sigma_{\alpha \chi}(E_{a}) \!=\! \sum_{J,\pi}  \sigma_{\alpha}^{CN}(E_{ex},J,\pi) \; G_{\chi}^{CN}(E_{ex},J,\pi) \; W_{\alpha\chi}(J)\; .
\label{eq:HF}
\end {equation}
\noindent
Here $\sigma_{\alpha}^{CN}(E_{ex},J,\pi)$ describes the formation of the CN at excitation energy $E_{ex}$, spin $J$, and parity $\pi$, by fusing a projectile $a$ with a target nucleus $A$ (in entrance channel $\alpha$), and $G_{\chi}^{CN}(E_{ex},J,\pi)$ describes the decay of the CN into the channel $\chi$ ($=c+C$).   
The product reflects the fact that the reaction proceeds in two stages, namely, formation and decay of the intermediate CN.  The sum runs over all spins and parities $(J,\pi)$ of the CN. Corrections to the factorization are encoded in the width fluctuation correction factor, $W_{\alpha\chi}(J)$~\cite{Hilaire:03}.
For a neutron capture reaction, $A(n,\gamma)B$, we have $\alpha$ = $n+A$ and ${\chi} = \gamma+B$.

Multiple inputs are needed to calculate the components in the Hauser-Feshbach formula: Optical models are used to determine the transmission coefficients (TCs), which quantify the fusion of neutrons or charged particles with the target to form the CN, and hence determine the formation cross section $\sigma_{\alpha}^{CN}(E_{ex},J,\pi)$.
They are also used to calculate the evaporation of these particles from the CN.  The probability for the decay into various competing decay channel, $G_{\chi}^{CN}$, requires similar TCs for additional particles (\eg deuterons, $\alpha$ particles, \etcx), as well as $\gamma$-ray strength functions ($\gamma$SFs) for the emission of photons, and information on fission barriers for decay by fission.  Also needed are level densities in the intermediate and residual nuclei occurring in the reaction.

Hauser-Feshbach calculations play a central role in nuclear data evaluations. Often, phenomenological descriptions are used to reproduce measured data~\cite{Capote:09}.  Experience and skill is required to select the most appropriate models and use experimental data to constrain the model parameters. Over the past two decades, much effort has been devoted to developing microscopic approaches, in order to achieve more predictive Hauser-Feshbach calculations.  This is particularly relevant for applications involving unstable isotopes for which no data exists.

Approaches based on both shell-model and density-functional theories 
have been used to predict level densities and $\gamma$-ray strength functions~\cite{Koning:08ld, Goriely:08ld, Hilaire:12ld, alhassid:15, shim2016, Goriely:19, Karampagia:20, Ormand:20, Goriely:22ld, Hilaire:2023ux, alhassid_2017_book, alhassid_PRL_2008, alhassid_PRC_2015, alhassid_PRL_2017, shim2016, Karampagia:20, Ormand:20, Johnson:23}. Multiple efforts are underway to develop microscopic optical models (for recent reviews, see~\cite{FRIBTAwhite2018,Hebborn:23omp}). It is in these areas where the SA-NCSM framework can continue to make important contributions.

\section{\textit{Ab initio} symmetry-adapted no-core shell model with continuum}
\label{sec:SANCSM}

Exploiting the SA-NCSM concept represents a major advance -- it has opened new domains of the nuclear chart for {\it ab initio} descriptions \cite{LauneyDD16}. {\it Ab initio} descriptions of spherical and deformed nuclei up through the Calcium region are now possible within a practically exact method \cite{LauneyMD_ARNPS21}. The approach provides exact solutions in the limit of the infinite-size model space  (infinitely many shells), extrapolated from finite-size calculations with \emph{controlled} approximations, while accounting for important correlations in the wavefunction, including the collective and clustering correlations that present an enormous challenge to approaches that scale polynomially.  The unique feature of the SA-NCSM is that solutions are presented in terms of a physically relevant basis that reflects the collectivity and associated symmetries inherent to the nuclear dynamics, hence dubbed a symmetry-adapted basis.
Such near symmetries were first recognized by Bohr \& Mottelson (1975 Physics Nobel Prize) \cite{BohrMottelson69}, followed by the  seminal work of Elliott  \cite{Elliott58,Elliott58b,ElliottH62}  and the microscopic no-core formulation by Rowe \& Rosensteel \cite{RosensteelR77,Rowe85}.
The relevant symmetries are \SU{3} and \SpR{3}.
The \SU{3} group is the exact symmetry  of the three-dimensional spherical harmonic oscillator (HO).
The \SpR{3} symplectic symmetry, which is the embedding symmetry of \SU{3}, is the dynamical symmetry  of the three-dimensional spherical HO. 

Most importantly, each \SpR{3}-preserving subspace  describes \emph{a microscopic nuclear shape}, including its deformation, rotations, and energetic surface vibrations \cite{RosensteelR77,Rowe85,Rowe_book16}. Furthermore, several physically relevant operators do not mix shapes (or symplectic subspaces), including the many-particle kinetic energy and HO potential (equivalently, the monopole operator $r^2$), the $Q$  mass quadrupole moment operator, and the total $L$ orbital angular momentum operator, as well as spin and isospin operators.

\vspace{8pt}
The SA-NCSM, based on the concept of the nuclear shell model \cite{BrussardG77,Shavitt98,BarrettNV13,CaurierMNPZ05}, solves the many-body Schr\"odinger equation for $A$ particles,
$
H \Psi(\vec r_1, \vec r_2, \ldots, \vec r_A) = E \Psi(\vec r_1, \vec r_2, \ldots, \vec r_A),
$
and obtains the eigenenergies $E$ and eigenfunctions $\Psi$, each of which describes a nuclear state. The intrinsic non-relativistic  Hamiltonian is given as
$
H = T_{\rm rel} + V_{NN}  + V_{3N} + \ldots + V_{\rm C}, 
$
which includes the relative kinetic energy $T_{\rm rel} =\frac{1}{A}\sum_{i<j}\frac{(\vec p_i - \vec p_j)^2}{2m}$ ($m$ is the nucleon mass), the Coulomb interaction  $V_{\rm C}$ between protons, and the nuclear interaction expanded in particle rank: $V_{NN}=\sum_{i<j}^A (V_{NN})_{ij}$ is the nucleon-nucleon interaction, with 3-nucleon interaction $V_{3N}=\sum_{i<j<k}^A (V_{NNN})_{ijk}$ and higher-rank interactions possibly included. The Hamiltonian may include additional terms, such as, for example,  magnetic dipole-dipole terms and higher-order electromagnetic interactions.  
For {\it ab initio} approaches that aim to retain predictive capability, adopting controlled many-body approximations is coupled with NN (NN+3N) potentials that reproduce NN phase shifts to high precision (often referred to as ``high-precision" potentials) and are informed by few-nucleon physics only. These potentials are typically derived in the chiral EFT framework (see, e.g., \cite{BedaqueVKolck02,EntemM03,Epelbaum:2014sza,RevModPhys.92.025004}), which starts from nucleon/pion degrees of freedom and accounts for
the symmetry and symmetry-breaking patterns of the underlying QCD. At low energy, the overall effect of the quark-gluon physics is encapsulated in the so-called low energy constants (LECs).  
For example, an NNLO chiral NN potential (such as NNLO$_{\textrm{opt}}$~\cite{Ekstrom13}) includes fourteen LECs as following: 
11 contact couplings  $C^{(\textrm{LO})}_{^1S_0(pp, np, nn),^3S_1}$ at leading order (LO) and $C_{^1S_0,^3S_1,^3S_1-^3D_1,^3P_{0,1,2},^1P_1}$ at next-to-leading order (NLO) for a given NN partial wave $^{2S+1}\ell_J$, along with $c_{1,3,4}$ at NNLO which parametrize the subleading two-pion exchange interaction. The nuclear force thus partly depends on these unknown parameters, which are typically fit to experimental few-nucleon data, in particular, to NN scattering phase shifts and properties of the deuteron (and  triton) (see, e.g., \cite{Machleidt01,GazitDQN09}).

As done in shell-model calculations, we adopt a complete orthonormal many-body basis $\psi_i$, such that the expansion  $\Psi(\vec r_1, \vec r_2, \ldots, \vec r_A) = \sum_{k} d_k \psi_k(\vec r_1, \vec r_2, \ldots, \vec r_A)$ in terms of  unknown coefficients $d_k$ casts the Schr\"odinger equation into a matrix eigenvalue equation, $ \sum_{k'} H_{k k'} d_{k'} = E d_k$,
with many-body Hamiltonian matrix elements
$H_{k k'} =\braketop{\psi_k}{H}{\psi_{k'}} $. To numerically solve this equation, a finite subset of the many-body basis is used. This subset is referred to as a ``model space", and its size is defined by the number of basis states, which is also the dimension of the many-body Hamiltonian matrix to be solved.
Each many-body basis state is an antisymmetrized product of  single-particle states (Slater determinant).
In this study, the single-particle states  of a three-dimensional spherical HO are adopted:
 $\phi_{\eta (\ell \half) j}(\mathbf{r};b)=R_{\eta \ell}(r;b)\mathcal{Y}_{(\ell \half)j}(\hat{r})=\sum_{m\sigma} C_{\ell m \half \sigma}^{j m_j} R_{\eta \ell}(r;b)Y_{\ell m}(\hat r)\chi_{\half \sigma}$, with radial wavefunctions $R_{\eta \ell}(r;b)$ and spin functions $\chi_{\half \sigma}$,
where $\eta=2n_r+\ell$ is the HO shell number, $\ell$ is coupled with spin-{1/2}~ 
to $j$, and $b=\sqrt{ \frac{\hbar }{m\Omega} }$ is the characteristic length for oscillator frequency $\hw$ (an additional quantum number $t_z$ is added to distinguish between protons and neutrons). 

In addition, the many-body basis can be organized into  \SU{3} subspaces of given deformation, and further into \SpR{3}$\supset$\SU{3} subspaces of given nuclear shape, referred to as SA bases. This means that each basis state is labeled according to  \SU{3}$_{(\lambda\,\mu)}\times$\SU{2}$_S$ by the total intrinsic spin $S$ and $(\lambda\,\mu)$ quantum numbers with $\lambda=N_z-N_x$ and $\mu=N_x-N_y$, where $N_x+N_y+N_z=N$ is the total number of the HO quanta distributed in the $x$, $y$, and $z$ direction (in addition to other quantum numbers that are needed for complete labeling). When further organized into  subspaces that preserve the \SpR{3} symmetry and describe microscopic shapes, basis states within each shape are defined by the shape deformation, its rotations and energetic vibrations. 
These monopole and quadrupole vibrations are of the giant-resonance type, and are largely different from low-energy quadrupole vibrations and $\beta$- and $\gamma$-vibrations. Namely, these high-energy vibrations are multiples of monopole and quadrupole 2\hw~1-particle-1-hole (1p-1h) excitations (for a typical $\hw=15$ MeV, the lowest excitation is at 30 MeV).
While vibrations along the symmetry axis are most favorable, vibrations perpendicular to the symmetry axis may appear in excited states but much higher in energy, since they are shown to increase the total kinetic energy \cite{castel1990121,BeckerLEDLSD25}. We note that while the SA basis is a complete basis that is related to the NCSM basis via a unitary transformation, the SA basis states themselves are not constructed using such transformation but utilize an efficient group-theoretical algorithm~\cite{LangrDDLT19}. 

This basis organization allows traditional complete model spaces to be augmented for large $N$ by a subset of the SA basis states (referred to ``SA model spaces"). Since small complete model spaces (up to a small $N$ value) are often sufficient to describe less deformed shapes, the additional SA basis states in high $N$ subspaces are exactly those needed for the full description of spatially expanded (very deformed or clustering) configurations.

In calculations, the model space is finite and is provided for given $N_{\max}$ (sufficiently large $N$), which is the total HO excitations above the nuclear configuration of the lowest HO energy. 
It is important that in the conventional NCSM with complete model spaces truncated by $N_{\rm max}$ (see the review \cite{BarrettNV13}) and in SA model spaces of the SA-NCSM (see the review \cite{LauneyDD16}), the center-of-mass (c.m.) wavefunction can be factored out exactly. The reason is that the c.m. operator ($\hat N_{\rm c.m.}$) does not mix c.m. states with different HO excitations, and in addition, being an SU(3) scalar, it does not mix SU(3) subspaces of the SA-NCSM \cite{VERHAAR1960508,Hecht:1971xlg,Millener92}.
Such a model space allows for preservation of translational invariance of the nuclear self-bound system and provides solutions in terms of single-particle HO wavefunctions that are analytically known. With larger model spaces utilized in the  no-core shell-model theory, the eigensolutions  converge to the exact values, and the corresponding observables become independent from the \hw~basis parameter.

The SA-NCSM approach can accommodate larger model spaces, and  reach heavier nuclei than traditional practically exact methods, including 
$^{20}$Ne \cite{DytrychLDRWRBB20,DreyfussLESBDD20}, $^{21}$Mg \cite{Ruotsalainen19}, $^{22}$Mg \cite{Henderson:2017dqc}, $^{28}$Mg \cite{PhysRevC.100.014322},  as well as $^{32}$Ne and $^{48}$Ti \cite{LauneySOTANCP42018,LauneyMD_ARNPS21}, and $^{40}$Ca \cite{Burrows_2025}. The SA approach with continuum, applicable to structure \cite{LauneyDD16,DytrychLDRWRBB20} and reactions \cite{MercenneLEDP19,DreyfussLESBDD20,LauneyMD_ARNPS21}, exploits new concepts that take maximum advantage of  the existence of symmetries of the nuclear many-body dynamics. In doing so, it resolves the scale explosion problem in nuclear structure and reaction calculations, {\em i.e.}, the explosive growth in computational resource demands with increasing number of particles and size of the spaces in which they reside. Achieving large model spaces is essential to accommodate important collective modes and to couple to continuum degrees of freedom within the interaction range, whereas beyond this, continuum Coulomb exact eigenfunctions are used. The aim is to solve this problem with unprecedented accuracy and beyond the lightest systems,
while  providing nuclear structure and reaction information of interest to theoretical and experimental studies.

\vspace{8pt}
In the present studies, the $A$-particle wavefunctions are calculated using the NNLO$_{\rm opt}$  NN chiral potential \cite{Ekstrom13} (unless otherwise stated) in the \textit{ab initio} SA-NCSM many-body approach, which has yielded energy spectra and observables (radii, quadrupole moments, $E2$ transitions, beta decays, charge form factors, sum rules and responses) in close agreement with experiment \cite{LauneyDD16,DytrychLDRWRBB20,LauneyMD_ARNPS21,sargsyanlbgs2022}. 
The NNLO$_{\rm opt}$ is used without three-nucleon forces, which have been shown to contribute minimally to the three- and four-nucleon binding energies \cite{Ekstrom13}. Furthermore, the NNLO$_{\rm opt}$ NN potential has been found to reproduce various observables and yield results equivalent to those obtained from chiral potentials that require three-nucleon forces. These observables include the $^4$He electric dipole polarizability \cite{BakerLBND20}, the nuclear matter compression modulus \cite{Burrows_2025}, along with $A=8$ energy spectra and quadrupole moments \cite{sargsyanlbgs2022}. We note that in cases where the 3N forces are  significant, such as for, e.g., the N3LO-EM and NNLO$_{\rm sat}$ potentials, they are included in the SA-NCSM as a mass-dependent monopole interaction \cite{LauneyDD12}, which  has a considerable effect on binding energies. For example, for the $^{16}$O ground-state energy,  
the  7-shell 3N contribution from NNLO$_{\rm sat}$ is 20.46 MeV, resulting in $-127.97$ MeV total energy for $N_{\rm max}=8$ and \hw=16 MeV, which closely agrees with the experimental value of $-127.62$ MeV.

Importantly, with the advent of high performance computing (HPC) facilities,  the SA-NCSM framework has revealed the emergence of symplectic symmetry from first principles \cite{DytrychLDRWRBB20}, typically with one or two shapes (symplectic subspaces) composing most of the nuclear wavefunction (ground and excited states \cite{DytrychLDRWRBB20,LauneyDSBD20}). The existence of this symmetry, coupled with HPC capabilities to manage hundreds or thousands of shapes, allows the SA-NCSM to reproduce important nuclear properties and nuclear observables, including:

\begin{figure}[th]
    \centering
    \includegraphics[width=0.6\linewidth]{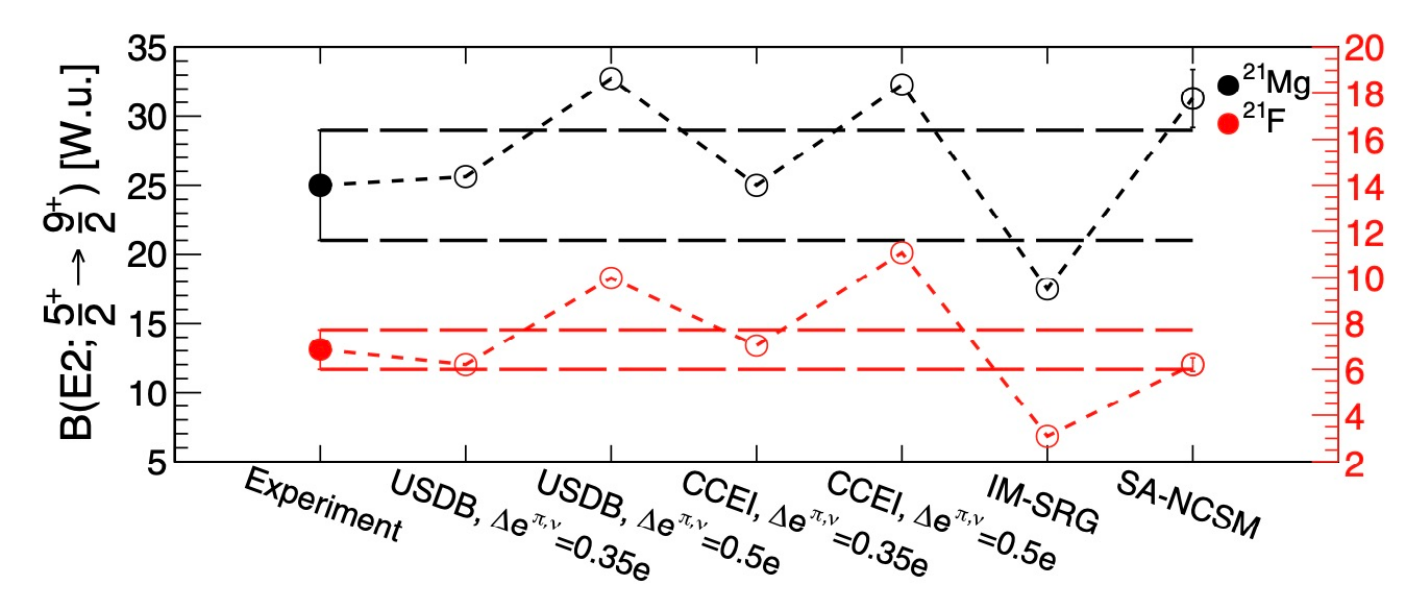}
     \includegraphics[width=0.35\linewidth]{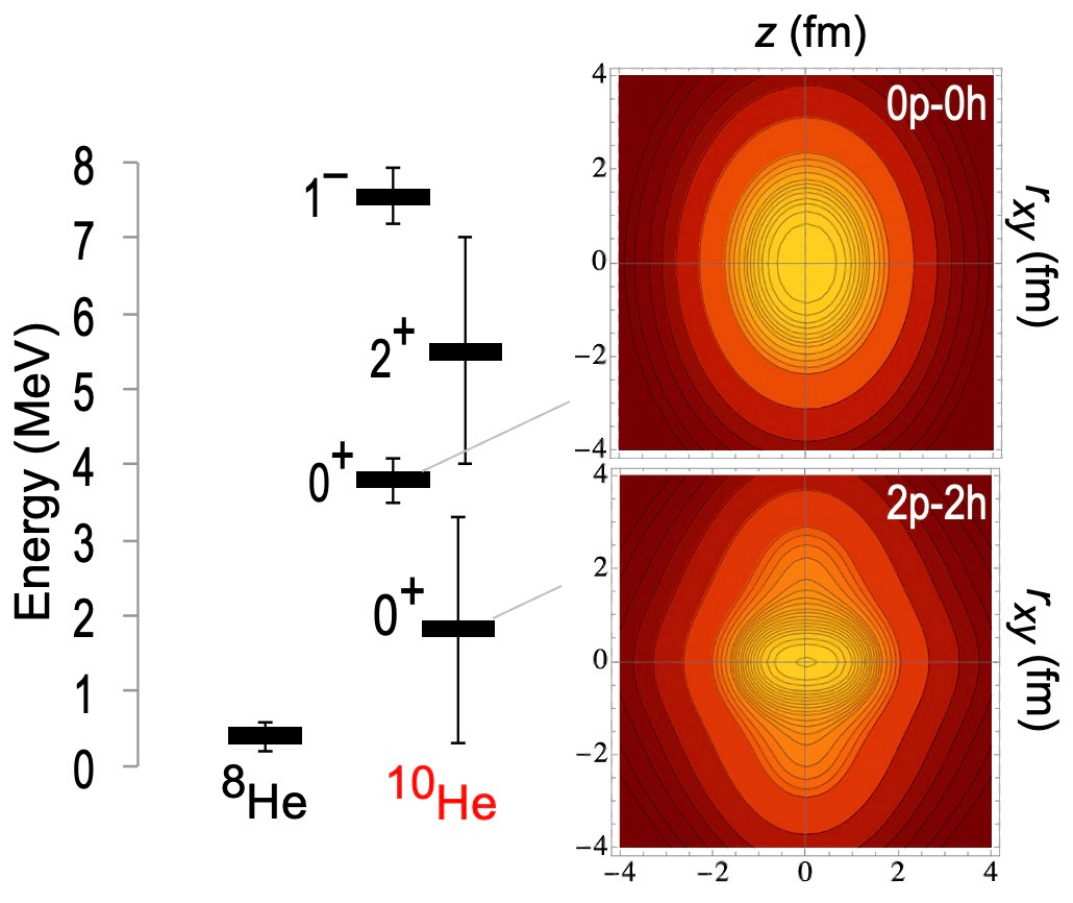}\\
     \hspace{0.15\linewidth} (a) 
     \hspace{0.45\linewidth} (b)
    \caption{SA-NCSM capabilities enable calculations beyond the oxygen isotopes, computations of B(E2) transition strengths \emph{without effective charges}, and descriptions of states dominated by multi-particle excitations, with illustrative examples: (a) $B(E2)$ transition strengths  for $^{21}$Mg and $^{21}$F in the SA-NCSM, compared to the experimental data of Ref.~\cite{Ruotsalainen19} and to the results of other theoretical approaches (models that adopt proton and neutron effective charges are also labeled by the corresponding $\Delta e^{\pi, \nu}$; see Ref.~\cite{Ruotsalainen19} for details). Figure from \cite{Ruotsalainen19}, with permission. (b) Energy spectrum of $^{8}$He (ground state) and $^{10}$He showing a deformed 2p-2h configuration as energetically favored and dominant in the $^{10}$He ground state, with the corresponding one-body density profiles.}
    \label{fig:21MgBE2}
\end{figure}

\begin{figure}[th]
    \centering
    \includegraphics[width=0.8\linewidth]{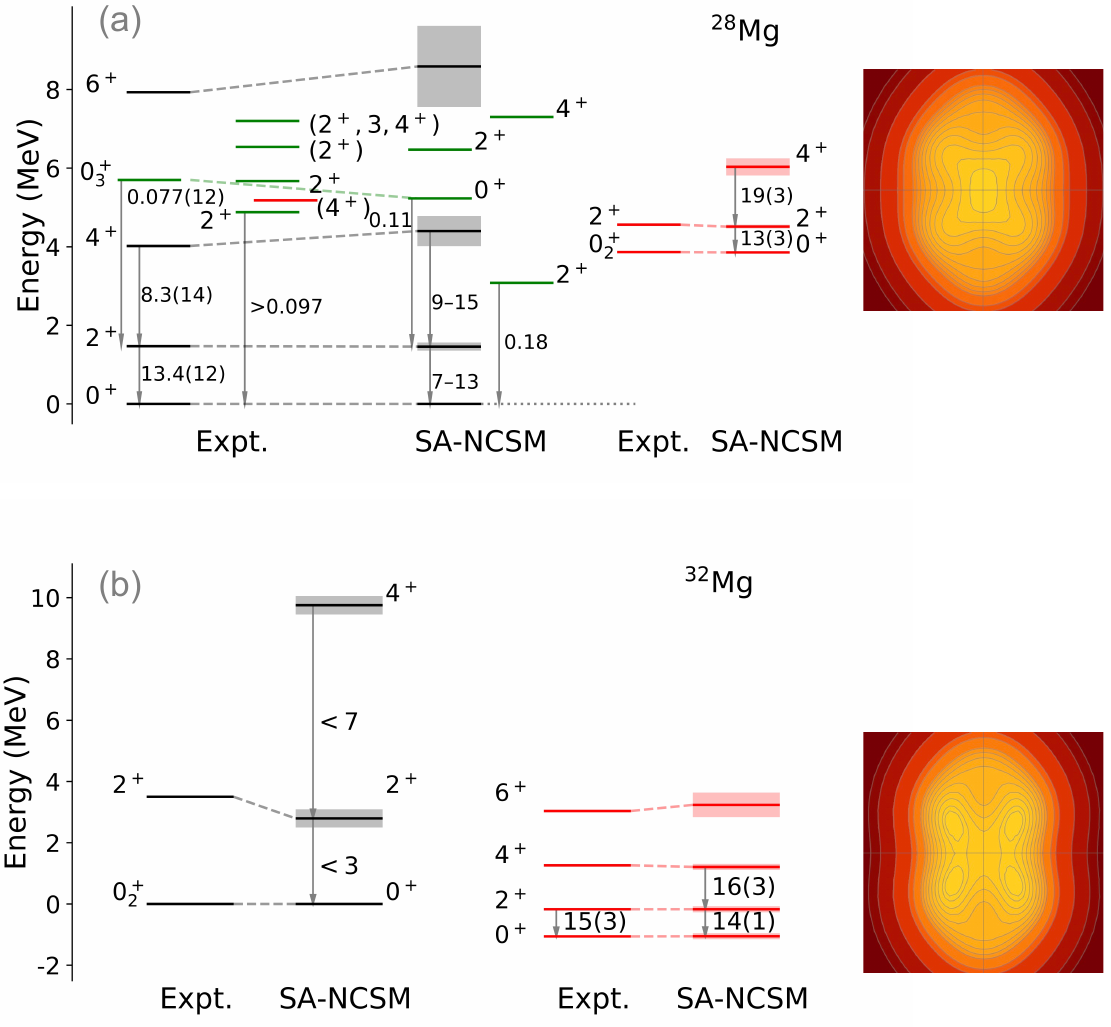}
    \caption{Energy spectra (in MeV) and $B(E2)$ transition strengths (in W.u.) for (a) $^{28}$Mg and (b) $^{32}$Mg calculated in the SA-NCSM with the NNLO$_{\rm opt}$ chiral potential, with uncertainties reported from extrapolations to the infinite-size model space, as compared to experiment (``Expt."): \cite{PhysRevC.100.014322} for $^{28}$Mg and \cite{physics4030048} for $^{32}$Mg. Calculated states with 0p-0h dominance are shown in gray and with 2p-2h dominance in red. For $^{28}$Mg, calculated levels with no uncertainties (green) are reported in \cite{PhysRevC.100.014322}; an experimental $(4^+)$ state is shown in red as a possible candidate that belongs to the 2p-2h rotational band. The one-body density profiles are shown to the right for the 2p-2h-dominated band (red). 
}
    \label{fig:Mg_spectra}
\end{figure}

\begin{figure}[th]
    \centering
    \includegraphics[width=0.7\linewidth]{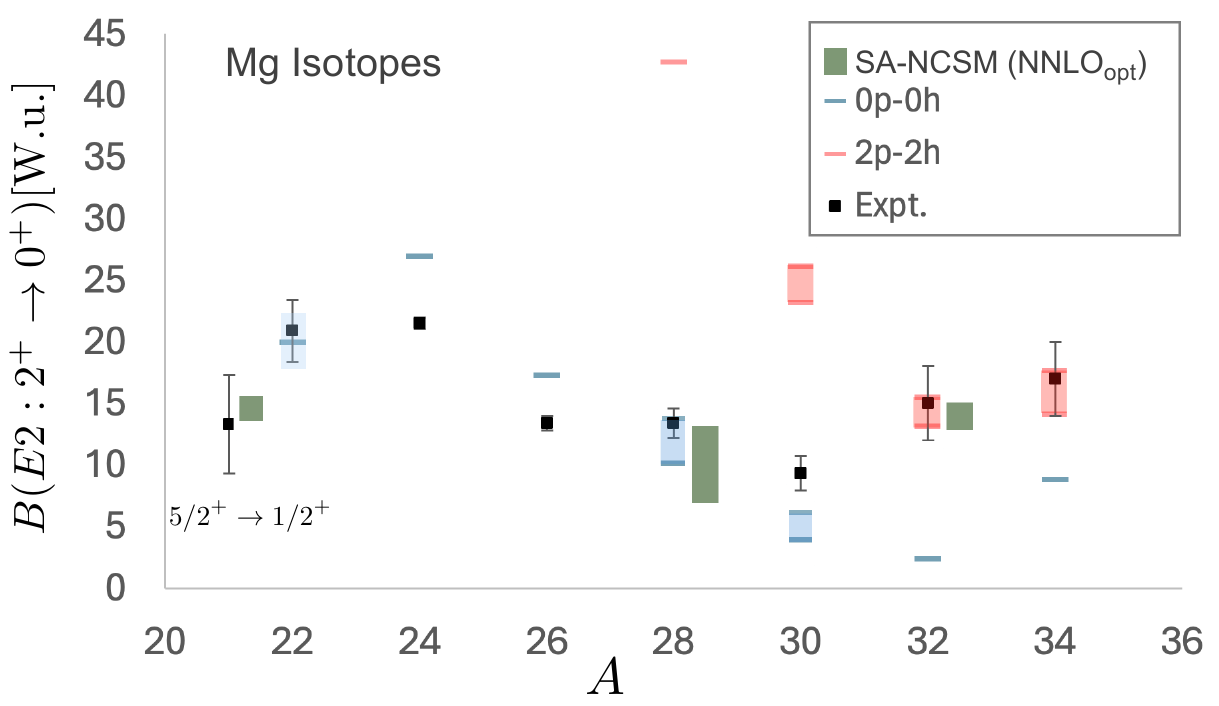}
    \caption{
    $B(E2)$ estimates for Mg isotopes calculated without effective charges using the microscopic interaction of Refs.~\cite{DreyfussLDDBp12,TobinFLDDB14} for the most deformed 0p-0h (blue) and 2p-2h (red) configurations (with no mixing between configurations) in the model space of 16 HO shells, as compared to experiment (black squares) and estimates from the SA-NCSM with NNLO$_{\rm opt}$ where available (green).  
    Results for $^{21}$Mg are from~\cite{Ruotsalainen19}; results
     for $^{22}$Mg are from~\cite{Henderson:2017dqc}, where theoretical results are reported with model uncertainties and validated in smaller model spaces against the SA-NCSM with NNLO$_{\rm opt}$. Note that B(E2) estimates for $Z=12$ and $A>30$ are consistent with dominant 2p-2h configurations. 
}
    \label{fig:Mg_BE2estimates}
\end{figure}

\begin{description}
    \item[B(E2) transition strengths] The SA-NCSM provides $B(E2) $ transition strengths \emph{without effective charges}. For example, the $B(E2) $ transition strengths for $^{21}$Mg and $^{21}$F  \cite{Ruotsalainen19} (Fig.~\ref{fig:21MgBE2}a) closely agree with the experiment and constitute the first \textit{ab initio} calculations of $B(E2)$ without effective charges in this mass region, which has been enabled by the SA basis.
    \item[Localized clusters] The SA-NCSM can include cluster degrees of freedom \cite{DreyfussLESBDD20,sargsyanlbgs2022}, such as the ones used in the NCSMC \cite{BaroniNQ13,HupinQN19}, thereby, providing estimates for resonance widths, ANCs, and reaction rates (see Sections~\ref{sec:alpha} \& \ref{sec:UQ}, and Fig. \ref{fig:SAfeatures}e for $\alpha+^{16}$O). 
    \item[Multi-particle excitations, or ``intruder states"]  
    Large-scale calculations have further revealed that the first-excited $0^+$ state in $^8$Be is dominated by 2p-2h and 4p-4h configurations (for further details, see Sec.~\ref{sec:A8beta}). While such a state is often referred to as an ``intruder" within a shell-model perspective, it has been argued that for many nuclei such ground states or low-lying $0^+$ states are practically the norm \cite{Wood16}. The situation is even more interesting in neutron-rich isotopes, especially in the proximity of the drip line, where the 2p-2h state is more energetically favored than the 0p-0h and becomes the lowest state, as for example in $^{10}$He (Fig.~\ref{fig:21MgBE2}b). In fact, these features only manifest themselves in sufficiently large model spaces (whereas 2p-2h and 4p-4h states lie higher in the energy spectrum even within $N_{\rm max}=12$ model spaces). Furthermore, the neutron-rich Mg isotopes reveal a remarkable pattern of 2p-2h dominance in the ground state based on measured and calculated B(E2) transitions 
    (Figs.~\ref{fig:Mg_spectra} \& \ref{fig:Mg_BE2estimates}). These features are further discussed next.
\end{description}

In particular, the SA-NCSM calculations for $^{10}$He (Fig.~\ref{fig:21MgBE2}b) reveal two $0^+$ states that lie close in energy but have different dominant shapes: the almost spherical shape appears as the $0^+_2$ excited state and the deformed shape with a dominant 2p-2h configuration is favored in the lowest state. This may explain the discrepancy in the $^{10}$He $0^+$ energy measured in different experiments (see, e.g., \cite{Kohley12,PhysRevC.91.044312}). This further points to the important role of multi-particle excitations in neutron-rich nuclei in the proximity of the drip line (see also \cite{MCCOY2024138870} for $^{12}$Be). The SA-NCSM calculations of Fig.~\ref{fig:21MgBE2}b are performed with the soft-core JISP16 interaction, for which the 2p-2h mode drops below the 0p-0h one at about $N_{\rm max}=30$. Fig.~\ref{fig:21MgBE2}b reports energies and their uncertainties that are extrapolated to the infinite-size model space, following the procedure of Ref. \cite{JurgensonNF09}, which is modified to also track with the size of the SA model spaces and is based on large-scale computations up through 12 HO shells across $\hw=15$--$25$ MeV.

As for the Mg isotopes, we show two examples in the SA-NCSM with NNLO$_{\rm opt}$, one for $^{28}$Mg (Fig.~\ref{fig:Mg_spectra}a), suggesting a 2p-2h nature of the 3.86-MeV $0^+_2$ state, and the other for $^{32}$Mg (Fig.~\ref{fig:Mg_spectra}b), which confirms the experimental identification of the $0^+_2$ state as dominated by 2p-2h configurations  \cite{PhysRevLett.105.252501,physics4030048}. 
This agrees with the outcome of Fig.~\ref{fig:Mg_BE2estimates}, which compares $B(E2)$ data to theoretical estimates obtained in the SA framework. Estimates with NNLO$_{\rm opt}$ are reported as extrapolated values and take full advantage of configuration mixing in the SA-NCSM (Fig.~\ref{fig:Mg_BE2estimates}, green). Complementary to these, calculations in 16 HO shells consider one to two most deformed 0p-0h and 2p-2h shapes without mixing (Fig.~\ref{fig:Mg_BE2estimates}, blue and red, respectively); these calculations use the microscopic interaction of the Hoyle-state study in Ref.~\cite{DreyfussLDDBp12}, along with the no-core symplectic shell-model (NCSpM) code~\cite{BahriR00,DreyfussLDDBp12,TobinFLDDB14}, which by utilizing analytical group-theoretical expressions enables the reach of ultra-large model spaces. Remarkably, the outcome of Fig.~\ref{fig:Mg_BE2estimates} suggests that the ground states of Mg isotopes for $A>30$ are consistent with dominant 2p-2h configurations, while lighter Mg isotopes are likely dominated by 0p-0h modes. The experimental $B(E2)$ value for $^{30}$Mg being intermediate between the calculated NCSpM 0p-0h and 2p-2h values suggests that the $^{30}$Mg ground state may have significant 0p-0h and 2p-2h mixing.

Indeed, the $^{28}$Mg analysis suggests that while the ground-state rotational band is dominated by a 0p-0h nuclear shape, 2p-2h configurations are found dominant in the first excited $0^+_2$ state (Fig.~\ref{fig:Mg_spectra}a). Both of these shapes are prolate, with the $0^+_2$ shape having much larger deformation. Energy calculations of the yrast band (Fig.~\ref{fig:Mg_spectra}a, gray) and other low-lying states, along with $E2$ transition strengths, are in remarkable agreement with experiment. An experimental $(4^+)$ state that lies closely to the calculated $4^+$ member of the 2p-2h dominated rotational band (Fig.~\ref{fig:Mg_spectra}a, red) is suggested as a possible candidate for this band. In addition, there are two  $2^+$ states with very close experimental energies, 4.56 MeV and 4.88 MeV. The 4.88-MeV state (Fig.~\ref{fig:Mg_spectra}a, the lowest $2^+$ state in green under ``Expt.") has a $B(E2:2^+\rightarrow 0^+_{\rm g.s.})$ lower limit similar to the SA-NCSM one from the $2^+_2$ member of the g.s. rotational band ($0.18$ W.u.). However, it has not been assigned to this band, since its energy is higher than the SA-NCSM predictions. The tentative assignment of the lower 4.56-MeV $2^+$ state (Fig.~\ref{fig:Mg_spectra}a, red) to the rotational band dominated by 2p-2h configurations is based on energies only. Additional calculations and measurements of $E0$ and $E2$ transition strengths, and especially $B(E2:J\rightarrow J-2)$ for the 2p-2h band, would be indeed important for further advancing our understanding of the $^{28}$Mg spectrum. 

To provide a level of specificity for the SA-NCSM calculations of $^{28}$Mg (Fig.~\ref{fig:Mg_spectra}a) [and of $^{32}$Mg (Fig.~\ref{fig:Mg_spectra}b)], we note that they use the NNLO$_{\rm opt}$ chiral potential and are performed up through 9 HO shells. Based on these calculations, a 2-parameter  extrapolation is used that addresses the infrared cutoff. This cutoff
can be understood as the effective size of the model space (``box size") in which the nucleus resides (cf. Ref.~\cite{JurgensonNF09,WendtFPS15}). For each nucleus, the exponential attenuation coefficient $k_{\inf}$ is kept fixed for all the states, for  a given \hw. In the limit of the infinite-size model space (infinite box size), one of the parameters of the extrapolation determines the extrapolated energy (or $E2$ transition strengths), with uncertainties reported across $\hw=13$--$17$ MeV and ensuring nonsignificant dependence on $\hw$. We note that the $B(E2)$ strengths in $^{28}$Mg are in close agreement with the experimental data, with uncertainty bands consistent with the decrease in the $E2$ transitions with increasing spin. We note that the smaller SA-NCSM $B(E2: 2^+ \rightarrow  0^+)$ value reported in Ref.~\cite{PhysRevC.100.014322} is  from calculations with single $N_{\rm max}$ and single $\hw$. In fact, without the extrapolations, the $0^+_2$ rotational band lies higher in energy.

Importantly, the SA-NCSM calculations confirm the 2p-2h nature of the  $^{32}$Mg ground state as identified and discussed in Refs.~\cite{PhysRevLett.105.252501,physics4030048}. The agreement to experiment is remarkable, including energies and the $B(E2:2^+ \rightarrow 0^+)$ transition strength. In addition, the $0^+_2$ rotational band is found to have very small deformation, while the ground state is largely prolate. Similarly,  the outcome of Fig.~\ref{fig:Mg_BE2estimates} suggests a largely deformed and prolate shape for the $^{34}$Mg ground state.

\vspace{8pt}
\noindent
\textit{Overlap Functions and Asymptotics.} To recover the asymptotics, one utilizes cluster wavefunctions for the partitionings in consideration. For distance $r$ between the center-of-mass of two clusters, a cluster wavefunction, the so-called spectroscopic amplitude,  is given by:
\begin{eqnarray}
 u_{\nu I \ell}^{ J^{\pi}}(r) &=& \sum_{n_r}  R_{n_r \ell}(r)\braket{\Psi_{(A+a)\alpha}^{ J^\pi (M)}}{\Phi_{ (A)\alpha_1 I_1; (a)\alpha_2 I_2; I n_r \ell} ^{J^{\pi}(M)}} \cr
 &=& \sum_{n_r}  R_{n_r \ell}(r)\braket{\Psi_{(A+a)\alpha}^{ J^\pi (M)}}{ (\mathcal{A} \left\{ \Psi_{(A)\alpha_1 } ^{I_1} \otimes \Psi_{ (a)\alpha_2 } ^{I_2} \right\}^I \otimes \psi_{n_r \ell})^{ J^\pi (M)}}, 
 \label{eq:overlapdef}
\end{eqnarray}
which is calculated through the translationally invariant (t.i.) spectroscopic overlap, as detailed in Refs. \cite{sargsyan:23,DreyfussLESBDD20} (cf. \cite{Navratil04,NavratilBC2006}),
\begin{eqnarray}
    u_{\nu I  n_r \ell}^{ J^{\pi}} = \braket{\Psi_{(A+a)\alpha}^{ J^\pi (M)}}{ (\mathcal{A} \left\{ \Psi_{(A)\alpha_1 } ^{I_1} \otimes \Psi_{ (a)\alpha_2 } ^{I_2} \right\}^I \otimes \psi_{n_r \ell})^{ J^\pi (M)}},
     \label{eq:overlapdefconfig}
\end{eqnarray}
where $\ket{\Phi_{ (A)\alpha_1 I_1; (a)\alpha_2 I_2; I n_r \ell}}$ are cluster states, $\Psi_{(A)}$ and $\Psi_{(a)}$
are the target and projectile  wavefunctions, respectively, $\psi$ describes their relative motion, and $\mathcal{A}$ is the antisymmetrization operator 
that acts on nucleons belonging to different clusters and ensures the Pauli exclusion principle. The cluster system is defined for a partial wave $\ell$, the orbital angular momentum of the relative motion of the clusters, and for a channel $\nu$, which is given by the spin and parity of each of the clusters $\nu  = \{ \alpha; (A)\alpha_1 I_1^{\pi_1}; (a)\alpha_2 I_2^{\pi_2}\}$ (the labels $\alpha$, $\alpha_1$ and $\alpha_2$ denote all other quantum numbers needed to fully characterize their respective states). It has a good total angular momentum and parity, $J^{\pi}$, given by the coupling of the channel spin  $I$ (the total angular momentum  of the clusters) to $\ell$. $R_{n_r \ell}(r)$ are the HO radial functions, with $n_r$ denoting the radial HO quantum number. 
We note that for overlaps calculated in the SA-NCSM, $R_{n_r \ell}(r)$ are defined as \emph{positive at infinity} (this convention is adopted throughout the paper).

The norm of the spectroscopic overlap,
\begin{eqnarray}
 S_{\nu I \ell}^{ J^{\pi}} &=& \int_0^\infty |u_{\nu I \ell}^{ J^{\pi}}(r)|^2 r^2 dr = \sum_{n_r} |u_{\nu I  n_r \ell}^{ J^{\pi}}|^2 ,
\label{eq:SF}
\end{eqnarray}
is called the spectroscopic factor (SF).

We use the SA-NCSM cluster wavefunction of Eq.~\eqref{eq:overlapdef} in the interior and the exact Coulomb function at large distances, both of which are matched using logarithmic derivatives within an R-matrix framework~\cite{Descouvemont:10}, as often done in many-body approaches (see, e.g.,\cite{Brida:11,Kravvaris:24,mercennemp19,DreyfussLESBDD20}). This prescription benefits from the use of SA-NCSM to calculate the interior wavefunction, since  sufficiently large model spaces are critical to achieve accurate descriptions of the interior wavefunction and corresponding dynamical observables. 

At large  distance $r$, the resonances are described by the outgoing spherical Hankel function solution to the Coulomb equation, $H_{\ell}^{+}(\eta,kr)$, defined by the Sommerfeld parameter $\eta=\frac{Z_{A}Z_{a}\mu_{A,a} e^2}{\hbar^2 k}$ for two clusters of charge $Z_{A}$ and $Z_{a}$, and the momentum $k$ corresponding to the energy in the c.m. frame
$|E|=\frac{\hbar^2k^2}{2\mu_{A,a}}$ with reduced mass
\begin{equation}
\mu_{A,a} = {m_{A}m_{a}\over m_{A}+m_{a} },
\label{eq:redmass}
\end{equation}
for clusters with masses $m_{a}$ and $m_{A}$. The partial decay widths $\Gamma_a$ is then determined as (see, e.g., \cite{Descouvemont:10,DreyfussLESBDD20}):
\begin{equation}
    \label{eq:width}
    \Gamma_{a} = \frac{k\hbar^2}
    {\mu_{A,a}} \left|\frac{ru_{\nu Il}^{J^{\pi}}(r)}{H_{\ell}^{+}(\eta,kr)} \right|^{2}_{r=r_{c}} ,
\end{equation}
calculated at a channel radius $r_c$. In microscopic calculations, converged $\Gamma_{a}$ widths are practically independent from the choice of $r_c$.

At large  distance $r$, the cluster wavefunction  $u_{\nu I \ell}^{J^\pi}(r)$ for bound states should asymptotically approach the exact Coulomb Whittaker function, $W_{-\eta, \ell+\frac{1}{2}}$:
\begin{equation}
  r u_{\nu I \ell}^{J}(r) \rightarrow C_{\nu I \ell}^J  W_{-\eta, \ell+\frac{1}{2}}(2kr),
  \label{eq:ANC}
\end{equation}
where the $C_{\nu I \ell}^J$ coefficient is the ANC, which is determined from this equation. While various methods exist for calculating ANCs (see, e.g., \cite{Brune_PRC66_2002,NollettW2011,Timofeyuk2010}), we use the  prescription of Ref. \cite{sargsyan:23}, which utilizes Eq.~\eqref{eq:ANC}, while retaining the microscopically calculated SF.

\section{Single-nucleon spectroscopic factors}
\label{sec:SFovlps}

Historically, spectroscopic factors have been used to approximately account for the intrinsic structure of the nuclei, whereas 
single-nucleon SFs have served as a measure of the single-particle content of a nuclear state. 
Experimentally, they are obtained from direct reaction measurements and serve as a normalization factor in the reaction cross section. However, SFs deduced from experiments also depend on the reaction type and the reaction model employed. 
SFs that are directly derived from \textit{ab initio} many-body solutions at infinite-size model spaces can be considered as a true measure of the single-nucleon  
partitioning in nuclei.  In fact, many calculations of direct-reaction cross sections often use well-informed SFs and global optical potentials as the only inputs. Furthermore, unlike cross sections, which can vary significantly based on the energy of the projectile and the mass range of the target, 
the spectroscopic factor may be regarded as a more straightforward quantity 
for comparing different systems. Although SFs, in general, are model dependent, their extraction from various types of reactions can serve as indicators of the physical significance of the underlying model
 (see the comprehensive review \cite{Aumann:21} and references therein, as well as, e.g., \cite{KaySF2013}).
Even further, SFs can be directly used in constructions of optical potentials as they can be interpreted as a measure of coupling strengths for different reaction channels \cite{SargsyanPKE2024} (see Sec.~\ref{sec:selfie}). 
\begin{figure}[btp]
    \centering
    \includegraphics[width=0.99\textwidth]{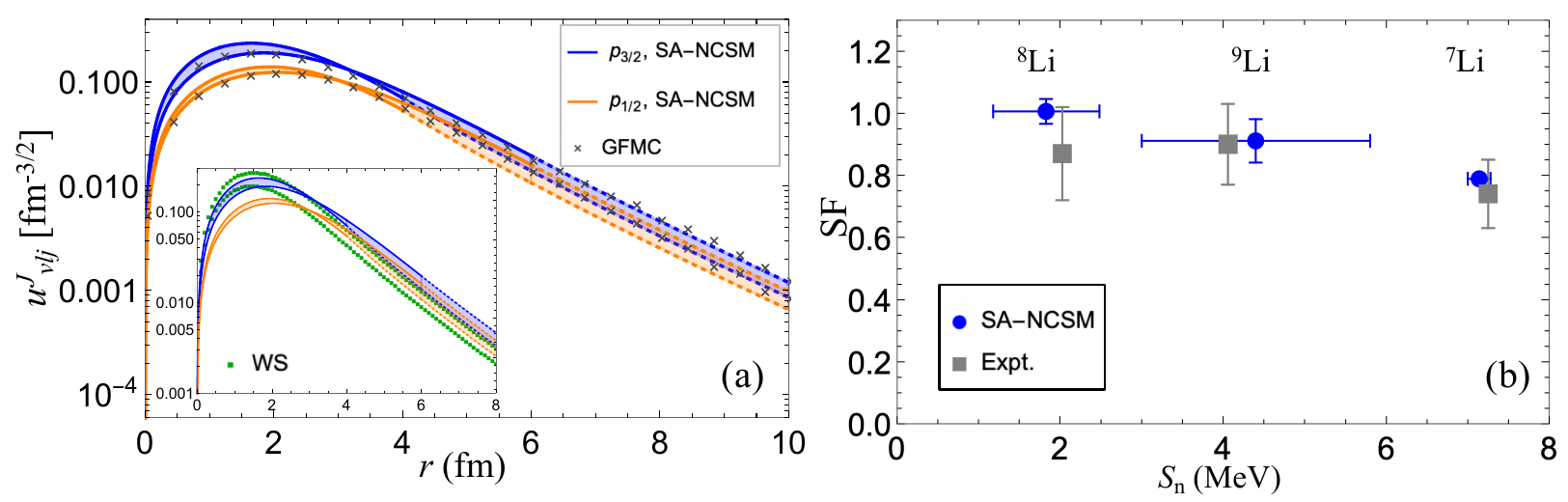}
    \caption[Single-nucleon overlaps and spectroscopic factors of $\braket{^7\mathrm{Li}}{^6\mathrm{Li}+{\rm n}}$]{(a) Single-nucleon overlaps of the $^7$Li ground state with the $^6\mathrm{Li}+{\rm n}$ in $N_{\rm max}=12$ for partial waves $p_{1/2}$ and $p_{3/2}$ vs. the separation between $^6\mathrm{Li}$ and the neutron, $r$, compared to the GFMC results from Ref. \cite{QMC_database}. The dotted lines correspond to the exterior Whittaker function. The shaded bands indicate the uncertainty due to the \hw~variance from 10 to 20 MeV (interior to exterior wavefunction matching radii differ for different \hw, see text for details). For all but the first two points, the GFMC uncertainties are smaller than the marker size on the plot. Inset: SA-NCSM overlaps compared to a typical Woods-Saxon parameterization (see text for details). (b) Calculated SFs in $N_{\rm max}=12$ model space vs. the extrapolated neutron separation energies ($S_{\rm n}$) and the experimentally deduced SFs from Refs. \cite{WuosmaaSRGH2008, Schiffer8Li1967, Wuosmaa9Li2005}. Uncertainties on extrapolated $S_{\rm n}$ are from variances in \hw~and the use of SA model spaces. Uncertainties on the experimental $S_{\rm n}$ are smaller than the marker size.
    Figure adapted from Ref. \cite{sargsyan:23}, with permission.
    }
   \label{fig:6Li_7Li_Overlap}
\end{figure}

We employ the SA-NCSM framework to compute the spectroscopic overlap for a single-nucleon projectile ($a=1$) interacting with a composite nucleus of mass $A+1$. 
To calculate single-nucleon spectroscopic overlaps, we follow Eq. (\ref{eq:overlapdef}) and its notations, but expressed in the single-particle $j$ coupling scheme \cite{Navratil04,NavratilBC2006}:
\begin{eqnarray}
 u_{A \alpha_1 I_1; \ell {1\over 2}j}^{ J^\pi}(r) 
 &=&\sum_{n_r} R_{n_r \ell}(r) \frac{1}{\Pi_J}  \frac{\RedME{\Psi_{(A+1) \alpha}^J}{\adagg_{n_r lj}}{\Psi_{(A)\alpha_1}^{I_1} }
 }{\braket{n_rl00l}{00n_rll}_{1/A}}.
 \label{eq:overlapj}
\end{eqnarray}
Here,
the matrix element, $\RedME{}{}{}$, is reduced with respect to angular momentum, but calculated in the \SU{3} basis, that is, using the eigenvectors of the initial and final many-body states expressed in the SU(3)-coupled basis, and $a^\dagger_{n_rlj}$ is the creation operator that creates a particle in the $\{n_rlj \}$ single-particle state.  
The orbital momentum $\ell$ of the nucleon coupled with its spin-1/2 yields $j$ (since we work in a proton-neutron formalism, isospin is not a good quantum number of the basis). The bra and ket eigenstates correspond to the laboratory-frame wavefunctions of the composite and target nuclei, respectively, calculated from the many-body theory. A Lawson procedure ensures that both eigenfunctions can be factorized to an intrinsic wavefunction  and a c.m. wavefunction that is in the lowest HO state ($\eta_{\rm c.m.}=0$ and $\ell_{\rm c.m.}=0$) \cite{Lawson74}. The Talmi-Moshinsky bracket in the denominator of Eq.~\eqref{eq:overlapj}, which transforms the  c.m. coordinates of the two clusters in the laboratory frame to a relative distance $r$ between the clusters (with the c.m. coordinate of the N-A system integrated out in the calculation of $u$), is given as
$\braket{n_rl00l}{00n_rll}_{a/A} = (-1)^\ell \Big(\frac{A}{A+a} \Big)^{(2n_r+\ell)/2}$,
where $a$ and $A$ are the numbers of nucleons in the clusters (see, e.g., \cite{Navratil04}). This ensures that the spectroscopic overlaps are translationally invariant.

Ref.~\cite{sargsyan:23} employs the SA-NCSM to calculate single-neutron SFs for the ground states of $^7$Li, $^8$Li and $^9$Li isotopes. The ground states of Li isotopes that differ in mass by one nucleon have opposite parities. The parity of a nucleon in a certain orbital is given by $(-1)^\ell$, hence $\ell$ of the added nucleon must be odd to preserve the parity between the ground states. Additionally, the $\ell=3$ overlaps are much smaller than the $\ell=1$ overlaps as it was shown in Ref. \cite{NollettW2011} and observed in our calculations. Thus, we present only the overlaps for $p_{1/2}$ and $p_{3/2}$ partial waves. 

We compare the SA-NCSM overlaps (Fig. \ref{fig:6Li_7Li_Overlap}a) to those calculated in the \emph{ab initio} GFMC framework with the AV18 + IL7 interaction, as well as to calculations with a typical Woods-Saxon (WS) potential 
\begin{equation}
    V_{\rm WS}(r)=\frac{V_0}{1+\mathrm{exp}[(r-R_0)/a]}.
    \label{eq:WS_pot}
 \end{equation}
Both SA-NCSM and GFMC approaches yield very similar overlaps despite the fact that they use different NN interactions. In contrast, the WS overlaps peak at higher values for both partial waves and drop below the \emph{ab initio} overlaps at long distances.  In these calculations, the WS potential depth has been adjusted to reproduce the experimental neutron separation energy of $^7$Li for each partial wave: $V_0=-71.95$ MeV for $p_{1/2}$ and $V_0=-61.31$ MeV for $p_{3/2}$; we use a spin-orbit depth of $V_\mathrm{SO}=6$ MeV, as well as radius of $R_0=1.25A^{1/3}$ fm  and diffuseness of $a=0.65$ fm in both central and spin-orbit terms. The overlaps from the WS solutions have been normalized to reproduce the same SFs as the SA-NCSM ones.

In the cases of $^7$Li, $^8$Li and $^9$Li, we use Eq. (\ref{eq:SF}) to obtain the SFs for each of the partial waves, with the total SF given by the sum of the SFs for both partial waves. The calculated total SFs for the $N_{\rm max}=12$ model spaces reproduce the experimentally deduced values within uncertainties (Fig. \ref{fig:6Li_7Li_Overlap}b). We note that for all these nuclei the SFs change slowly with increasing model-space size.  Hence, we do not expect the SFs at the $N_{\rm max}\rightarrow \infty$ limit to be significantly different from the $N_{\rm max}=12$ values. In all cases the calculated SFs are approximately unity, with the exception of the $^6$Li+n case (Fig. \ref{fig:6Li_7Li_Overlap}b). This discrepancy indicates a more complex structure for the ground state of $^7$Li, potentially linked to the low-lying $\alpha + t$ threshold that is positioned closer to the ground state than the $^6$Li+n threshold. In comparison, for both $^8$Li and $^9$Li, the neutron channel is the lowest in energy. 
By examining the dependence of the SFs on the neutron threshold, we observe a gradual decline in the SFs as the separation energy increases (Fig. \ref{fig:6Li_7Li_Overlap}b), similar to Fig. 18 in Ref. \cite{Aumann:21}. This indicates a more pronounced clustering of single particles in scenarios of weaker neutron binding. 

We note that the neutron thresholds reported in Fig. \ref{fig:6Li_7Li_Overlap}b are determined from $N_{\rm max} \rightarrow \infty$ extrapolations of the binding energies, with uncertainties that take into account  variances in \hw~and the use of SA model spaces. All the thresholds are in agreement with the experimental values. 

The single-nucleon spectroscopic factors presented in this section provide initial information on the dependence of the SFs on the neutron separation energy, which is promising and motivates further explorations. We reserve a comprehensive analysis for future studies utilizing the SA-NCSM, which will encompass a wider range of isotopes, from light to medium-mass nuclei.

\section{Neutron and proton elastic scattering: from alpha to Calcium}
\label{sec:scatt}

In this section, we present differential and total cross sections for neutron and proton elastic scattering for targets from alpha to Calcium. We discuss several approaches that utilize the SA concept for targets in the mass range from Helium to Calcium isotopes.

\subsection{SA-NCSM Green's function approach: Cross sections and optical potentials}
\label{sec:GF}

Optical potentials or inter-cluster effective interactions are a main ingredient in reaction theory. Using chiral potentials, microscopic nonlocal dispersive optical potentials have been derived in Refs.~\cite{RotureauDHNP17,PhysRevC.98.044625,idini19,BurrowsLMBSDL24} at low projectile energies ($\lesssim20$ MeV per nucleon), as well as in Refs.~\cite{BurrowsBEWLMP20,VorabbiGFGNM2022} in the intermediate-energy regime ($\gtrsim 65$ MeV per nucleon). Similarly, optical potentials for light to heavy nuclei have been derived from two- and three-nucleon chiral forces in nuclear matter \cite{WhiteheadLH2019}. 

Recently, in the SA-NCSM framework, Ref.~\cite{BurrowsLMBSDL24} has provided the first \textit{ab initio} nucleon-nucleus (NA) nonlocal dispersive optical potentials that are translationally invariant and energy dependent, derived from finite nuclei, and applicable to a broad range of open-shell spherical and deformed targets. In addition, they can  accommodate nuclear features of single-particle, collective, and clustering nature, through the use of the SA-NCSM. This is achieved by combining the Green's function (GF) approach with SA-NCSM (dubbed SA-NCSM/GF~\cite{BurrowsLMBSDL24}). The GF technique proffers important advantages: the NA optical potential includes the information about all near reaction channels  through the GF calculations in the $(A\pm1)$ systems; the imaginary part of the optical potential allows one to account for the loss of flux to open channels;  and such an effective potential constitutes a good mean-field inter-cluster potential~\cite{Escher:02PRC}. 

For a particle projectile and a target in its ground state $\ket{\Psi_{(A)\rm{g.s.}}^{J_0}} $ with energy $E_{\rm{g.s.}}^A$ and total angular momentum $J_0$, the Green's function is calculated for an orthonormal cluster basis in a ``particle-hole" space. The cluster basis state can be given symbolically as $\ket{\Phi}=[a+a^\dagger] \ket{\Psi_{(A)\rm{g.s.}}^{J_0}}$. The first term corresponds to the $A-1$ system (or ``hole" space), while the second term corresponds to the A+1 system (or ``particle" space). This second state is the same cluster basis state (not normalized) as the one used in the RGM approach (Sec.~\ref{sec:RGM}), and is the antisymmetrized product  $\mathcal{A} ( \Psi_{(A)\rm{g.s.}}^{J_0} \otimes \psi_{a} )  $ of the wavefunctions of the target  and projectile in the $a \equiv \{ \eta_a (\ell_a \half) j_a \}$ single-particle state~\cite{BurrowsLMBSDL24}. An additional feature of the GF basis is the inclusion of the ``hole" space, with which the norm for each cluster basis state is unity by default,  $\braket{\Phi}{\Phi} = \bra{\Psi_{(A)\rm{g.s.}}^{J_0}} a^\dagger a \ket{\Psi_{(A)\rm{g.s.}}^{J_0}}+\bra{\Psi_{(A)\rm{g.s.}}^{J_0}} a a^\dagger \ket{\Psi_{(A)\rm{g.s.}}^{J_0}}=1$. Here again, the second term is the norm used in the RGM approach to orthogonalize the RGM cluster basis states defined in the ``particle" space (see also the Appendix in Ref.~\cite{BurrowsLMBSDL24}). Importantly, as it is also evident from the similarity of the outcomes presented by the SA-NCSM/GF and the SA-RGM, the reaction observables are independent from the choice of the space employed (``particle-hole" or ``particle" only).

In this orthonormal cluster basis the Green's function
is given as (cf. Ref. \cite{BurrowsLMBSDL24,Birse:1981ghl}):
\begin{eqnarray}
G^J_{ab}  
&=&G^{J+}_{ab}+ (-1)^{2J_0+1}\sum_{J'}\Pi^2_{J'}
 \sixj{j_a}{J_0}{J'}{j_b}{J_0}{J}G^{J'-}_{ba}, \nonumber
 \label{GFconfig2nonzeroJ0}
\end{eqnarray}
with the ``particle" ($G^+$) and ``hole" ($G^-$) Green's function defined as: 
\begin{equation}
   G^{J\pm}_{J_0;ab}(E)= \lim_{\epsilon \rightarrow 0} \bra{\Phi^{J\pm}_{ J_0 a }} \frac{1}{E \mp (H - E^A_{\rm{g.s.}}) \pm \iu\epsilon} \ket{\Phi^{J\pm}_{J_0 b } }, 
   \label{eq:Gpm}
\end{equation}
where $E$ is the energy for the reaction in the c.m. frame, $\hat H$ is the ($A\pm 1$)-body realistic Hamiltonian, 
and $a$ and $b$ are single-particle quantum numbers associated with the projectile. To provide a level of specificity, the cluster basis states with good total angular momentum are defined for the ``particle" ($+$) and ``hole" ($-$) cases  through the single-nucleon overlaps for laboratory particle coordinates (L): 
\begin{eqnarray}
	\ket{ \Phi^{J(M)+}_{ J_0a} }_{\rm L} &\equiv& (-1)^{\phi_a}\left\{ a_{a}^{\dagger} \otimes \ket{\Psi_{(A)\rm{g.s.}}^{J_0}} \right \}^{J(M)} =-\sum_{t}\frac{1}{\Pi_{J}}\ket{tJ(M)}\RedME{tJ}{a_{a}^{\dagger}}{\Psi_{(A)\rm{g.s.}}^{J_0}}, 
    \\
		\ket{ \Phi^{J(M)-}_{ J_0 a }}_{\rm L} &\equiv& (-1)^{\eta_a+\phi_a}\left\{ \tilde a_{a} \otimes \ket{\Psi_{(A)\rm{g.s.}}^{J_0}} \right \}^{J(M)} =\sum_{t}\frac{(-1)^{1+\eta_a}}{\Pi_{J}}\ket{tJ(M)}\RedME{tJ}{\tilde a_{a}}{\Psi_{(A)\rm{g.s.}}^{J_0}}, \nonumber
\label{Eq:pivots}
\end{eqnarray}
where $\phi_a=j_a+J_0-J$, $t$ is the complete many-body $A\pm 1$ basis, and $\Pi_J = \sqrt{2J+1}$. In the SA-NCSM, we use SU(3) proper tensors for the creation and annihilation operators, $a_{(\eta_a\,0) \ell_a j_a m_a}^\dagger \equiv \hat{a}_{a m_a}^\dagger$ and $\tilde a_{(0\,\eta_a) \ell_a j_a -m_a} =(-)^{\eta_a+j_a-m_a} \hat{a}_{a m_a} $, respectively.
In this work, the eigenfunctions $\ket{\Psi_{(A)\rm{g.s.}}^{J_0}}_{\rm L}$ and the single-particle overlaps  for the cluster basis states for each $a$ and $J$ in Eq.~\eqref{Eq:pivots} are calculated in the SA-NCSM, but any many-body approach, which utilizes intrinsic particle coordinates or exact c.m. factorization, can be adopted. In the SA-NCSM/GF,  the $\ket{\Phi^{J\pm}_{J_0a}}_{\rm L}$ cluster basis states are then used to calculate $\braketop{\Phi^{J\pm}_{J_0a}}{\hat{G}(E,\epsilon)}{\Phi^{J\pm }_{J_0b}}$ matrix elements for the t.i. Green's function through the Lanczos method, discussed in Ref.~\cite{BurrowsLMBSDL24}. We note that a special care is needed to remove the spurious c.m. motion in the Green's function formalism, and a prescription to achieve this is detailed in~\cite{BurrowsLMBSDL24}. Namely, it requires the explicit removal of the spurious c.m. motion for each cluster basis, and employs a Lawson procedure, similar to the one used in no-core shell-model calculations, that pushes states with spurious c.m. motion to high energies \cite{Lawson74}. In addition, the transformation from the laboratory cluster coordinates to the relative distance $r$ and the c.m. coordinate of the composite target-projectile system through a Talmi-Moshinsky bracket leads to a mass-dependent factor that multiplies the $\braketop{\Phi^{J\pm}_{J_0a}}{\hat{G}(E,\epsilon)}{\Phi^{J\pm }_{J_0b}}_{\rm L}$ matrix element, namely, $\left(\frac{A+1}{A}\right)^{\frac{\eta_a+\eta_{b}}{2}}$ for the $(+)$ case and $\left(\frac{A}{A-1}\right)^{\frac{\eta_a+\eta_{b}}{2}}$ for the $(-)$ case (for details, see Ref.~\cite{BurrowsLMBSDL24}). 

Alternatively, instead of employing the Lanczos method to calculate the GF matrix elements, one can use the completeness of the many-body Hamiltonian eigenfunctions for the $A\pm 1$ systems in Eq. (\ref{eq:Gpm}), the so-called Lehmann representation:
\begin{equation}
   G^{J\pm}_{J_0;ab}(E)= \lim_{\epsilon \rightarrow 0}  \sum_{\alpha}
   \frac{\braket{\Phi^{J\pm}_{ J_0 a }}{\Psi_{(A+1)\alpha }^{J}} \braket{\Psi_{(A+1)\alpha }^{J}}{\Phi^{J\pm}_{J_0 b } } }{E \mp (E^{A+1}_{\alpha} - E^A_{\rm{g.s.}}) \pm \iu\epsilon}  , 
   \label{eq:GpmLehmann}
\end{equation}
where $\braket{\Phi^{J\pm}_{ J_0 a }}{\Psi_{(A+1)\alpha }^{J}} $ are the spectroscopic amplitudes as defined in Eq.~\eqref{eq:overlapdef}.

With the s.p. HO radial wavefunctions $R_{\eta \ell}(r)$, used in this study, one obtains for  $G$ (similarly, for  $G^{-1}$):
\begin{eqnarray}
    G_{J_0;\ell j \ell' j'}^{J}(r,r';E) &=&\sum_{\eta_a \eta_b}R_{\eta_a \ell}(r) R_{\eta_b \ell'}(r')G_{J_0;\eta_a \ell j, \eta_b \ell' j'}^J(E).
\end{eqnarray}

\begin{figure}[th]
    \centering
    {\includegraphics[width=0.49\linewidth]{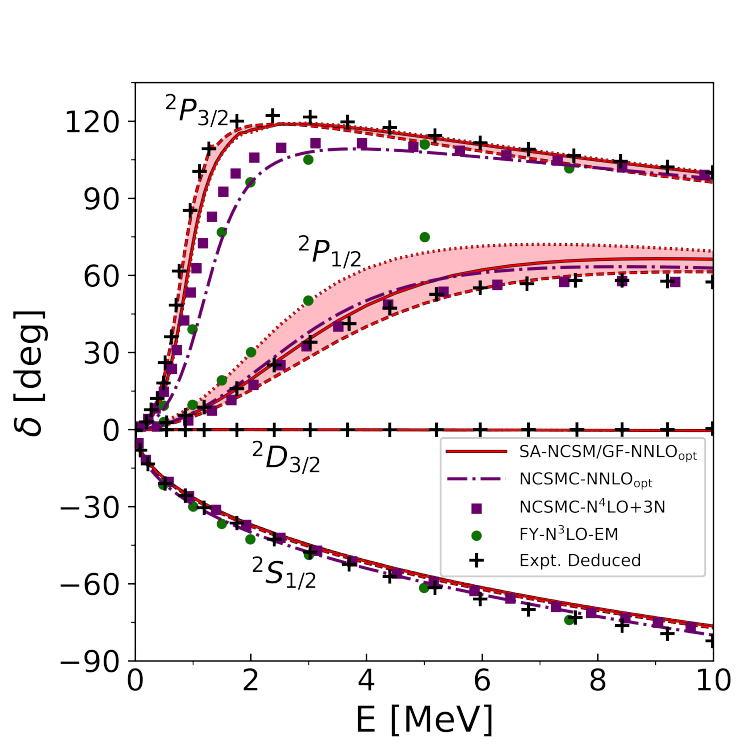}}
    {\includegraphics[width=0.49\linewidth]{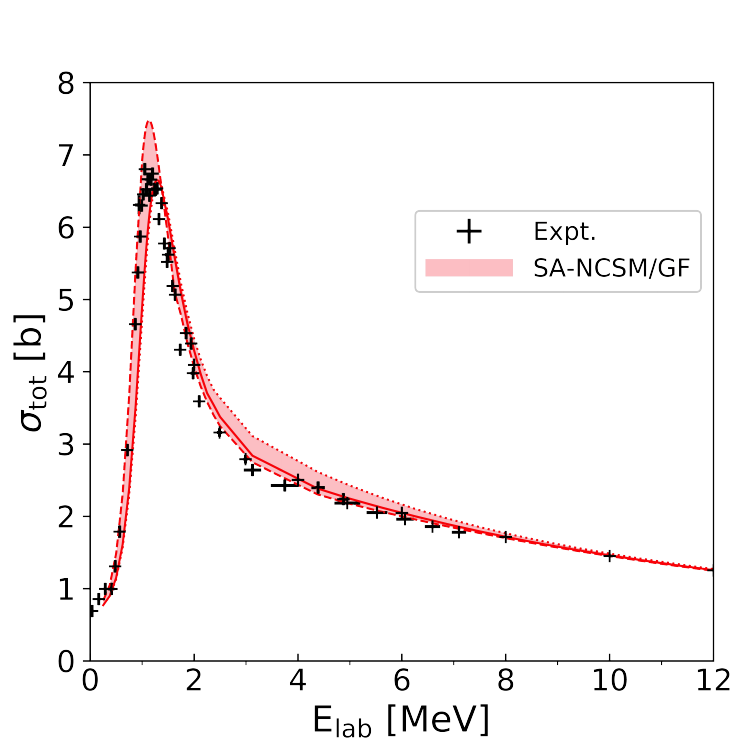}}\\
    \hspace{0.1\linewidth} (a) \hspace{0.45\linewidth} (b)
    \caption{(a) Neutron-$^{\rm 4}$He phase shifts as a function of the energy in the c.m. frame using the \textit{ab initio} SA-NCSM/GF for $N_{\rm max}=13$ with NNLO$_{\rm opt}$ (red). These are compared to the outcomes of other approaches and the experimentally deduced values (black crosses) obtained using an R-matrix analysis.
    (b) $^{\rm 4}$He(n,n)$^{\rm 4}$He total cross sections vs. the laboratory-frame projectile kinetic energy obtained using the \textit{ab initio} SA-NCSM/GF and compared to three sets of experimental data \cite{Total_Cross_Section_Expt1, Total_Cross_Section_Expt2, Total_Cross_Section_Expt3} (labeled as ``Expt."). In (a) and (b), the  red bands show the $\hw$ spread to guide the eye, with $\hbar\Omega=12$ (red dashed), 16 (red solid), and $20$ MeV (red dotted). 
    There are energy ranges where curves are indistinguishable from each other.
    Figures adapted from \cite{BurrowsLMBSDL24}, with permission.}
    \label{fig:SANCMGF}
\end{figure}

As a first application, we employ the SA-NCSM/GF approach to calculate optical potentials, phase shifts, and cross sections for neutron elastic scattering off a $^4$He target, since those could be benchmarked against earlier calculations obtained in complementary approaches (Fig. ~\ref{fig:SANCMGF}). We use the effective potential for the channels $c \equiv \{(A) {\rm g.s.}, J_0; \ell j \}$ and $c'$, given as (for further details, see Ref.~\cite{BurrowsLMBSDL24}):
\begin{eqnarray}
    V_{c c'}^J(r,r')
    =[E-T_{\rm rel}(r)]\delta_{\ell \ell'}\delta_{j j'}\frac{\delta(r-r')}{r r'} 
    -(G_{c c'}^{J})^{-1}(r,r',E).
    \label{eq:GFoptpot}
\end{eqnarray}

The SA-NCSM/GF largely benefits from the use of extrapolated energies, that is, energies computed in finite model spaces in the SA-NCSM and subsequently extrapolated to the infinite-size model space. Indeed, using extrapolated energies, phase shift for n+$^4$He perform remarkably well when compared to both experimentally deduced values and earlier theoretical calculations carried out in alternative frameworks (Fig.~\ref{fig:SANCMGF}a). The experimentally deduced phase shifts are calculated from experimental total cross sections using the R-matrix method (for details, see Ref.~\cite{quaglionin2009}). 
Ref.~\cite{BurrowsLMBSDL24}  finds that the $^2P_{\frac32}$ phase shifts from the SA-NCSM/GF, e.g., for \hw=16 MeV, yield a  threshold energy that agrees with the experimental one within 230 keV. This is important since reaction observables are very sensitive to the threshold energy. Furthermore, there is close agreement with earlier \textit{ab initio} many-body approaches based on alternative frameworks, including the NCSMC with multiple channels using the chiral NNLO$_{\rm opt}$ NN interaction  \cite{PhysRevLett.125.112503} and the chiral N$^4$LO interaction including 3N forces \cite{Kravvaris:2020lhp}, the Faddeev-Yakubovsky (FY) approach \cite{Lazauskas2018} using the chiral N$^3$LO-EM NN interaction \cite{EntemM03}, and the single-state harmonic oscillator representation of scattering equations \cite{PhysRevC.98.044624} using different NN interactions. 
\begin{figure}[t]
    \centering
    {
    \includegraphics[width=0.4\textwidth]{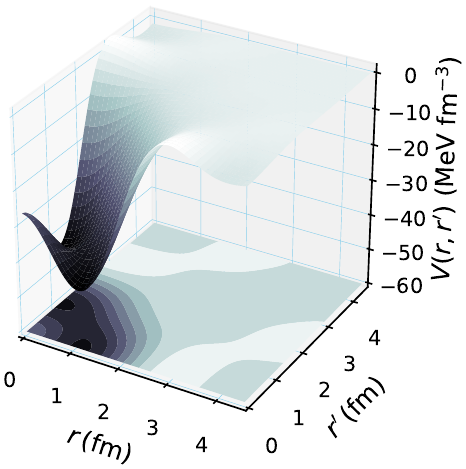}}
    {\includegraphics[width=0.4\textwidth]{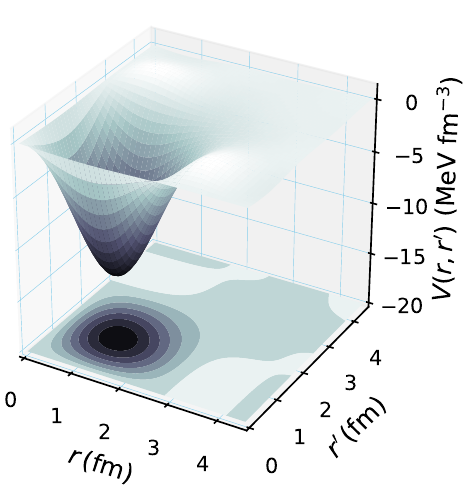}}\\
    (a) \hspace{2.3in } (b)
    \caption{Translationally invariant nonlocal n+$^4$He optical potentials for the (a) $^2S_{\frac12}$ and (b) $^2P_{\frac32}$ partial waves for $E=0$ MeV and $\epsilon=0$ MeV, calculated in the SA-NCSM/GF with NNLO$_{\rm opt}$  for $\hbar\Omega=16$ MeV and $N_{\rm max}=13$.
}
    \label{fig:nonlocal_potentials}
\end{figure}

Most importantly, the  
total cross sections for the SA-NCSM/GF calculations  agree remarkably well with experiment, as shown in Fig. \ref{fig:SANCMGF}b for projectile laboratory kinetic  energies $E_{\rm lab} \le 12$ MeV. As expected from the good description of the $^2P_{3/2}$ phase shifts, the cross-section peak energy is well reproduced in the SA-NCSM/GF approach. Notably, the spread of the calculated cross section arising from the \hw~variation is very small even though a significant \hw~range  is considered (12--20 MeV). This spread further decreases across a smaller range of \hw=12--16 MeV, while remaining in agreement with the data.
\begin{figure}[th]
\centering
\includegraphics[width=0.44\linewidth]{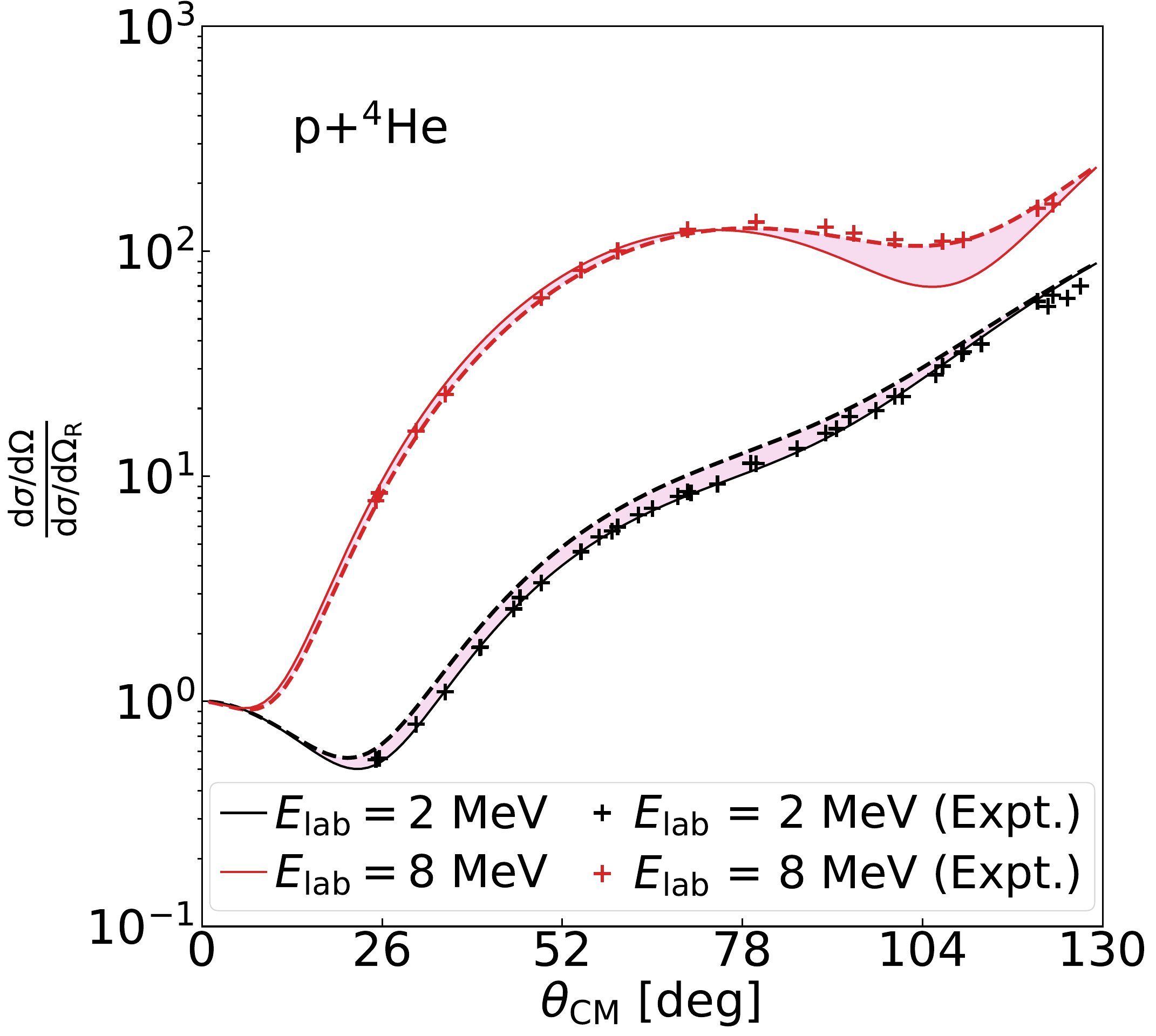}
\includegraphics[width=0.4\linewidth]{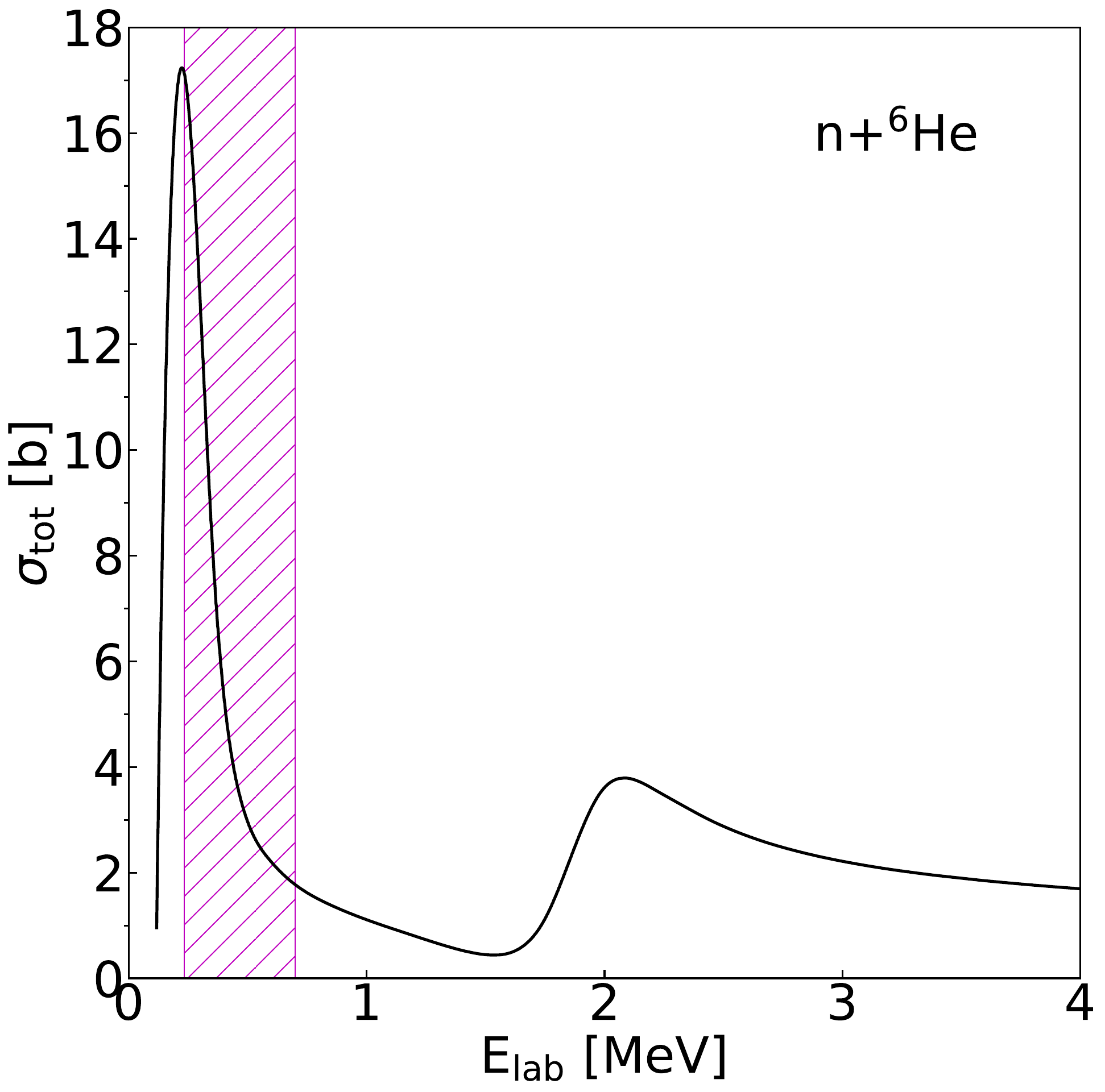}
\\
    \hspace{0.05\linewidth} (a) \hspace{0.35\linewidth} (b)
\caption{\label{fig:5Li_DiffCS_logRuth} (a) Differential cross section vs. scattering angle in the c.m. frame, relative to the Rutherford cross section, for the $\alpha$(p,p)$\alpha$ reaction. Calculations include $\ell \leq 3$ partial waves, and use $N_\mathrm{max} = 13$, $\epsilon = 0$ MeV, and $\hbar\Omega = 12$--$20$ MeV, with $\hbar\Omega = 12$ MeV indicated by dashed lines and $\hbar\Omega = 20$ MeV indicated by solid lines. Experimental data from Refs.~\cite{BARNARD1964604,kraus_2024_34e8x-n0z43}. (b) Total elastic cross section (from partial waves up through $D_{3/2}$), calculated in the SA-NCSM/GF with $N_{\rm max}=9$ and $\hw=12$ MeV, where the experimentally deduced $P_{3/2}$ resonance energy~\cite{PhysRevLett.18.611,CAO201246} is indicated with a vertical hashed bar. Figures from Ref.~\cite{Mumma25}.
}
\end{figure}

The corresponding \textit{ab initio} SA-NCSM/GF optical potentials $V(r,r')$ of Eq.~\eqref{eq:GFoptpot} for n+$^4$He are highly nonlocal  (Fig.~\ref{fig:nonlocal_potentials} for $E=0$ MeV). In general, they depend on the scattering energy $E$, however, for all $^2S_{1/2}$, $^2P_{1/2}$, $^2P_{3/2}$, and $^2D_{3/2}$ partial waves for this system, there is almost no dependence for $E \le 12$ MeV, except when $E$ is very close to a pole of the Green's function for $\epsilon=0$ (cf. Fig. 4 of Ref.~\cite{BurrowsLMBSDL24}).
Although optical potentials are not observables and cannot be compared exactly between methods and inter-nucleon interactions used, similarities may exist in some features.
For example, there is similarity 
for $^2S_{1/2}$ and $^2P_{3/2}$, such as an attractive well in the target interior and repulsive nonlocal peaks close to the surface, 
not only between the potentials calculated in the SA-NCSM/GF and SA-RGM, but also when comparing two spherical targets, such as $^4$He and $^{40}$ Ca (cf. Fig.~\ref{fig:RGM_40Ca}). 

Furthermore, we find that the proton elastic scattering differential cross sections from the SANCSM/GF calculations agree remarkably well with experiment, as shown in Fig. \ref{fig:5Li_DiffCS_logRuth} for p$+^4$He for projectile laboratory kinetic energies $E_\mathrm{lab}=2$ MeV and 8 MeV. At forward scattering angles, the spread in the calculated differential cross section arising from the $\hbar \Omega$ variation is very small even though a significant $\hbar \Omega$ range is considered. For backward angles at 8 MeV, a tighter $\hbar \Omega$ spread may be achieved by including more partial waves.

In addition, open-shell targets can be studied in the SA-NCSM/GF framework. For example, the peak location in the neutron elastic scattering cross section calculated in the SA-NCSM/GF for a $^6$He target, being only about $\sim 500$ keV above the threshold, is found to closely agree with the $3/2^-$ resonance energy (Fig.~\ref{fig:5Li_DiffCS_logRuth}b).

\begin{figure}[th]
    \centering
    {\includegraphics[width=0.45\columnwidth]{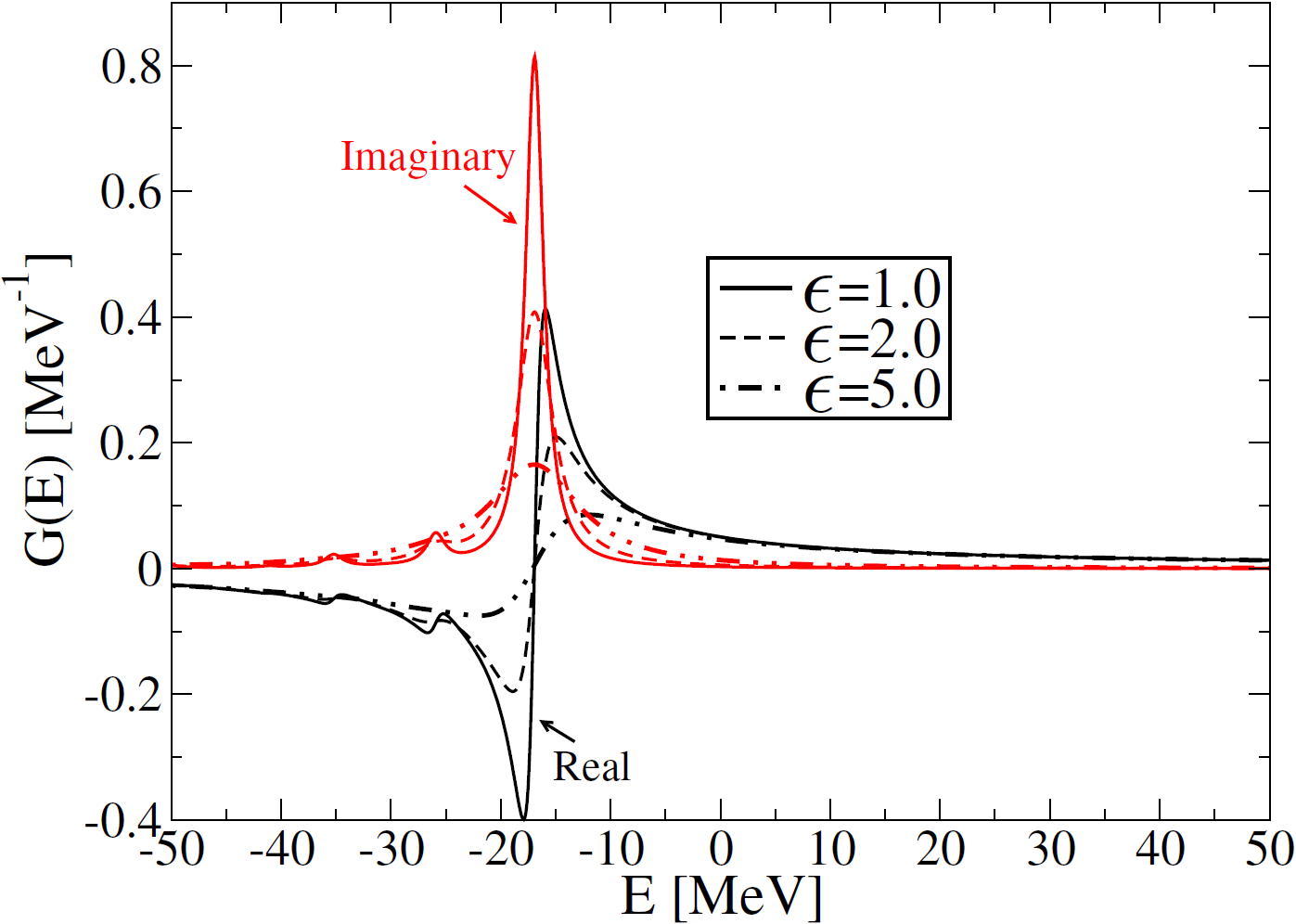}}
    \caption{Translationally invariant spectral functions $S_h$ for $^4$He ($\times \pi$), which is ${\rm Im} G(E,\epsilon)$ (red), along with ${\rm Re} G(E,\epsilon)$ (black),  summed over $\eta_a$ for the $s_{1/2}$ single-neutron levels plotted against the energy in the c.m. frame, for $\epsilon= 1$ MeV (solid curves), $2$ MeV (dashed curves), and  $5$ MeV (dotted dashed curves). Figure from~\cite{BurrowsLMBSDL24}, with permission.}
    \label{fig:spectral}
\end{figure}

An important outcome of this approach is that the imaginary part of the Green's function can be used to calculate the spectral function \cite{DickhoffBook} for $^4$He, $\pi \sum_{\eta_a}S_h(a,E)={\rm Im}(G^{J-}_{(J_0=0)aa})$ (see Fig.~\ref{fig:spectral} for the $\ell_a=0$ and $j_a=1/2$, or $s_{1/2}$, single-particle states  for $\epsilon$ values of 1, 2, and 5 MeV). 
The integral from $-\infty$ to the single-nucleon threshold $\varepsilon_F^- \equiv E_{\rm g.s.}^A- E_{\rm g.s.}^{A-1}$ is independent of $\epsilon$ and equivalent to the $\ell=0$ single-neutron spectroscopic factor $S^-$ for a particle removal from the $^4$He ground state [cf. Eq.~(\ref{eq:SF}) for $S^+$ in the case of particle addition]. 
In this particular case, Ref.~\cite{BurrowsLMBSDL24} finds for the $s_{1/2}$ single-particle levels $S^-_{\ell=0,j=\half}=0.897$ when using the spectroscopic overlaps,  which can be calculated directly from the SA-NCSM wavefunctions; indeed, this closely agrees with  $S^-_{\ell=0,j=\half}=0.885 \,($0.879$)$ when using the integrals of ${\rm Im} G(E,\epsilon)$ for $\epsilon=1.0$ ($2.0$) MeV.

\subsection{Symmetry-adapted resonating group method}
\label{sec:RGM}

The resonating group method (RGM) \cite{WildermuthT77,rgm_tang_1978} is a many-body approach to nuclear reactions in a coupled-channel formalism. 
The microscopic wavefunction of each cluster is used to construct an antisymmetrized many-body wavefunction that describes the cluster dynamics in each channel, rendering this formalism suitable
for unifying nuclear structure and reactions. As such, the RGM can be conveniently formulated to accommodate the translationally invariant wavefunctions of practically any many-body approach. 
For example, modern implementations of the RGM approach 
are based on
the combination of the \textit{ab initio} NCSM with RGM, which has allowed successful predictions for light nuclei of scattering, transfer, and capture reactions with nucleons and light composite projectiles \cite{quaglionin2009,navratil:12,Kravvaris:24, Atkinson:25}. 
More recently, the RGM has been combined with the Gamow shell model \cite{mercennemp19} with applications to deuteron elastic scattering, and extended to intermediate-mass targets with the use of the SA-NCSM framework \cite{MercenneLEDP19,LauneyMD_ARNPS21}.

In its most general formulation, the RGM wavefunction is expressed as a superposition of antisymmetrized cluster states, characterized by the intrinsic properties of each cluster, including spin, parity, and other quantum numbers, as well as by the wavefunction describing their relative motion.
In particular, 
in the case of two clusters of mass $A$ and $a$, the cluster states for a channel $c$ are defined as [cf. Eq.~\eqref{eq:overlapdef}]:
\begin{align}
    \mathcal{A} \ket{ { \Phi }_{ c r }^{J^\pi} } & = \mathcal{A}{ \left[ { \left\{ \ket{ \Psi_{(A) \alpha_1}^{ I_1^{\pi_1}} } \otimes \ket{ \Psi_{(a) \alpha_2}^{ I_2^{\pi_2} } } \right\} }^{I} Y_{ \ell } ({ \hat{ r } }_{ A,a }) \right] }^{J^\pi}  \frac{ \delta(r - { r }_{ A,a }) }{ r { r }_{ A,a } } .
    \label{eq:rgm_su2_channel}
\end{align} 
Here, $\vec{r}_{A,a} = r_{A,a}\hat{r}_{A,a}$ is the relative coordinate between the center-of-mass of the $a$ and $A$ clusters,  $\ket{ \Psi_{(A) \alpha_1}^{ I_1^{\pi_1}} }$ and $\ket{ \Psi_{(a) \alpha_2}^{ I_2^{\pi_2} } }$ are eigenstates of the intrinsic Hamiltonians describing the $A$ and $a$ clusters, respectively. Following the notations defined in Eq.~\eqref{eq:overlapdef}, we remind that $\alpha_{1} I_{1}^{\pi_1}$ and $\alpha_{2} I_{2}^{\pi_2}$ denote all relevant quantum numbers needed to identify the respective states, the channel spin $I$ and the relative angular momentum $\ell$ are coupled to total spin and parity $(J^\pi)$, and $Y_{\ell}(\hat{r}_{ A,a })$ are the spherical harmonics.
The nuclear wavefunction is given in terms of the cluster states 
  \begin{equation}
    \ket{ { \Psi }^{J^\pi} } = \sum_{c} \int_{r} dr { r }^{ 2 } {g}_{c}^{J^\pi}(r)
    { \mathcal{A} } \ket{ { \Phi }_{ c r }^{J^\pi} } \;.
    \label{RGM_ansatz}
  \end{equation}
Here, the unknown amplitudes ${ { g }_{ c }^{J^\pi}(r) }$ describe the relative motion between the projectile and the target in a channel $c$ and are determined  by solving the Hill-Wheeler equations (similarly to the generator coordinate method \cite{Griffin_PR_1957}), 
$
    \sum_{c} \int dr { r }^{ 2 } \left[ { H }_{ c'c }^{J^\pi} (r',r) - E { \mathcal{N} }_{ c'c }^{J^\pi}(r',r) \right] {g}_{c}^{J^\pi}(r) 
    = 0,
$
where $\mathcal{N}_{ c'c }^{J^\pi}(r',r)={ \bra{ { \Phi }_{ c' r' } } { \mathcal{A} } { \mathcal{A} } \ket{ { \Phi }_{ c r } } }$ and $H _{ c'c }^{J^\pi}(r',r) ={ \bra{ { \Phi }_{ c' r' } } { \mathcal{A} } H { \mathcal{A} } \ket{ { \Phi }_{ c r } } }$ are the norm and the Hamiltonian kernels, respectively. 
This is a generalized eigenvalue problem, and the norm here reflects the nonorthogonality induced by the inter-cluster antisymmetrizer in the so-called ``particle" space (we remind that in the Green's function approach, detailed in Sec.~\ref{sec:GF}, the interactions are derived in the ``particle-hole" space, where the total norm is unity by default). 

Typically, it is advantageous to employ an orthonormalized channel basis for practical calculations, as the process will conveniently take care of Pauli forbidden states \cite{Kamada2000}.
One can then proceed with solving the coupled-channel equations, given for a single-nucleon projectile as: 
\begin{equation}
 \left( T_{\rm rel}(r) + V_{\rm C}(r) - (E - E_{\alpha_1}^{I_1^{\pi_1}}) \right) u_{c}^{J^\pi}(r)
+ \sum_{c'} \int_{} dr' {r'}^2 W_{cc'}^{J^\pi} (r,r') u_{c}^{J^\pi}(r)
= 0,
\label{eq:CC_eq}
\end{equation}
with $W = \mathcal{N}^{-1/2} T_{\rm rel} \mathcal{N} ^{1/2} - T_{\rm rel} + \mathcal{N}^{-1/2} V \mathcal{N}^{-1/2}$, where $\mathcal{N}$ is the norm, $T_{\rm rel}$ is the relative kinetic energy, $V$ is the nucleon-nucleus NA interaction, and $W$ is the nonlocal potential.
 Accordingly, the wavefunctions $u_c^{J^\pi}(r)$ presented in Eq.~\eqref{eq:CC_eq} correspond to the wavefunctions $g_c^{J^\pi}(r)$, but expressed within the orthonormalized framework. Importantly, since the antisymmetrization effects vanish at large distances, the channels become naturally orthonormal in this asymptotic region, resulting in $u_c^{J^\pi}(r)$ sharing identical asymptotic behavior with $g_c^{J^\pi}(r)$. Their asymptotic contains the scattering amplitude, from which one can extract the phase shifts to compute the cross sections.
 
The most challenging task is the computation of the Hamiltonian kernel $H _{ \nu'\nu }^{J^\pi}(r',r)$, which depends upon the intrinsic microscopic wavefunctions of the clusters. Typically, these, along with $\mathcal{N}$, $T_{\rm rel}$, $V$, and $W$ in Eq.~\eqref{eq:CC_eq}, are solved in finite configuration spaces.
Once the kernels are computed, Eq. \eqref{eq:CC_eq} can then be solved using a microscopic R-matrix approach \cite{BayeB00,Descouvemont:10}, the code for which is publicly available \cite{Descouvemont16}.

There are several methods for computing the Hamiltonian kernel. In the case of nucleon-nucleus collisions, it is often sufficient to express the kernels in terms of the one- and two-body densities of the target nucleus and the inter-nucleon interaction matrix elements \cite{quaglionin2009,mercenne:ldeqsd21} by explicitly accounting for the exchange of the projectile nucleon with those in the target. For reactions involving light projectiles, similar techniques may be employed \cite{Navratil_PRC_2011}. However, alternative approaches based on configuration-interaction and clustering methods have been developed to facilitate the calculation of Hamiltonian kernels for a broader range of projectiles \cite{Kravvaris_PRL_2017,mercennemp19,Kravvaris:24}.
\begin{figure}[th]
\centering
\includegraphics[width=0.65\linewidth]{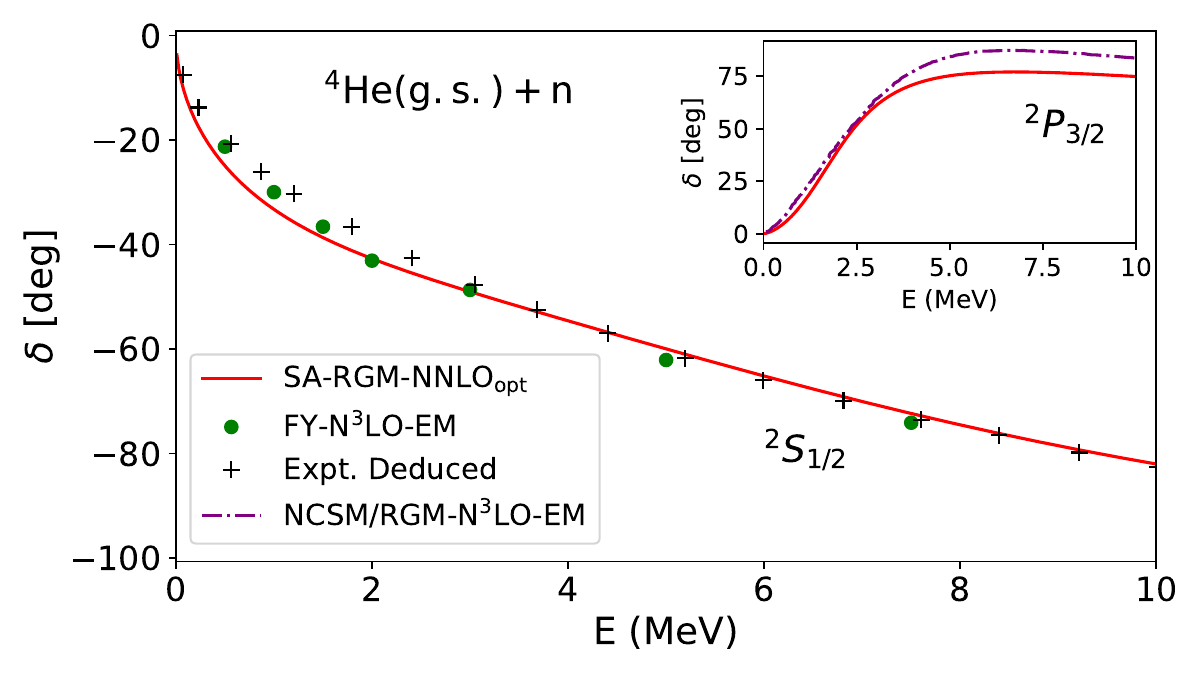}
\caption{\label{fig:RGM_4He} Calculated  n-${}^{4}$He phase shifts  for the ${}^{2}S_{1/2}$ and (inset) ${}^{2}P_{3/2}$ partial waves, using the SA-RGM approach. The ${}^{4}$He $0_{\rm g.s.}^{+}$ ground state is calculated with the SA-NCSM in an $N_{\rm max}=14$ model space, with the NNLO$_{\rm opt}$ chiral potential and $\hw=15$ MeV. The experimentally deduced phase shifts are reported in Ref.~\cite{quaglionin2009}. 
}
\end{figure}

Most RGM applications employ a standard $LS$ or $jj$ coupling scheme, as it is naturally defined in the HO basis.
Alternatively, one can explore other coupling schemes, such as those based on the SU(3) symmetry.
This idea was first explored by Hecht and collaborators in a series of seminal papers \cite{Hecht77,HechtZ79,SuzukiH86}, and recently utilized in Ref.~\cite{DreyfussLESBDD20} within an \textit{ab initio} SA-NCSM framework. 
In this RGM formulation, the intrinsic structure of each cluster, as well as their relative motion, are now expressed in terms of the SU(3) basis states, whose quantum numbers inform about the cluster deformation. 
Although these early studies used a phenomenological Hamiltonian, it is now possible to adapt this SU(3) formulation by merging the RGM with the \textit{ab initio} SA-NCSM approach, dubbed SA-RGM \cite{mercenne:ldeqsd21,LauneyMD_ARNPS21}. 
This method utilizes
the SU(3)-scheme
and wavefunctions calculated in the SA-NCSM.
Such a wavefunction for an $A$-nucleon cluster is  written as:
\begin{equation}
    \ket{\Psi_{(A)\alpha_1}^{I_1^{\pi_1}}} = \sum_{\substack{ b_1 \nu_1} } d_{b_1}^{\nu_1 I_1} \ket{b_1 \nu_1 I_1^{\pi_1}},
\end{equation}
where $\nu = \{ \omega \kappa (L S) \}$,
$\omega \equiv (\lambda\, \mu)$ are the \SU{3} quantum numbers that signify intrinsic deformation, 
$L$ denotes the total orbital angular momentum, $\kappa$ is the multiplicity index that distinguishes multiple occurrences for  the same set of $\omega$ and $L$. The quantum number $S$ corresponds to the total intrinsic spin,
the $b$ labels gather all other quantum numbers necessary for labeling a complete set of SU(3) basis states, and $d$ are the expansion coefficients.
With this notation, we can define the two-cluster RGM channel state \eqref{eq:rgm_su2_channel} in the SU(3)-scheme as:
 \begin{equation}
    \ket{ { \Phi }_{ \nu_1 I_1; \nu_2 I_2 }^{ \nu J^\pi M } } = \sum_{ { b }_{ 1 }  { b }_{ 2 }} 
    { d }_{ { b }_{ 1 } }^{ \nu_1 I_1} { d }_{ { b }_{ 2 } }^{ \nu_2 I_2}
    { \left\{ \ket{ { b }_{ 1 } { \omega }_{ 1 } { S }_{ 1 } } \times \ket{ { b }_{ 2 } { \omega }_{ 2 } { S }_{ 2 } } \right\} }^{ \rho \nu J M } ,
    \label{SU3RGMstatesgen}
  \end{equation}
where the SU(3) basis states $\ket{ { b }_{ i } { \omega }_{ i } { S }_{ i } }$ of each cluster are coupled to $\omega S$ with SU(3) outer multiplicity $\rho$.
An important consequence of the use of SU(3) is that there is no dependence on the relative orbital momentum.
Furthermore, the summation over ${ b }_{ 1 }$ and ${ b }_{ 2 }$ implies that the SA-RGM basis requires only a part of the information present in the SA basis.
In the case of a nucleon projectile~\cite{mercenne:ldeqsd21}, the basis of (\ref{SU3RGMstatesgen}) reduces to $\ket{ { \Phi }_{ \nu_1 I_1; \eta }^{ \nu J^\pi M } }$ with  $\ket{ 1(\eta\, 0) \half }$ replacing $ { d }_{ { b }_{ 2 } }^{ \nu_2 I_2} \ket{ { b }_{ 2 } { \omega }_{ 2 } { S }_{ 2 } }$, where
$\eta$ is the number of the shell occupied by the projectile.
The expression for the norm and Hamiltonian kernels in the case of nucleon projectile reactions has been derived in Ref.~\cite{mercenne:ldeqsd21}.
To compute reaction observables, we use Eq.~(\ref{eq:rgm_su2_channel}) defined in the \SU{3}-scheme as:
\begin{align}    
    \ket{\Phi_{c r}^{J^\pi}} & = \sum_{\eta} R_{\eta \ell} (r) \sum_{j} \Pi_{Ij} {(-1)}^{I_1 + J + j} \Wigsixj{I_1}{\half}{I}{\ell}{J}{j}  \cr 
    & \times \sum_{\nu \nu_1} \Pi_{LS I_1 j} \RedCG{\omega_1 \kappa_1 L_1}{(\eta\, 0)\ell}{\omega \kappa L} \Wigninej{L_1}{S_1}{I_1}{\ell}{\half}{j}{L}{S}{J} \ket{ { \Phi }_{ \nu_1 I_1; \eta }^{ \nu J^\pi M } }
    \label{eq:trans_su3_su2},
\end{align}
where $\RedCG{\omega_1 \kappa_1 L_1}{(\eta\, 0)\ell}{\omega \kappa L}$ is an \SU{3} reduced Clebsch–Gordan coefficient.

Ref.~\cite{mercenne:ldeqsd21} shows that SA model spaces yield practically the same SA-RGM norm kernels as those obtained in complete model spaces, and similarly for the SA-RGM kernel of the interaction between the projectile and a single nucleon in the target. It illustrates this for $^{4}$He, $^{16}$O, and $^{20}$Ne targets.
In addition,  the SA-RGM  computational advantages 
stem from the significantly reduced number of SU(3) basis states needed to describe the target, as well as the manageable number of the SA-RGM basis states for the target+N system with increasing model space sizes. This cost-effective scalability 
means that the memory resources needed for these calculations remain manageable and do not grow exponentially (for further details, see Ref.~\cite{mercenne:ldeqsd21}). The demonstrated efficacy of the SA basis and its scalability with particle numbers and model space dimensions opens the path to \textit{ab initio} calculations up through the medium-mass region of NA interactions that enter nucleon scattering and nucleon capture reactions.
\begin{figure}[th]
\centering
\includegraphics[width=0.9\linewidth]{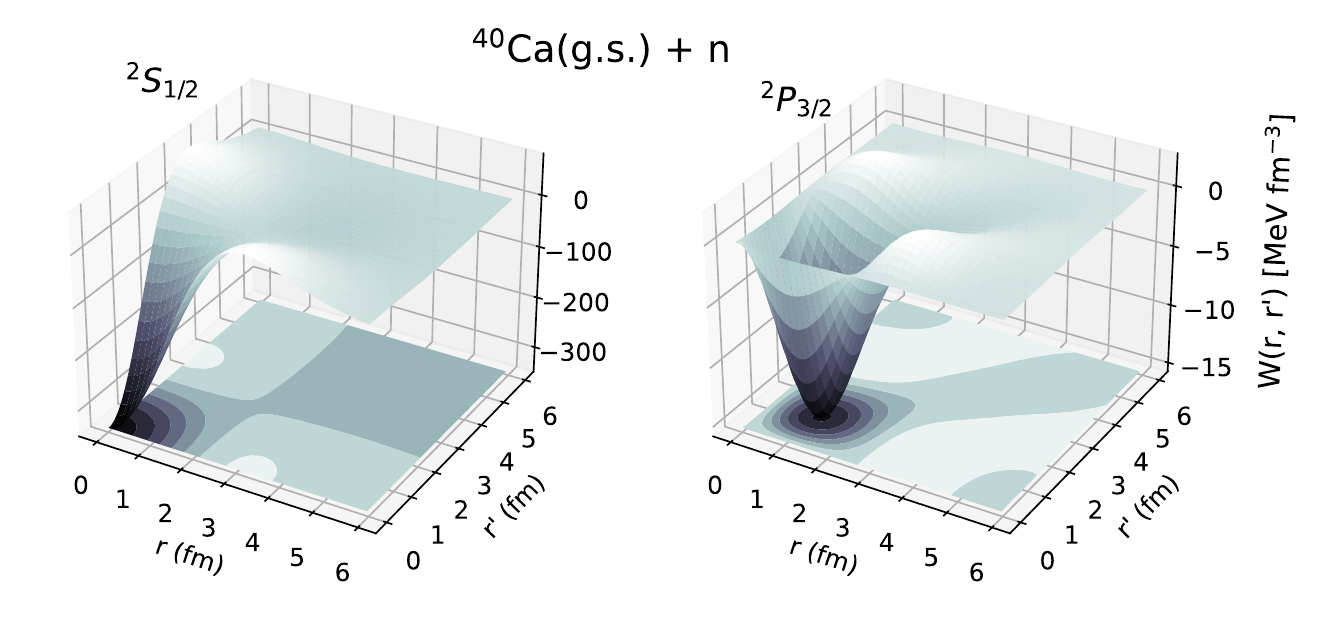}
\caption{\label{fig:RGM_40Ca} $^{40}$Ca(g.s.)$+$neutron nonlocal potential $W(r',r)$ for ${}^{2}S_{1/2}$ and ${}^{2}P_{3/2}$ partial waves,  using the SA-RGM approach. The ${}^{40}$Ca ground state is calculated with the SA-NCSM in model space of 8 HO shells, with the NNLO$_{\rm opt}$ chiral potential and $\hw=15$ MeV.
}
\end{figure}

The results presented in Figs.~\ref{fig:RGM_4He} and \ref{fig:RGM_40Ca} illustrate the capability of the SA-RGM to provide descriptions of neutron elastic scattering for both light and medium-mass nuclei, exemplified here for $^4$He and $^{40}$Ca targets. Calculations use SA-NCSM wavefunctions for the targets with the NNLO$_{\rm opt}$ chiral potential.
We show that the calculated phase shifts for neutron scattering off the ground state of ${}^{4}$He target reproduce the experimental data in the ${}^{2}S_{1/2}$ partial wave, even when only the target ground state is included in the calculations (Fig.~\ref{fig:RGM_4He}). It also closely agrees with the exact Faddeev-Yakubovsky  approach \cite{Lazauskas2018} using the N$^3$LO-EM NN interaction \cite{EntemM03}, as well as with the SA-NCSM/GF results with NNLO$_{\rm opt}$ shown in Fig.~\ref{fig:SANCMGF}. The phase shifts for the ${}^{2}P_{3/2}$ partial wave capture essential features of the resonance and compare well with the corresponding NCSM/RGM calculations of Ref.~\cite{quaglionin2009} using the N$^3$LO-EM NN interaction. However, in this case, as shown in~\cite{quaglionin2009}, additional channels that include excited states of ${}^{4}$He are necessary to fully reproduce the experimental data.  Furthermore, the nonlocal potential $W(r',r)$ computed for neutron scattering off ${}^{40}$Ca represents a significant advance, as it demonstrates the feasibility of SA-RGM to reach medium-mass targets, including Ca isotopes (Fig.~\ref{fig:RGM_40Ca}).

\subsection{Feshbach projection for optical potentials}
\label{sec:selfie}

 Microscopic nuclear structure information can be incorporated into the development of optical potentials via the Feshbach formulation \cite{Feshbach:58, SargsyanPKE2024}. Particularly, the method of Ref.~\cite{SargsyanPKE2024} connects the optical potential to the underlying nucleon-nucleon interaction, allowing for the construction of a nonlocal dispersive potential that takes into account excitations  of the target. The general expression for the optical potential within this framework can be presented as follows:
\begin{eqnarray}
\mathcal V(\mathbf{r,r'},E)&=& U_{00}(\mathbf r)+\sum_{i,j \neq 0} U_{0i}(\mathbf{r})G^{Q}_{ij}(\mathbf{r,r'},E)U_{j0}
(\mathbf{r'}) \nonumber \\
&=&U_{00}(\mathbf r)+V_\text{dpp}(\mathbf{r,r'},E),  
\label{eq:OP}
\end{eqnarray}
where $V_{dpp}$ stands for dynamic polarization potential, which modifies the ground-state potential by including the effect of the target excited states, and the coupling potentials $U_{ij}$ between the target states are defined as 
\begin{align}\label{eq:coupl_pot}
  U_{ij}(\mathbf r)=\int\Psi^*_j(\xi)V(\mathbf r,\xi)\Psi_i(\xi)\,d\xi,
\end{align}
with $\mathbf{r}$ being the relative coordinate and $\xi$ the coordinates of the $A$ nucleons in the target. $V(\mathbf r,\xi)$ is the interaction between the projectile and $A$ nucleons of the target, $\Psi_i(\xi)$ are the eigenstates of the target, with $i=0$ corresponding to the ground state. $G^{Q}$ is the Green's function restricted to the excited states of the target nucleus (the $Q$-space in Feshbach terminology).
\begin{figure} [t]
    \centering
    \includegraphics[width=0.6\linewidth]{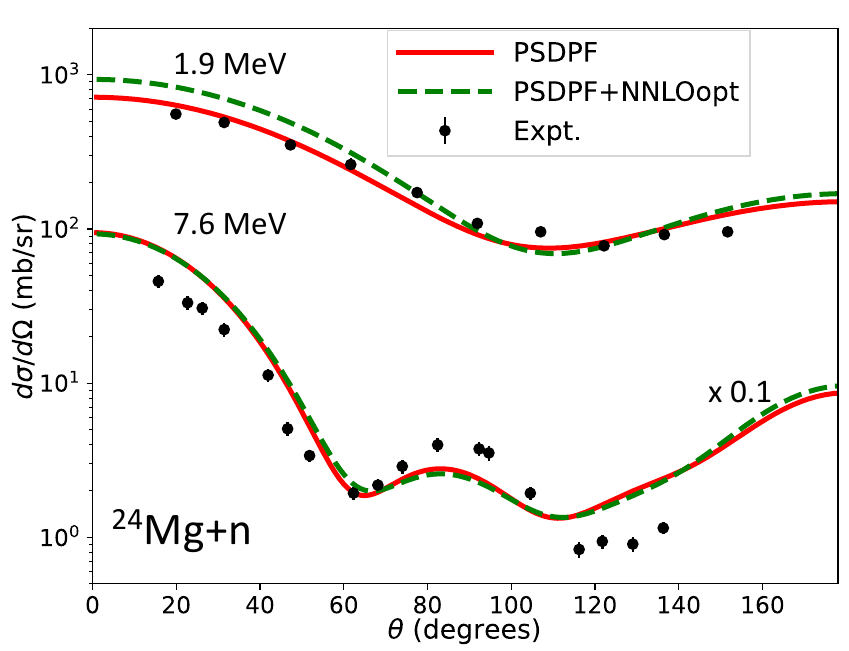}
    \caption{ Elastic scattering cross sections of neutrons 
    on a $^{24}$Mg target for 1.9 MeV and 7.6 MeV laboratory projectile energies calculated using the Feshbach projection approach and compared to the experiment. The solid red curves correspond to the calculations with shell-model states, while the dashed green curves use structure input for
    the lowest 24 states from
    SA-NCSM with the NNLO$_{\rm opt}$ chiral potential
    (see text for details). Experimental data is from \cite{ThomsonCL1962, Koning03}.  }
    \label{fig:24Mg_n}
\end{figure}

 The formulation in Eq. (\ref{eq:OP}) has been used with various approximations to the central potential ($U_{00}$), coupling potentials ($U_{0i}$), and propagator ($G_{ij}^Q$). Many of these models are reviewed in Ref. \cite{Hebborn:23omp}. A recent study in Ref. \cite{SargsyanPKE2024} presents an approach to construct optical potentials using the Feshbach formulation taking as inputs the excitation energies, along with the single-nucleon overlaps of the states of the composite $A+1$ nucleus with those of the target. These overlaps can be calculated in configuration-interaction models, such as the valence-shell model, NCSM, and SA-NCSM.

 Ref. \cite{SargsyanPKE2024} applies this method to study n+$^{24}$Mg  scattering by constructing an optical potential based on shell-model calculations of $^{25}$Mg. In particular, around 600 states of $^{25}$Mg  (up to about 15 MeV of excitation energy) have been calculated using the PSDPF interaction \cite{BouhelalHCN2011}, which assumes an inert $^4$He core. This interaction allows for the description of both positive and negative parity states since the valence nucleons occupy the \textit{p}, \textit{sd} and \textit{pf} shells. The extended model space ensures an exact separation of the   c.m.
 coordinate by allowing for excitations from the $p$ to $sd$ shell, and the $sd$ to $pf$ shell. A simple real Woods-Saxon shape, given in Eq.~\eqref{eq:WS_pot}, is used for the static potential $U_{00}(r)=V_{\rm WS}(r)$ with parameters $V_0=-48.5$ MeV, $R_0=3.69$ fm,  $a=0.65$ fm. These values bind a single neutron in the $1d_{5/2}$ state to 7.3 MeV, thus reproducing the experimental neutron separation energy of $^{25}$Mg. The integral in Eq.(\ref{eq:coupl_pot}) is not calculated explicitly. Instead, similarly to $U_{00}(r)$, the coupling potentials $U_{0i}(r)$ that connect the ground state to each excited state $i$ are chosen to be real-volume Woods-Saxon type with the same radius ($R_0$) and diffuseness ($a$) as $U_{00}(r)$, but with a depth ($V_0$) adjusted to reproduce the calculated energy of each state $i$. These potentials are also multiplied by coupling strengths taken to be the spectroscopic amplitudes of the corresponding $^{25}$Mg states with the ground state of $^{24}$Mg. The propagator is taken to be diagonal, $G_{ii}^Q$ (weak coupling approximation \cite{Feshbach:58}) and is calculated using the Lehmann representation, given in Eq.~\eqref{eq:GpmLehmann}. 

Both the shell-model calculation with the PSDPF interaction and the SA-NCSM with NNLO$_{\rm opt}$ interaction exhibit larger single-neutron spectroscopic factors for the lower-lying positive and negative parity states of $^{25}$Mg, which makes their contribution in the construction of the optical potential the largest. The spectroscopic factors for the states at higher energies become orders of magnitude smaller; however, the density of states increases exponentially with energy, and hence the total contribution of higher-lying states may still be significant. 

Using the valence-shell model, Ref. \cite{SargsyanPKE2024} obtains an energy-dependent nonlocal dynamic polarization potential, $V_\mathrm{dpp}$, that has real
and imaginary terms. This potential is calculated for a given neutron scattering energy. The static local $U_{00} (r)$ part is added to obtain the optical potential.  Given that these potentials do not contain any phenomenological imaginary terms (fitted to nuclear data), the calculations of elastic scattering cross sections compare remarkably well with the experimental data (Fig. \ref{fig:24Mg_n}, red curves). Furthermore, these optical potentials can be constructed using overlaps calculated in the SA-NCSM with the NNLO$_{\rm opt}$ potential.
Fig. \ref{fig:24Mg_n} shows the similarity between the cross sections, when some of the overlaps from shell-model states with the strongest contribution are replaced by the corresponding SA-NCSM overlaps. In particular, we have replaced the lowest  four states of each $J=$1/2, 3/2 and 5/2 positive and negative parity, for a total of 24 states (dashed green lines in Fig. \ref{fig:24Mg_n}). These states have some of the largest spectroscopic factors and exhaust about the third of the spectroscopic factors sum rule. 

In these calculations, all processes that contribute to the formation of an excited state of the $^{25}$Mg compound nucleus are taken into account in the imaginary component of the optical potential. This, in turn, influences the reaction cross section and the reduction of the elastic channel. Some of these compound-nucleus states may decay by emitting a neutron with energy equal to that of the incident particle, thus contributing to the compound elastic cross section. Given that these events cannot be experimentally distinguished from direct elastic scattering, we have incorporated the compound elastic contribution, determined by the Hauser-Feshbach code YAHFC \cite{Ormand2021YAHFC}.

\subsection{Intermediate energies for $^{20}$Ne and $^{40}$Ca targets} 
\label{sec:mst}

\begin{figure}[th]
    \centering
    \includegraphics[width=0.47\linewidth]{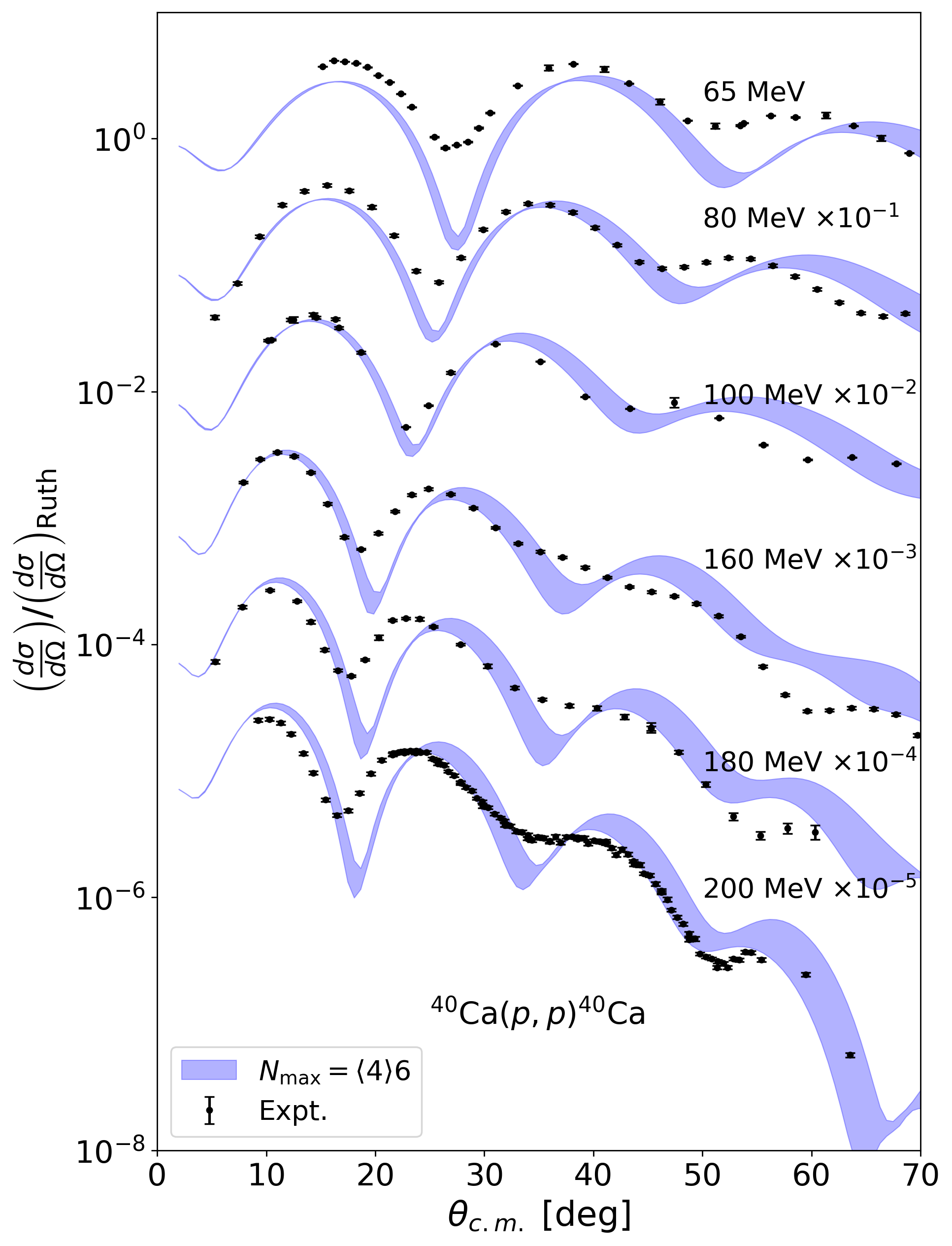}
    \includegraphics[width=0.46\linewidth]{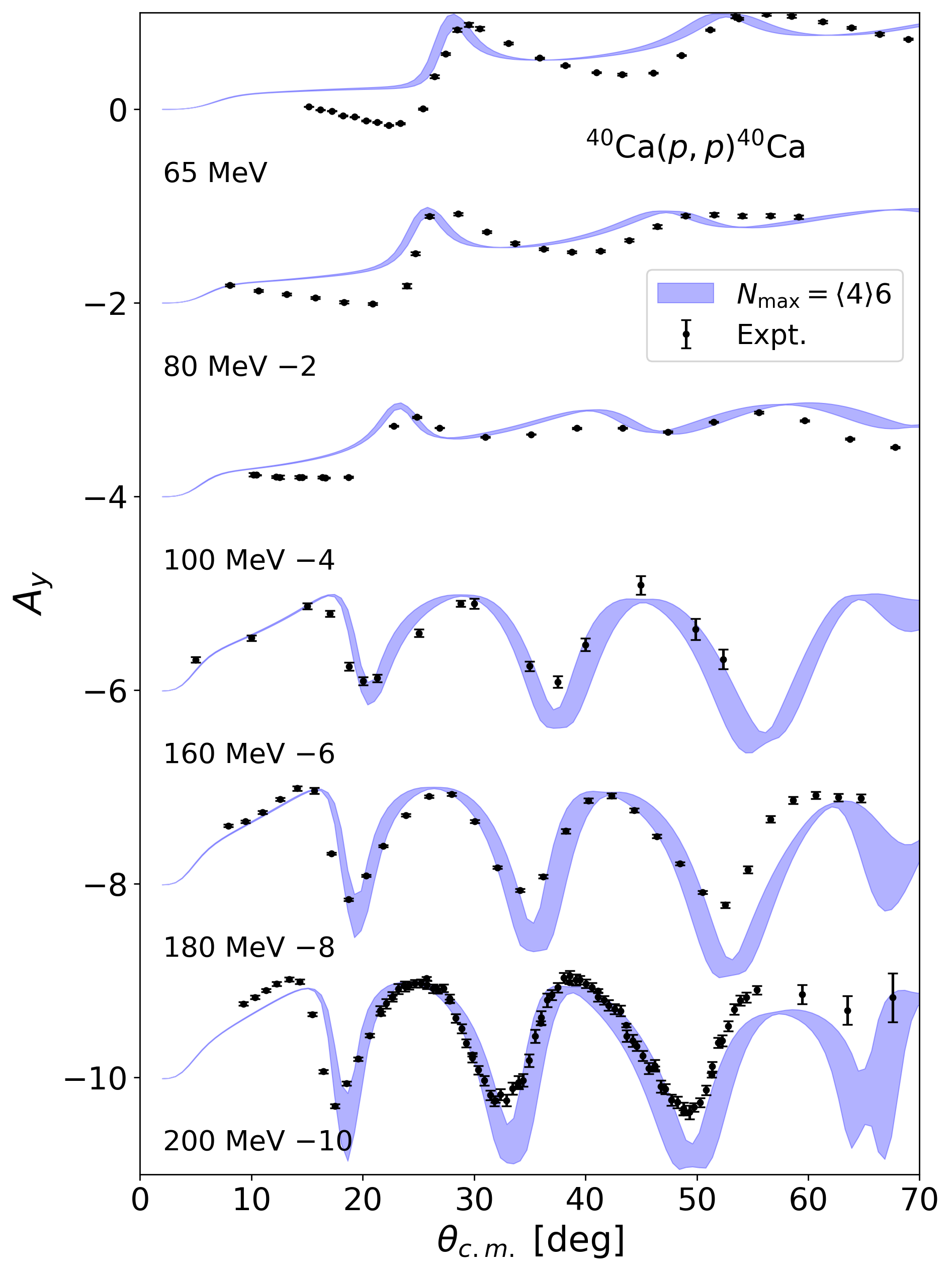}
    \caption{Elastic proton scattering on $^{40}$Ca from 65 to 200~MeV laboratory projectile energy in the leading-order multiple scattering framework: angular distribution of (a) the differential cross section divided by the Rutherford cross section, and (b) the analyzing power ($A_y$). The calculations use the SA-NCSM with NNLO$_{\rm opt}$ in 10 HO shells, with band indicating variations across $\hbar\Omega=11-13$ MeV. Data from Ref.~\cite{Sakaguchi:1979fpk,Schwandt:1982py,Nadasen:1981ep,Seifert:1990,vanOers:1971es,Stephenson1985,Seifert:1993zz}.
    Figure from \cite{Baker24_PhysRevC.110.034605}, with permission.
    }
    \label{fig:pCa40}
\end{figure}

For the intermediate-energy region ($\gtrsim 65$ MeV per nucleon),  which corresponds to many current experimental studies at rare isotope beam facilities, the spectator expansion of the multiple scattering theory \cite{Elster:1996xh,Dussan:2014vta}  has recently offered a fully consistent \textit{ab initio} approach to nucleon scattering at leading order that accounts for the microscopic structure of the target \cite{Burrows:2017wqn} and the spin of the struck nucleon in the target in the reaction dynamics \cite{BurrowsBEWLMP20}. In a series of studies, the leading-order \textit{ab initio} effective nucleon-nucleus potential, which is nonlocal and energy-dependent, has been constructed \cite{Burrows:2017wqn,BurrowsEWLMNP19,BurrowsBEWLMP20}, and applied to calculations of cross sections and analyzing power $A_y$.
 In the approach, the effective potential operator is expanded  in terms of active particles~\cite{Chinn:1993zza}. For the leading order (two active particles), a consistent treatment requires an NN interaction to be used to calculate, first, the NN transition amplitude describing the interaction between the projectile and the struck target nucleon, and second, the microscopic structure of the target nucleus that enters by means of  one-body nuclear densities (for details, see \cite{BurrowsBEWLMP20}).
Refs. \cite{BurrowsEWLMNP19,BurrowsBEWLMP20} have studied light targets, for which one-body densities are available in the NCSM.  The SA framework can extend these calculations to intermediate- and medium-mass nuclei, namely, the $^{20}$Ne(p,p)$^{20}$Ne differential cross section at 100 MeV and 200 MeV, and for $^{40}$Ca(p,p)$^{40}$Ca in an energy range of 65--200 MeV \cite{LauneyMD_ARNPS21,Baker24_PhysRevC.110.034605} (see Fig.~\ref{fig:pCa40} for a $^{40}$Ca target). The outcome of these studies indicates that even for intermediate- and medium-mass targets, the differential cross section and $A_y$  as a function of the c.m. angle, or equally the momentum transfer $q$, exhibit reasonable agreement with the experimental data  when the chiral NNLO$_{\rm opt}$ NN potential is employed. It is interesting to observe that, for 100 MeV, the overall agreement for $\theta_{c.m.} \lesssim 50$ deg 
is remarkable, whereas the deviations at other energies may stem from rescattering effects not included at leading order in the scattering theory (at lower energies) and properties of the NN interaction (at higher energies).

\section{Role of alpha clustering in reactions and decays}
\label{sec:alpha}

For the $^{20}$Ne ground state and the first $1^-$ resonance, the spectroscopic amplitude (or cluster wavefunction) is calculated through the spectroscopic overlap using Eq.~\eqref{eq:overlapdef}, as detailed in Ref. \cite{DreyfussLESBDD20}. The formation of clusters in the ground state and the lowest $1^-$ resonance of $^{20}$Ne is clearly evident from the one-body density profiles in Fig.~\ref{fig:SAfeatures}b \& c, respectively. In addition, the enhanced surface clustering in both states is manifested by the last large peak in the corresponding cluster wavefunctions (Fig.~\ref{fig:SAfeatures}e). These cluster features and the way they emerge from the underlying chiral EFT interaction have been further explored in light of alpha capture reactions \cite{DreyfussLESBDD20} and alpha knock-out reactions \cite{SargsyanYOLELT25}, as discussed next.

\subsection{Alpha capture reactions}
\label{sec:alphacapture}

Ref. \cite{DreyfussLESBDD20} has provided the first calculations of translationally invariant $\alpha$-A cluster wavefunctions in the intermediate-mass region that utilize states in the  composite system, calculated in an \textit{ab initio} framework. The efficacy of the approach is based on the use of the raising \SpR{3} symplectic operator $A_L$, which generates $2\hw$ 1p-1h  monopole ($L=0$) and quadrupole ($L=2$) excitations (2 shells up), and its conjugate $B_L$ for de-excitations 2 shells down. These are the same operators that compose the quadrupole moment  and total kinetic energy operators: $T_{\rm rel}/\hw=-\sqrt{3/8}(A_0+B_0)+H_0/2$ and $Q_2=\sqrt{3}(A_2+B_2+C_2)$, where $H_0$ is the HO Hamiltonian operator and $\sqrt{3}C_2$ is the single-shell quadrupole moment operator, sometimes referred to as the Elliott (or ``algebraic") quadrupole moment tensor \cite{Elliott58b}. 
For calculations of the  $u^{J^\pi}_{\nu I \eta \ell }$ alpha spectroscopic overlap~\eqref{eq:overlapdefconfig}, we  use the $B_L$ operators, 
which as symplectic generators do not mix symplectic subspaces $\sigma_i$. This means that the overlaps can be calculated by utilizing the following idea: a $2\hw$ 1p-1h de-excitation within 
a symplectic subspace 
in the many-body $(A+a)$ wavefunction $\Psi_{(A+a)\alpha}^{ J^\pi (M)}$
implies a $2\hw$ de-excitation of the relative motion of the two clusters within the cluster wavefunction $\Phi_{ (A)\alpha_1 I_1; (a)\alpha_2 I_2; I n_r \ell} ^{J^{\pi}(M)}$ (up to higher-order contributions, including, e.g., excitations of the clusters).
That is, for an (A+a)-body wavefunction expressed in terms of the symplectic basis states, $\ket{\Psi^{J^\pi}}=\sum_i d_{\sigma_i N_i \omega_i} \ket{\Psi^{J^\pi}_{[\sigma_i N_i \omega_i]}}$ with $N_i$ counting the total number of HO quanta and $\omega_i$ labeling all additional quantum numbers, 
the $u^{J^\pi}_{\nu I \eta \ell }$ alpha spectroscopic overlap
 for each  $\sigma_i$ can be calculated recursively, for example for $L=0$, as 
\begin{eqnarray}
\sum_{\omega'_i} && \braketop{
\Psi^{J^\pi}_{\sigma_i N_i \omega_i}}{T_{\rm rel}}{\Psi^{J^\pi}_{\sigma_i, N_i+2, \omega'_i
}} u^{J^\pi}_{\nu [\sigma_i, N_i+2, \omega'_i] I,  \eta+2, \ell} \sim \braketop{
 \eta \ell }{T_{\rm rel}}{\eta+2, \ell} u^{J^\pi}_{\nu [\sigma_i N_i \omega_i] I \eta \ell },
\end{eqnarray}
with the base case of a unity overlap for $N_i=0$, which is the valence shell in the case of the $^{20}$Ne ground state (and similarly for $L=2$, but using $Q_2$ instead of $T_{\rm rel}$). In addition, this overlap is multiplied by the norm of the cluster state, derived in Refs.~\cite{HechtRSZ81,HechtZ79}.

\begin{figure}[th]
    \centering
\includegraphics[width=0.75\textwidth]{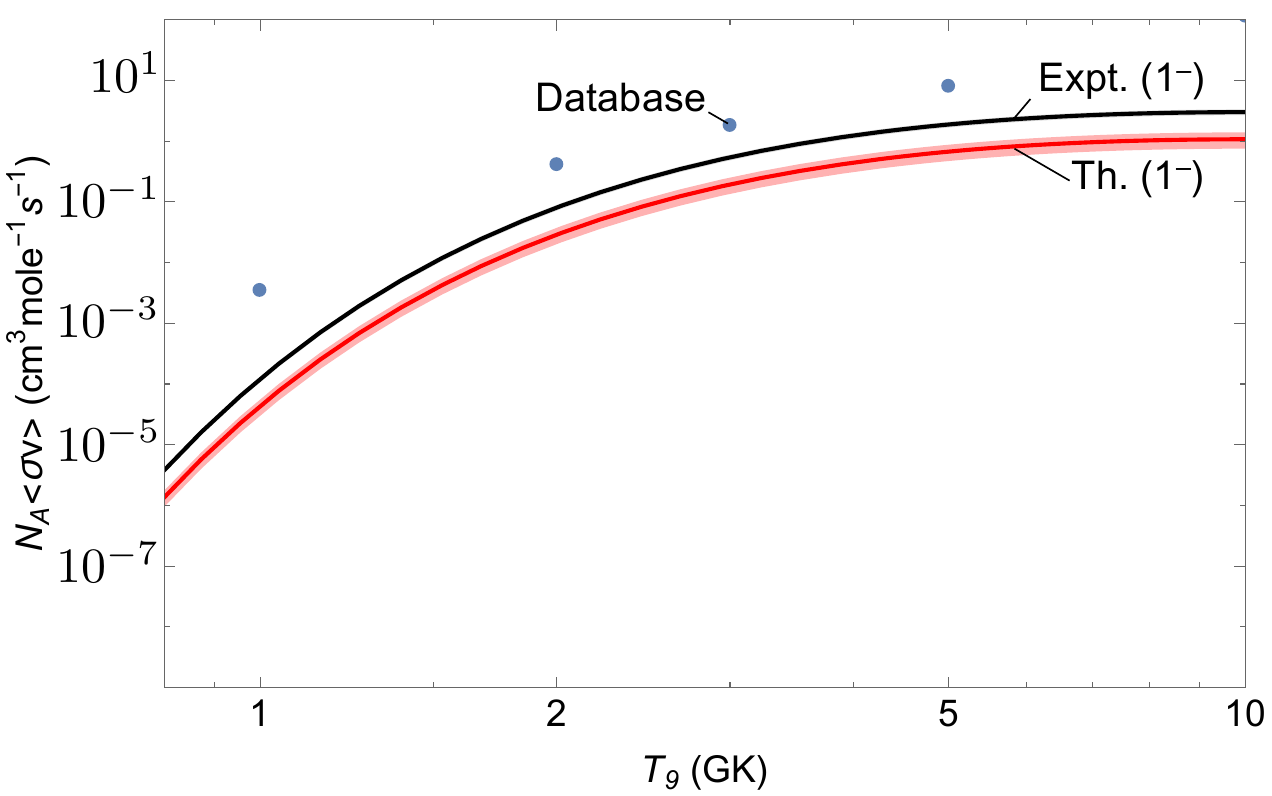}
    \caption
    {
        \label{fig:rxnrate} 
         $^{16}{\rm O}(\alpha,\gamma)^{20}{\rm Ne}$ reaction rate (in $\mathrm{cm}^{3}\mathrm{mol}^{-1} \mathrm{s} ^{-1}$) through the 1.06-MeV $1^{-}$ resonance in $^{20}$Ne   as a function of the temperature (in GK), based on the narrow-resonance formula~\eqref{eq:narrowres}, which takes as input the experimental alpha partial width (black) or the extrapolated alpha partial width $\Gamma_{\alpha}=10(3)$ eV derived from $\textit{ab initio}$ SA-NCSM wavefunctions (red). Also shown is the {\em total} reaction rate (blue dots) taken from the Joint Institute for Nuclear Astrophysics (JINA) REACLIB database \cite{Cyburt10}. The error in the rates is given by the thickness in the curves.
    }
\end{figure}

Ref.~\cite{DreyfussLESBDD20} computes the contribution of the 1.06-MeV $1^{-}$ resonance in $^{20}$Ne to the temperature-dependent  $^{16}{\rm O}(\alpha,\gamma)^{20}{\rm Ne}$ reaction rate.
Using the narrow resonance approximation, 
the reaction rate for a single resonance is given by
\begin{equation}
    \label{eq:narrowres}
    N_{A}\langle\sigma v\rangle_{r}
    =
    \frac{1.539\times10^{11}}{(\mu_{A,a} T_{9})^{3/2}}
    e^{-11.605E_{{\rm lab},r}/T_{9}}
    (\omega\gamma)_{r},
\end{equation}
with the resonance strength defined as $
    (\omega\gamma)_{r}
    =
    \frac{2J+1}{(2I_1+1)(2I_2+1)}
\frac{\Gamma_{\alpha}\Gamma_{\gamma}}{\Gamma}$,
where $T_{9}$ is the temperature (in GK), $\mu_{A,a}$ (\ref{eq:redmass}) is the reduced mass (in amu), and $E_{{\rm lab},r}$ is the laboratory resonance energy (in MeV). The resonance strength $(\omega\gamma)_{r}$ is dependent on the spins of the two clusters, $I_1 =0$ and $I_2=0$,  and the spin of the narrow resonance, $J=1$, for the $^{20}{\rm Ne}$ $1^-$ resonance and $\nu=\{(^{16}{\rm O}_{\rm g.s.})I_1; (\alpha_{\rm g.s.})I_2 \}$.  It also takes as input  the gamma decay branching ratio $BR_\gamma \equiv \Gamma_{\gamma}/\Gamma$. In this study, this branching ratio is extracted from experiment \cite{Constantini_PRC82_2010}, namely, we adopt $\Gamma_{\gamma}/\Gamma=1.9\times10^{-4}$, but can be determined within the  SA-NCSM framework through calculations of electromagnetic strengths. In addition, the resonance strength $(\omega\gamma)_{r}$ depends on the alpha partial width $\Gamma_{\alpha}$, which is computed from the SA-NCSM wavefunction for $^{20}$Ne~\cite{DreyfussLESBDD20}, or taken from the experiment for comparison purposes. The resulting reaction rates show comparatively small deviations, as illustrated in Fig.~\ref{fig:rxnrate}, considering that evaluations of $\alpha$ partial widths and alpha capture reaction rates are very challenging and may often miss the order of magnitude. Because the branching ratio $\Gamma_{\gamma}/\Gamma$ and the resonance energy are kept the same in both calculations, the differences in the reaction rates shown in Fig. \ref{fig:rxnrate} reflect the difference by approximately a factor of two between the experimental and calculated alpha partial widths.  

Nonetheless, to illustrate the impact of these calculations in astrophysics simulations, Ref.~\cite{DreyfussLESBDD20} further explores the abundance pattern produced during an x-ray burst. Specifically, the $\Gamma_\alpha$ partial width calculated in the SA-NCSM with continuum  is used in an initial exploration of XRB nucleosynthesis. Because the $^{16}{\rm O}(\alpha,\gamma)^{20}{\rm Ne}$ total  reaction rate (see Fig.~\ref{fig:rxnrate}, blue dots labeled as ``Database") is dominated by the contribution through the $1^{-}$ resonance at XRB temperatures, we use the calculated partial width to characterize this contribution. In a MESA simulation of XRB nucleosynthesis, we find almost no difference on nuclear abundances as compared to the XRB simulation results when the experimental rate is used. Similarly, this analysis can be extended to the $^{15}{\rm O}(\alpha,\gamma)^{19}{\rm Ne}$ reaction, to which XRB light curves may be most sensitive (see, e.g., \cite{Cyburt10}). Preliminary calculations yield an extremely small $\Gamma_\alpha=5.7 \times 10^{-6}$ eV for the first $3/2^+$ state above the alpha threshold, with possible considerable impacts on the XRB abundances.

These studies show that the present method, starting with \textit{ab initio} wavefunctions and without any parameter adjustments, enables reasonable predictions  for astrophysically relevant alpha-induced reaction rates and can be extended to reactions that cannot be measured, especially those along the  nucleosynthesis paths. In some cases, these estimates may represent a large improvement over existing database entries.

\subsection{Alpha knock-out reactions}
\label{sec:alphaknockout}

\begin{figure}[th]
    \centering
    {\includegraphics[width=0.65\columnwidth]{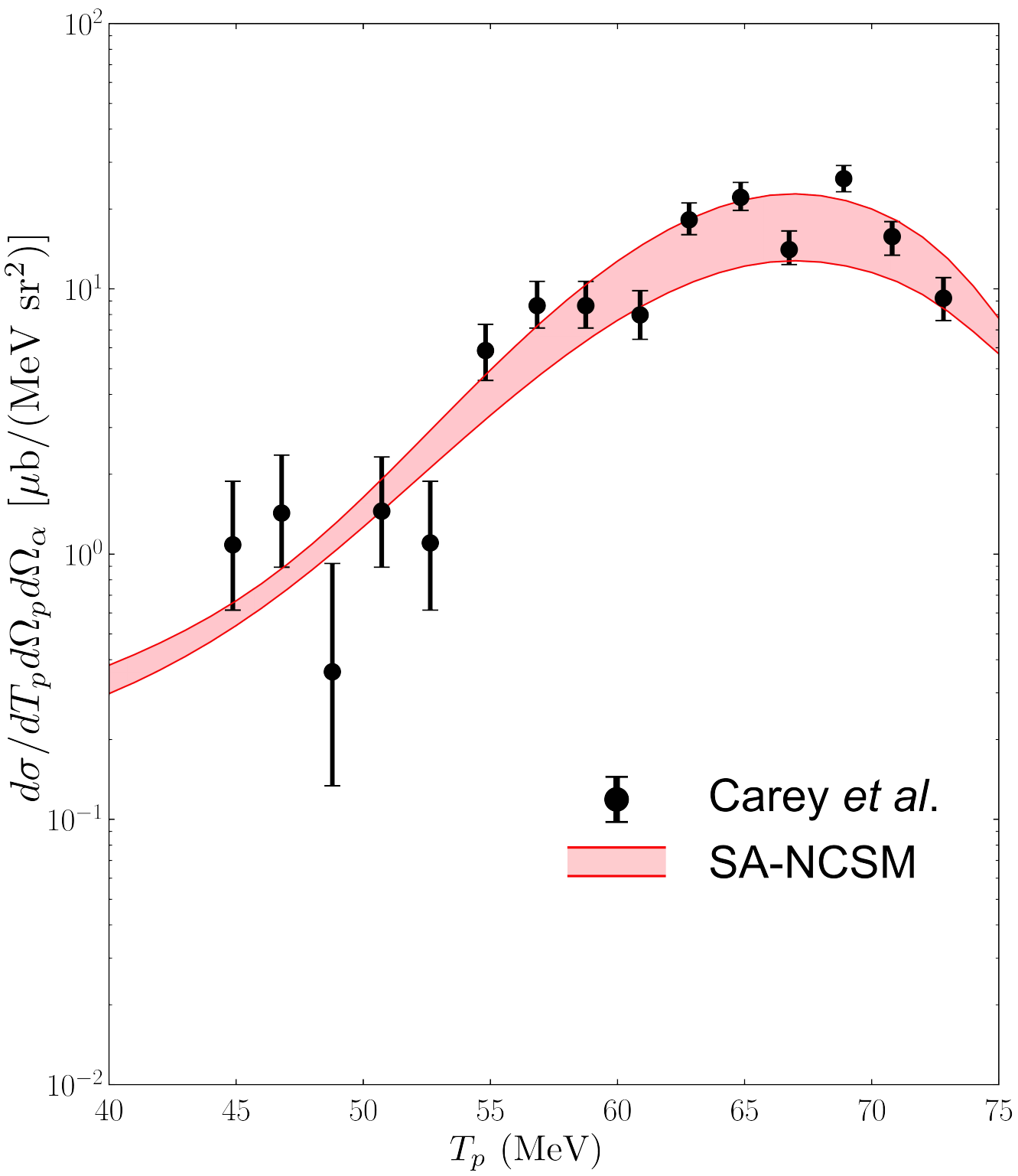}}\\
    \caption{Triple differential cross sections for the $\alpha$ knock-out reaction $^{20}$Ne(p, p$\alpha$)$^{16}$O vs. the laboratory $T_p$ proton kinetic energy, using the $\alpha+^{16}$O spectroscopic amplitudes (cf. Fig.~\ref{fig:SAfeatures}e), where the $^{20}$Ne ground state is calculated in the \textit{ab initio} SA-NCSM with the NNLO$_{\rm opt}$ chiral potential in 13 HO shells (labeled as ``SA-NCSM") with model uncertainties (red band), and compared to the experiment of Ref.~\cite{Carey84} (labeled as ``Carey et al."). Figure adapted from \cite{SargsyanYOLELT25}, with permission.
    }
    \label{fig:xsecSANCM20Ne}
\end{figure}

Ref.~\cite{SargsyanYOLELT25} has reported on the first \textit{ab initio} informed $\alpha$ knock-out reaction for the intermediate-mass $^{20}$Ne nucleus, which yields a triple differential cross section for 
$^{20}$Ne(p, p$\alpha$)$^{16}$O that reproduces the experimental data of Ref.~\cite{Carey84} (Fig.~\ref{fig:xsecSANCM20Ne}). The NNLO$_{\rm opt}$ chiral potential is used in the SA-NCSM many-body approach to calculate 
the $^{20}$Ne $0^+$ ground state, which in turn is used in Eq.~\eqref{eq:overlapdef} to provide the $\alpha+^{16}$O cluster wavefunction for the $S$ partial wave, $u_{\alpha(0^+_{\rm gs})+^{16}{\rm O}(0^+_{\rm gs}), I=0\, \ell=0}^{J^\pi=0^+_{\rm gs}}(r)$ (cf. Fig.~\ref{fig:SAfeatures}e) (to simplify notations, $u_{\ell=0}$ will be used henceforth). The $u_{0}(r)$ wavefunction is supplied to the few-body theoretical framework of the knock-out reaction dynamics outlined in Ref.~\cite{Yoshida19}. Specifically, the approach follows the reaction analysis for $^{20}$Ne(p,p $\alpha)^{16}$O of Ref.~\cite{Yoshida19} in the framework of the distorted-wave impulse approximation~\cite{Chant77,Chant83,Wakasa17}, using the {\sc pikoe} code~\cite{Ogata23}.  The triple-differential cross section (TDX) with respect to the laboratory kinetic energy of the emitted proton $T_p$, emission angle of the proton $\Omega_p$, and the $\alpha$ emission angle $\Omega_\alpha$ is given by
\begin{equation}
 \frac{d^3\sigma}{dT_pd\Omega_pd\Omega_\alpha} 
 = F_{\mathrm{kin}} \left( C_0 \frac{d\sigma_{p\alpha}}{d\Omega_{p\alpha}} \right)\left|\bar{T}\right|^2,
 \label{eq:dwia-tdx}
\end{equation}
where the reduced transition matrix is defined through the spectroscopic amplitude, $
 \bar{T} = \int d\bm{r}\, F(\bm{r}) u_{0}(r) Y_{00}(\hat{\bm{r}})$,
assuming $\ell=0$ based on the single-peak shape of the experimental TDX.  $F(r)$ is determined by the incoming and outgoing distorted waves (for further details, see  Ref.~\cite{Yoshida19}).
For the proton scattering, the global  optical potential by Koning and Delaroche~\cite{Koning03} is adopted, whereas the $\alpha$-$^{16}$O optical potential by Michel~\cite{Michel83} is adopted for the $\alpha$ scattering wave in the final state. 
The $p$-$\alpha$ elementary cross section, $d\sigma_{p\alpha}/d\Omega_{p\alpha}$, is calculated by the folding model~\cite{Toyokawa13} using the Melbourne $g$-matrix interaction, with a  $C_0$ constant that contains physical constants and factors related to the p-$\alpha$ elementary process.
$F_{\mathrm{kin}}$ consists of the phase volume factor and the Jacobian from the c.m. frame to the laboratory frame (for more details of the reaction framework, see Refs.~\cite{Yoshida19,Wakasa17}).

Indeed, as illustrated in Fig.~\ref{fig:xsecSANCM20Ne} and discussed in Ref.~\cite{SargsyanYOLELT25}, the agreement to data is remarkable given that the $\alpha$-cluster features in $^{20}$Ne emerge from the underlying chiral potential that is fitted to only two-nucleon systems, with no additional parameters to tweak. In this study, the $^{20}$Ne ground state is calculated in the \textit{ab initio} SA-NCSM with the NNLO$_{\rm opt}$ chiral potential in 13 HO shells.  The many-body model uncertainties are estimated across $\hw=11$--$13$ MeV, which provide the upper ($\hw=11$ MeV) and lower ($\hw=13$ MeV) bounds for the  $\alpha+^{16}$O  ANC ($C_0$) of the $^{20}$Ne ground state.

Importantly, alpha knock-out reactions are very suitable for probing clustering within the target nucleus. Specifically, given the beam energy, the proton and alpha share this energy after the collision. Therefore,
the region around the maximum cross section in Fig.~\ref{fig:xsecSANCM20Ne} (large $T_p$) involves slow alphas that leave the surface and probes  surface $\alpha$ clustering. Indeed, the good agreement for energies around 67 MeV suggests that the SA-NCSM with continuum describes reasonably well the cluster features in $^{20}$Ne, confirming the critical need for collective correlations and coupling to the continuum for reproducing the cross section. Furthermore, low proton kinetic energies probe, to a certain extent, higher-energy alpha particles that leave the nuclear interior. The reasonable agreement with data for $T_p$ below 55 MeV suggests that the NNLO$_{\rm opt}$ chiral potential captures the overall behavior within the nuclear medium, 
where clustering formation is suppressed compared to the surface region but remains significant. This is likely driven by the constructive spatial contribution of $s$- and $sd$-shell 2-proton-2-neutron substructures within the interior. We note that strong spin mixing can largely reduce in-medium clustering, however, Ref.~\cite{HellerSLJDD23} has demonstrated that SA-NCSM calculations with NNLO$_{\rm opt}$ show significantly weaker spin mixing in the $^{20}$Ne low-lying states compared to conventional valence-shell calculations.
The remarkable agreement to experiment further validates the properties of the NNLO$_{\rm opt}$ chiral
potential, coupled with the SA-NCSM, for predicting collective and clustering features in the intermediate-mass region. It further confirms the significance of the \textit{ab initio} SA-NCSM approach that is capable to facilitate the relevant many-body configurations and, thereby, to describe localized clustering from first principles beyond the lightest nuclear systems (for further details and discussions, see Ref.~\cite{SargsyanYOLELT25}). 

\subsection{Beta decays and probing beyond-the-standard-model physics}
\label{sec:A8beta}

\begin{figure}[ht]
    \centering
    \includegraphics[width=0.49\textwidth]{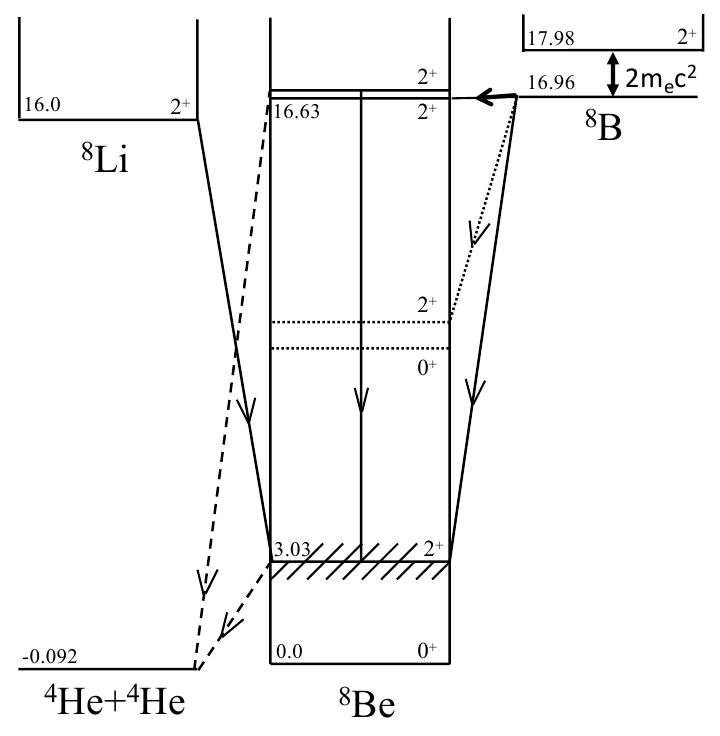}
    \caption{Level diagram of the $^8$Li and $^8$B $\beta$ decays to $^8$Be. All of $^8$Be states are above the $\alpha+\alpha$ separation threshold. The dotted levels correspond to the  states that have not been directly observed experimentally, but calculated in the SA-NCSM and proposed in earlier studies (see text for details). 
    }
        \label{fig:A8_levels}
\end{figure}

\begin{figure*}[t]
    \centering
     \includegraphics[width=0.47\textwidth]{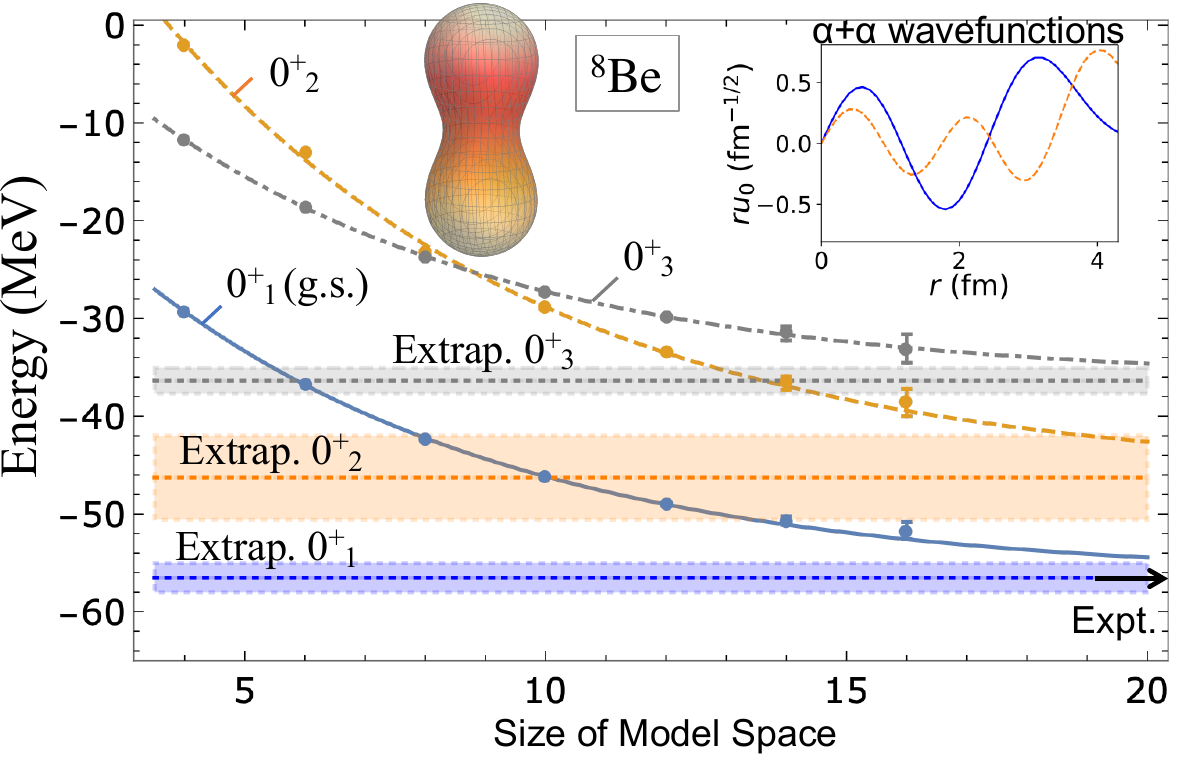}  \includegraphics[width=0.49\textwidth]{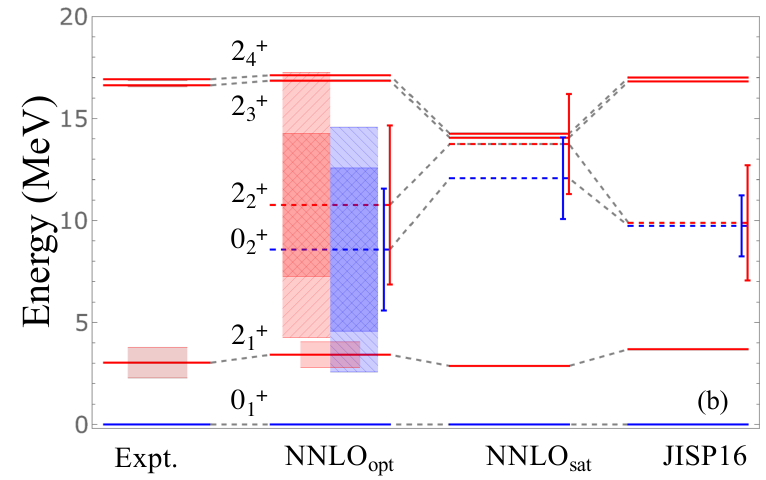}\\
     (a) \hspace{3in}(b)
\caption{(a) Calculated $^8$Be low-lying $0^+$ state energies illustrated for the NNLO$_\mathrm{opt}$ chiral potential (\hw=15 MeV) vs. the model-space size ($N_{\rm max}$), together with the extrapolated values (dotted lines) and uncertainties (bands). 
The experimental $0_{\rm g.s.}^+$ energy is indicated by a black arrow~\cite{HuangWKAN2021ame}. 
Inset: $\alpha+\alpha$ $S$-wave for $0^+_{\rm g.s.}$ (blue solid) and $0^+_2$ (orange dashed).
(b) \textit{Ab initio} low-lying states for $^8$Be, compared to experiment (Expt.). For the $0^+_2$ and $2^+_2$  states (dashed levels), $\alpha$ widths are given by shaded areas, with error bars for energy uncertainties and lighter shades for width uncertainties; 
the small $0_{\rm g.s.}^+$ width (not shown) is estimated to be 5.7 eV, compared to the experimental value of 5.57 eV \cite{TILLEY2004155}. Figure adapted from Ref.~\cite{sargsyanlbgs2022}, with permission from APS.
    }
    \label{fig:JJ0}
\end{figure*}

Alpha clustering plays a major role in some of the precision $\beta$ decay measurements that aim to probe physics beyond the standard model (BSM). In particular, a series of $^8$Li and $^8$B $\beta$ decay experiments have been conducted to measure existence of BSM tensor currents in the weak interaction \cite{SternbergSSSC2015,BurkeySGSC2022, GallantSSC2023,LongfellowGSB2024}. Both $^8$Li and $^8$B nuclei $\beta$ decay to $^8$Be, whose states are all above the two-alpha separation threshold (Fig. \ref{fig:A8_levels}). These $\beta$ decays primarily populate the lowest 2$^+$ state in $^8$Be~permitted by the $Q$ values and the selection rules of the allowed $\beta$ transitions.  However, the SA-NCSM  calculations of Ref.~\cite{sargsyanlbgs2022} indicate the existence of another $2^+$ state with a corresponding $0^+$ state below the $^8$Li and $^8$B $\beta$ decay $Q$ values (see Fig.~\ref{fig:A8_levels}, dotted levels for the $0^+_2$ and $2^+_2$  states).  
In the SA-NCSM calculations, these states rapidly decrease in energy as the model space increases (Fig. \ref{fig:JJ0}a) for various realistic interactions, similar to the Hoyle-state rotational band in $^{12}$C \cite{DreyfussLTDB13}, and are often referred to as ``intruder" states. 
The energies are extrapolated to the $N_{\rm max}\rightarrow\infty$ limit using the three-parameter exponential formula from Ref. \cite{MarisVS2009}, with uncertainties based on variations in the model-space size. The extrapolated excitation energy of the $^8$Be~$2_2^+$ state is between 5 and 15 MeV (dashed lines in Fig. \ref{fig:JJ0}b) making it accessible to the $^8$Li and $^8$B $\beta$ decays. Hence, it is necessary to include it in the analysis of high-precision $\beta$-decay measurements, as well as calculations of $\beta$-decay strengths. 

Indeed, using the $^8$Be
SA-NCSM calculations, recent precision measurements of $^8$Li and $^8$B $\beta$ decays
have provided some of the most precise BSM weak tensor current limits available to date \cite{BurkeySGSC2022,LongfellowGSB2024}.
These experiments have achieved remarkable precision that now requires confronting the systematic uncertainties that stem from higher-order corrections in nuclear $\beta$ decay. These corrections are difficult to measure experimentally but can be accurately calculated in the SA-NCSM. In addition, these studies 
have included in their analyses a $2_2^+$ state below 16 MeV in the $^8$Be~spectrum \cite{BurkeySGSC2022,LongfellowGSB2024}. 
The uncertainty from the existence of this state in $^8$Be~leads to an additional uncertainty on the weak tensor current limit. Hence, experimental confirmation regarding the existence of intruder states in the $^8$Be~ spectrum could further improve the precision on the latest limits on weak tensor currents.

The intruder 0$^+$ and 2$^+$ states in the low-lying spectrum of $^8$Be were proposed in the late 1960's by Barker from concurrent R-matrix fits to scattering, reaction, and decay data associated with the $^8$Be nucleus \cite{Barker1968,Barker1969}.  
The inclusion of an intruder 2$^+$ state with energy below 16 MeV in the R-matrix fits of $^8$Li and $^8$B $\beta$ decays significantly changes the Gamow-Teller (GT) matrix element to the $2_1^+$ state at 3 MeV. Indeed, the analysis  in \cite{Warburton1986} results in a decrease of the extracted $M_{\rm GT}$ for a decay to $2_1^+$ state by almost 1.5 times, as compared to when a low-lying 2$_2^+$  intruder is not present.
The energies from Barker's R-matrix fits for the intruder 0$^+$ and 2$^+$ states are  $\sim 6$ MeV and 9 MeV, respectively, with $\alpha$ widths $>7$ MeV. 
These excitation energies agree with the SA-NCSM extrapolated results given the error bars (Fig. \ref{fig:JJ0}b), as well as with the predicted widths.

To further investigate the structure of the low-lying states in $^8$Be, Ref.~\cite{sargsyanlbgs2022} calculates $\alpha$ widths for the lowest two $0^+$ and $2^+$ states (Fig. \ref{fig:JJ0}b) by projecting the $N_{\rm max}=16$ SA-NCSM wavefunctions onto $\alpha+\alpha$ cluster states and considering the exact continuum Coulomb wavefunctions outside of the interaction effective range, following the procedure of \cite{DreyfussLESBDD20}. This is done for the NNLO$_{\rm opt}$ interaction and the HO parameter that yields the fastest energy convergence ($\hw=15$ MeV).
We express the $^8$Be~and $^4\mathrm{He}$ states in the \SpR{3} basis, where the basis states are associated with nuclear intrinsic shapes \cite{DytrychLDRWRBB20}. For $^8$Be, the $\alpha+\alpha$ overlap computations include three predominant prolate shapes with contributions, determined from the $N_{\rm max}=16$ large-scale calculations, of 75\%, 4\%, and 3\% to  the $0^+$ ground state (totaling 82\%), and 46\%, 15\%, and 11\%  to the $0_2^+$ state (totaling 72\%), and similarly for the $2^+$ states.
These shapes consist of the three most deformed configurations among all those in the valence shell, 2\hw-, and 4\hw-excitations, namely, 0p-0h(4\, 0), 2\hw(6\, 0) and 4\hw(8\, 0), with vibrations thereof up through 18 HO shells in the \SpR{3} basis.
Except for the lowest $0^+$ width that uses  the experimental threshold of $-92$ keV relative to the $^8$Be~ground state, all the widths use the $\alpha$+$\alpha$ threshold of $-104$ keV obtained from the SA-NCSM  extrapolations of the $^4$He and $^8$Be binding energies using the NNLO$_{\rm opt}$ interaction. 
The width uncertainties arise from the upper and lower limits of the $^8$Be~excitation energies from the SA-NCSM calculations (Fig. \ref{fig:JJ0}b). 
Our calculated widths are in good agreement with experimentally deduced values \cite{TILLEY2004155} and earlier theoretical studies \cite{Kravvaris_PRL_2017, ElhatisariLRE15}.

\begin{figure*}[th]
    \centering
   \includegraphics[width=0.99\textwidth]{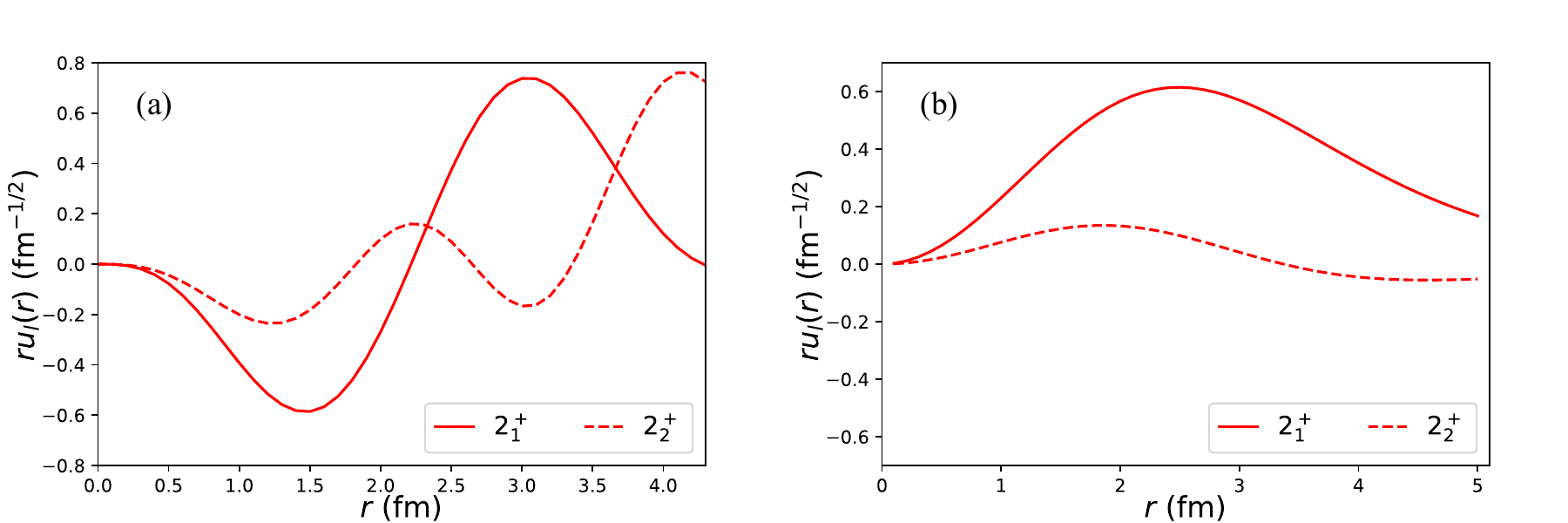}
 \caption{Cluster wavefunctions for the lowest two $^8$Be~ $2^+$ states for (a) $\alpha+\alpha$ $^1D_2$ wave, and (b) $^7\mathrm{Li}+p$ $^3P_2$ wave.
  SA-NCSM calculations use  NNLO$_{\rm opt}$, \hw = 15 MeV, along with (a) $N_{\rm max}=16$ and (b) $N_{\rm max}=10$. 
    }
    \label{fig:a+aJJ4}
\end{figure*}

The $0_2^+$ and $2_2^+$ states are more spatially expanded
than the lowest $0_1^+$ and $2_1^+$ states and have large overlaps with the $\alpha$+$\alpha$ system, making them extremely short lived (Fig. \ref{fig:JJ0}a, inset). Nonetheless, they have a nonnegligible impact on theoretical calculations, including $\beta$-decay strengths and the above-mentioned higher-order terms. To study the cluster structure of the lowest $2_1^+$ and $2_2^+$ states, we calculate the overlaps of the many-body $^8$Be wavefunctions with the $\alpha+\alpha$ and $^7\mathrm{Li}+$p systems, and present their cluster wavefunctions in Fig. \ref{fig:a+aJJ4}. The experimental threshold of these systems with respect to the $^8$Be~ground state are $-0.092$ MeV for $\alpha+\alpha$ and $17.255$ MeV for $^7\mathrm{Li}+$p, thus the lowest two $2^+$ states are below the proton separation energy. Both $2^+$ states have larger peaks at longer distances, suggesting alpha clustering, whereas the $2_2^+$ state is more spatially expanded. 
In addition, Ref.~\cite{sargsyanlbgs2022} finds a large overlap between the $^7\mathrm{Li}+$p and the lowest $2^+$ state, and a comparatively small but non-negligible overlap with the $2_2^+$ state (Fig. \ref{fig:a+aJJ4}b). This suggests that the intruder $2^+_2$ behaves almost as $\alpha+\alpha$, and differs from $2^+_1$ in its single-particle content.

It is noteworthy that the excitation energy of the $0_3^+$ state in Fig. \ref{fig:JJ0}a converges to $20.1\pm1.5$ MeV and has a structure similar to the doublet 2$^+$ states and an isospin of $T=1$. This state is likely to be the isobaric analogue of the lowest 0$^+$ state in $^8$Li and $^8$B nuclei predicted by recent \textit{ab initio} calculations \cite{NavratilRQ2010,McCrackenNMQH2021} and experimental measurements \cite{MitchellRJB2013}.  

In summary, the investigation of short-lived states in the low-lying spectrum of $^8$Be has important implications for $\beta$-decay studies of $^8$Li and $^8$B nuclei, and sheds new light on the discrepancy of calculated mass-8 $\beta$-decay GT strengths as compared to the experimentally deduced values \cite{sargsyanlbgs2022}. The SA-NCSM with continuum framework has allowed the study of the underlying structure of the $^8$Be nucleus 
and the $\alpha+\alpha$ system 
in unprecedentedly large model spaces, further corroborating earlier predictions and experimental results suggesting ``new" low-lying states. Ultimately, confirming the existence of these states could refine measurements of $\beta$~decays and advance the search for new physics.

\section{Uncertainty quantification}
\label{sec:UQ}

\subsection{Deducing local potentials with uncertainties for input to reaction models}

In recent years, the Bayesian methods have become an important tool to provide uncertainty quantification in nuclear physics \cite{WesolowskiKFP2016,MelendezWF2017,SchunckMHS2015,UtamaCP2016,LovellN2018,Pruitt:23}. One such application uses the overlaps calculated from the SA-NCSM (discussed in Sec.~\ref{sec:SFovlps}) as input to Bayesian analysis to obtain parameterization for local nucleon-nucleus (NA), deuteron-nucleus (dA), and alpha-nucleus ($\alpha$A) 
potentials of the Woods-Saxon type. These potentials typically include a central WS term~\eqref{eq:WS_pot} and a spin-orbit potential, and can be readily used as input to various reaction models, including those discussed in Sec.~\ref{sec:reactions}. 

The methodology has been first applied to \textit{ab initio} GFMC calculations in Ref.~\cite{Brida:11} for deducing nucleon-nucleus WS potentials,
but here we take full advantage of the Bayesian technique, coupled with the SA-NCSM framework and its treatment of single- and multi-nucleon clusters. 
This provides local inter-cluster potentials with uncertainties, rooted in \emph{ab initio} calculations.

The first applications have shown promising results when using the \textit{ab initio} SA-NCSM $\braket{^7\mathrm{Li}}{^6\mathrm{Li}+\rm{n}}$ overlaps for deducing local NA potentials \cite{Dudeck2021prc}. In particular, the Bayesian analysis allows for detecting correlations among the WS parameters, while the posterior distribution of these parameters provides, in turn, a probability distribution for the n+$^6$Li ANC of the $^7$Li ground state. For example, from this distribution, the study determines ANC of 1.48(3) fm$^{-1/2}$ for the $p_{1/2}$ wave and 1.90(3) fm$^{-1/2}$ for the $p_{3/2}$ wave, with a total ANC of 2.41(4) fm$^{-1/2}$, which agrees with the extrapolated SA-NCSM estimate of 2.5(1) fm$^{-1/2}$ \cite{sargsyan:23}, as well as with the experimentally deduced range of 1.26–2.82 fm$^{-1/2}$ \cite{GulamovMN1995}.

Similarly, Ref.~\cite{BECKER2025123203} discusses dA and $\alpha$A inter-cluster potentials, with a focus on the d$+\alpha$ and $\alpha+^{12}$C systems.  For d$+\alpha$, this study explores uncertainties in the parameterization first of the underlying chiral interaction and then in the WS potential deduced from the SA-NCSM $\braket{^6\mathrm{Li}}{\alpha+\rm{d}}$ overlaps.  These uncertainties are propagated to various
reaction observables, namely scattering phase shifts, cross sections, partial widths, and resonance
energies for d$+\alpha$. The outcomes suggest that for a fixed chiral parameterization, the  uncertainties in the deduced WS potential lead to reaction observables that are well constrained. However, there is a higher degree of sensitivity to the chiral parameterization, as also discussed next.
These studies open new opportunities for uncertainty quantification of reaction observables from light to medium-mass nuclei, starting with  \textit{ab initio} wavefunctions.

\subsection{Uncertainties in reaction observables from the chiral potential parameterization}

\begin{figure}[th]
    \centering
    \includegraphics[width=1\linewidth]{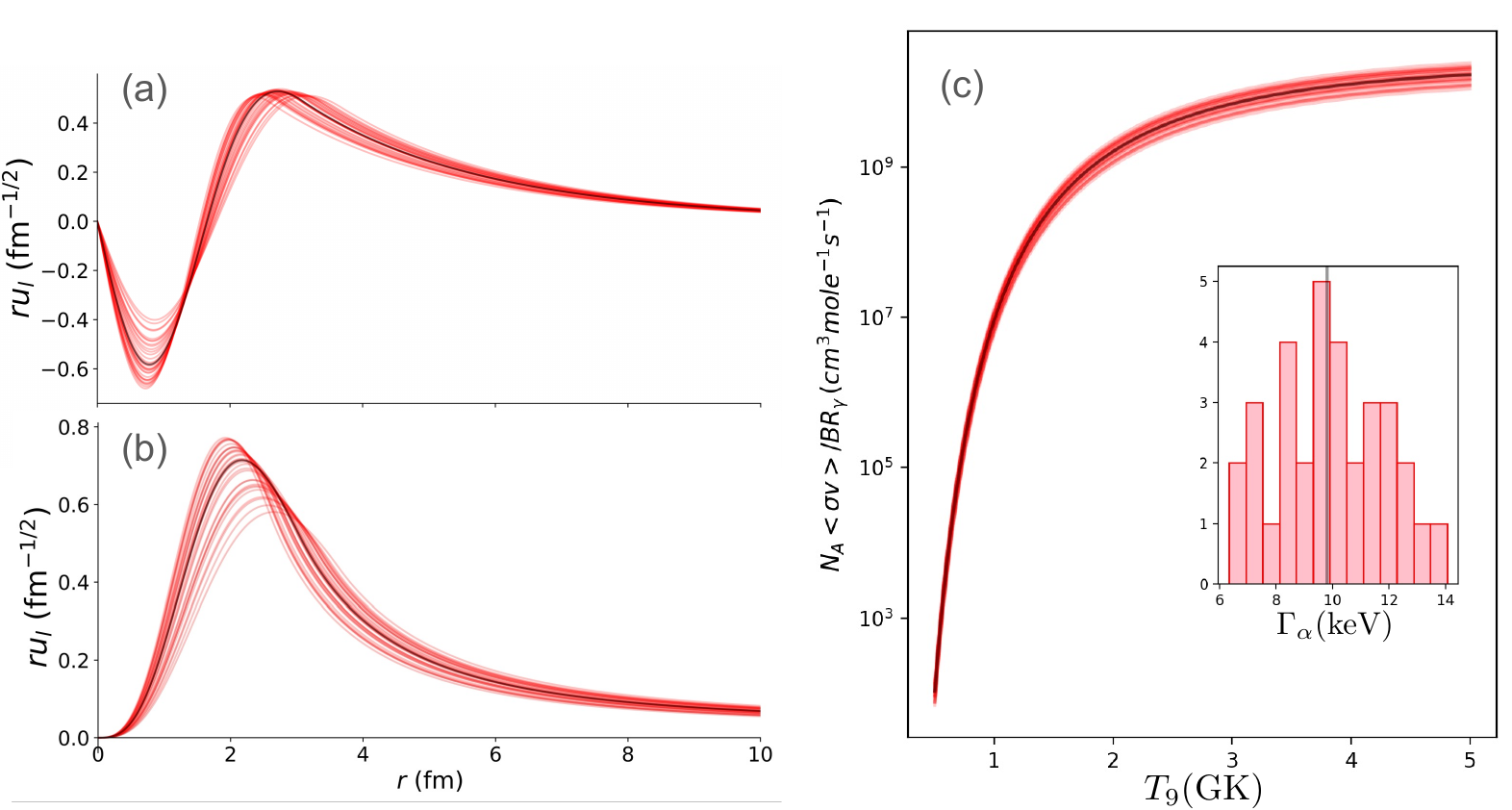}
    \caption{$\alpha+$d spectroscopic amplitude for the (a) $^3S_1$ partial wave in the $^6$Li ground state, and (b) $^3D_3$ partial wave in the $3^+_1$ resonance of $^6$Li, as functions of the cluster separation $r$. Figures adapted from~\cite{BeckerLED22}. (c) Capture reaction rate of d on $^4$He relative to the $\gamma$ branching ratio $BR_\gamma$ (in $\mathrm{cm}^{3}\mathrm{mol}^{-1} \mathrm{s} ^{-1}$) through the 0.7117-MeV $3^{+}_1$ resonance in $^{6}$Li  as a function of the temperature (in GK). The reaction rate is calculated with the narrow-resonance reaction-rate formula~\eqref{eq:narrowres}, which uses as input the alpha partial width $\Gamma_{\alpha}$, with probability distribution shown in the inset. All results are  obtained from \textit{ab initio} SA-NCSM wavefunctions computed with the NNLO$_{\textrm{opt}}$ chiral potential (black) and 32 sets of chiral potential parameterization (red).
    }
    \label{fig:da6Li}
\end{figure}

As mentioned above, the SA-NCSM adopts an inter-nucleon force, which is modeled in the chiral EFT framework and parameterized in terms of the low-energy constants (LECs) that encapsulate the high-energy physics unresolved at nuclear scales. At the next-to-next-to-leading-order, there are fourteen LECs (Sec.~\ref{sec:SANCSM}). As a first step toward a robust uncertainty quantification that considers the probability distribution of the LECs, Ref.~\cite{BeckerLED22} simultaneously varies all $14$ LEC parameters within $\pm 10\%$ of the corresponding NNLO$_{\textrm{opt}}$ parameterization, and uniformly draws $32$ samples in the 14-dimensional parameter space following a Latin hypercube design \cite{lhs}.  This means that only uncorrelated uncertainties in the values of the LECs are considered, providing an upper bound for the errors (we note that other uncertainties related to the nuclear force, for example, in the chiral expansion parameter and in truncating the chiral expansion, are not considered).  The goal is to analyze the impact of the chiral potential parameterization on $\alpha$ clustering in $^6$Li and on the deuteron capture reaction rate upon small perturbations of the nuclear force (Fig.~\ref{fig:da6Li}).

Ref.~\cite{BeckerLED22} calculates d-$\alpha$ spectroscopic amplitudes, following the procedure outlined in Ref.~\cite{DreyfussLESBDD20} and Sec.~\ref{sec:alpha}, where both clusters are kept frozen in the $0s$ shell. 
This study focuses on the $1^+_{\rm g.s.}$ ground state of $^6$Li, as well as its first $3^+_1$ excited state, which is a low-lying resonance at $0.71$ MeV above the $\alpha$-d threshold \cite{TILLEY20023}. The SA-NCSM wavefunctions utilized for these states are detailed in \cite{BeckerLED22}. Specifically, the $1^+_{\rm g.s.}$ state is comprised of all shapes corresponding to the valence shell, 1-shell 2-particle excitations and 2-shell 1-particle excitations, while thirteen dominant shapes are expanded up to $10$ HO shells. 
We note that for the NNLO$_{\rm opt}$ parameterization, the most dominant shape
comprises about 86\% of the $^6$Li ground state, whereas thirteen shapes recover about 97\% of the total probability, and that these probabilities remain practically the same across all parameterizations considered in \cite{BeckerLED22} (cf. Ref.~\cite{BeckerLEDLSD25}).

As discussed in Ref.~\cite{BeckerLED22}, for the $^3S_1$ partial wave, varying the LECs leads to about $20\%$ variations in the calculated ANCs ($C_{0}=1.45$--$2.07$ fm$^{-1/2}$) and $10\%$ variations in the spectroscopic factor (Fig. \ref{fig:da6Li}a). This tracks with the $\pm 10\%$ variation in the LECs. Interestingly, the height of the second peak, which is located near the nuclear surface and informs the probability of surface cluster formation, remains fixed for all the parameterizations (including the case of NNLO$_{\textrm{opt}}$), while only its position slightly varies across the LEC sets.
As for the $^3D_3$ partial wave (Fig. \ref{fig:da6Li}b),  the $^3D_3$ spectroscopic factors vary approximately at the $15\%$ level, which is practically the same as for the $^3S_1$ partial wave. However, $\alpha$ widths of the $3^+$ state range from $\Gamma_{\alpha}=6.34$ keV to  14.05 keV, which is about $\pm$40\% from the NNLO$_{\rm opt}$ value calculated for this particular channel (Fig. \ref{fig:da6Li}c, inset, dashed red line). We note that $C_0$ and $\Gamma_{\alpha}$ are reported for a single channel without taking excitations of the clusters into account (these were included, e.g., in Ref. \cite{PhysRevLett.114.212502}, which reports $C_0$=2.695 fm$^{-1/2}$ for $1^+_{\rm g.s.}$ and a width for $3^+_1$ of 70 keV). Also, the  calculated $\Gamma_{\alpha}$ should not be compared directly to the experimental width of 24(2) keV \cite{TILLEY20023}, which gives the  natural width that also includes $\gamma$ emissions. Overall, as evident from Fig.~\ref{fig:da6Li}b, varying the LECs affects not only the wavefunction tail, which informs how spatially extended the system is, but also impacts the magnitude of the peak and its location, thereby affecting the surface $\alpha$ clustering~\cite{BeckerLED22}.

Of particular interest to this study is that the LEC parameterizations further induce a change in the deuteron capture reaction rate $^{4}{\rm He}({\rm d},\gamma)^{6}{\rm Li}$ for temperatures $ \gtrsim 1.5$ GK (see Fig. \ref{fig:da6Li}c, for the rate relative to the $BR_{\gamma}$). The reaction rate is calculated using the narrow-resonance formula of Eq.~\eqref{eq:narrowres} along with the alpha partial width distribution determined from the $\pm10\%$ LECs variation. The results suggest that
if $\Gamma_\alpha << \Gamma_\gamma$, the reaction rate would be sensitive to the probability for alpha (deuteron) decay and the underlying  chiral parameterization.
This outcome along with the results discussed in this section provide illustrative examples that speak to the need for tighter constraints on realistic inter-nucleon interactions for predictions in the continuum.

\section{Conclusions}

In summary, we discussed recent \textit{ab initio} developments made possible in the SA-NCSM with continuum framework that can reach ultra-large shell-model spaces in light through medium-mass nuclei, representing a fully microscopic unified approach to determining structure and reaction observables. 
The underlying theme of this review is to expose the key role of symmetries in overcoming present computational resource limitations, thereby enabling solutions in ever larger model spaces and heavier nuclei. To illustrate this, 
we explored the role of multi-particle excitations that give rise to large deformation in low-lying $0^+$ states -- an interesting phenomenon,  especially important in neutron-rich isotopes and in the proximity of the drip line. Specifically, using a chiral EFT interaction, we explored structure properties of $^{4,6,8,10}$He, $^{21-34}$Mg, and $^{6-9}$Li isotopes, with a focus on nuclear collectivity, clustering, multi-particle excitations, and spectroscopic factors.

In addition, the ability of the SA-NCSM to describe spherical, deformed and cluster structures on the same footing is critical to calculating single-nucleon and alpha induced reaction cross sections, and to deriving nucleon-nucleus potential rooted in first principles (fully or partially). These applications were discussed here in light of the SA-NCSM/GF, SA-RGM, and Feshbach projection approaches for the astrophysically relevant energy regime, along with the  multiple scattering method at intermediate energies, for targets from alpha through the deformed $^{20}$Ne and $^{24}$Mg to the medium-mass $^{40}$Ca. For such descriptions, we utilized the SA-NCSM with continuum for determining the microscopic structure of reaction fragments and, in turn, for calculating reaction cross sections, illustrated here for neutron and proton elastic scattering, alpha and deuteron capture, and alpha knock-out reactions. In addition, we showed  the impact of alpha clustering on reactions of significance to nuclear astrophysics, as well as on beta decays and beyond-the-standard-model physics. 
Furthermore, as these approaches build upon first principles, they can probe features of the underlying inter-nucleon interaction that are relevant to reactions and can quantify their uncertainties.

The developments presented in this review 
have built upon the work of many pioneers and collaborators over many years. They 
are enabled by important features of the SA-NCSM with continuum framework, including:
\begin{itemize}
\item The SA-NCSM adopts a complete basis that is organized according to inherent symmetries in atomic nuclei, the deformation-related \SU{3} and the shape-related symplectic \SpR{3} symmetry: 
each \SpR{3}-preserving subspace describes a microscopic nuclear shape, including its deformation, rotations,
and energetic surface vibrations.

\item  Small model spaces are sufficient to describe less deformed shapes, but omit the vibrations of largely deformed/clustering shapes; this is what is remedied through the use of the SA basis through a well-prescribed  procedure detailed in Ref.~\cite{LauneyDSBD20}. Namely, the SA-NCSM includes all possible configurations in small model spaces, while including the necessary vibrations of largely deformed/clustering shapes in large model spaces, enabling the study and prediction of various observables
for spherical and deformed open-shell nuclei.

\item Single-particle and collective degrees of freedom enter on an equal footing. The reason is that all particle-hole configurations (or equivalently, all possible shapes) are included up to some number of HO excitations $N_{\rm C}$. This captures single-particle effects, manifested as mixing of shape configurations (often this mixing occurs predominantly in the valence shell). In addition, this captures less deformed shapes, and only partly largely deformed shapes. The collectivity enters through the  enhancement of the model space  to large $N_{\rm max}>N_{\rm C}$, which fully develops the largely deformed/clustering shapes.

\item For the SA basis, the center-of-mass motion can be factored out exactly. This is critical for modeling reactions. Indeed, in the SA-NCSM with continuum framework, the reach of large model spaces enables  couplings to the continuum, through otherwise inaccessible excitations. At the same time, the correct asymptotics are included through the exact Coulomb eigenfunctions. This provides translationally invariant inter-cluster interactions and the corresponding reaction observables, illustrated here for single-nucleon, deuteron, and alpha projectiles.

\item Only a few shapes dominate in low-lying nuclear states, thereby studying these shapes provides further insight into the physics of nuclei, their structure and dynamics \cite{DytrychLDRWRBB20, BeckerLEDLSD25}. Whereas typical SA-NCSM calculations include hundreds of shapes, while ensuring convergence of observables with model-space sizes and minimal dependence on $\hw$.
\end{itemize}

In short, with the help of HPC resources,  using the SA basis in {\it ab initio} studies represents a powerful tool to model the structure and reactions of nuclei. It is manageable and expandable -- with ongoing developments -- to larger model spaces imperative for reaction descriptions and to heavier nuclear systems, utilizing the predictive power of the {\it ab initio} approach at each stage. As such, the symmetry-adapted approach provides a unified framework to describe and predict structure and reaction phenomena for an ever-growing domain within the nuclear chart.

\section*{Acknowledgments}

We acknowledge very useful discussions with D.J. Rowe, J.P. Draayer, J. Wood,  G. Rosensteel, T. Dytrych, D. Langr,  Ch. Elster, K. Yoshida,  K. Ogata, G. Potel, S.T. Marley, M. Burrows, J.-P. Linares Fernandez, K.S. Becker, and N.C. Thompson. This work was supported by the U.S. National Science Foundation (PHY-1913728, PHY-2209060,OIA-2327385), as well as in part by
the U.S. Department of Energy (DE-SC0023532, DE-FG02-93ER40756), and in part by the National Nuclear Security Administration through the Center for Excellence in Nuclear Training and University Based Research (CENTAUR) under grant number DE-NA-0004150. 
This work performed in part under the auspices of the U.S. Department of Energy by Lawrence Livermore National Laboratory under Contract DE-AC52-07NA27344, with partial support from LDRD projects 19-ERD-017 and 24-ERD-023.
This material is based upon work supported by the U.S. Department of Energy, Office of Science, Office
of Nuclear Physics, under the FRIB Theory Alliance award DE-SC0013617.
This work benefited from high performance computational resources provided by LSU (www.hpc.lsu.edu), the National Energy Research Scientific Computing Center (NERSC), a U.S. Department of Energy Office of Science User Facility at Lawrence Berkeley National Laboratory operated under Contract No. DE-AC02-05CH11231, as well as the Frontera computing project at the Texas Advanced Computing Center, made possible by National Science Foundation award OAC-1818253.

\bibliographystyle{elsarticle/elsarticle-num}
\bibliography{sancsmc_refs}

\end{document}